\documentstyle[aps,epsfig]{revtex}

\begin{document}

\preprint{}

\draft 

\newcommand{\be}{\begin{equation}} 
\newcommand{\ee}{\end{equation}} 
\newcommand{\ba}{\begin{eqnarray}} 
\newcommand{\ea}{\end{eqnarray}} 

\newcommand{\etal}{{\em et al.}}

\title{ One-neutron removal reactions on light neutron-rich 
nuclei}

\author{E.~Sauvan$^{1}$\thanks{Present address: CPPM, Marseille, 
France.}, 
F.~Carstoiu$^{1,2}$, 
N.~A.~Orr$^{1}$\thanks{orr@lpccaen.in2p3.fr},
J.~S.~Winfield$^{1}$\thanks{Present address: INFN-LNS, via Sofia 44, 
I-95123, Catania, Italy.}, 
M.~Freer$^{3}$, 
J.~C.~Ang\'elique$^{1}$, 
W.~N.~Catford$^{4}$, 
N.~M.~Clarke$^{3}$, 
M.~Mac~Cormick$^{5}$\thanks{Present address: IPN, Orsay, France.}, 
N.~Curtis$^{3}$, 
S.~Gr\'evy$^{6}$\thanks{Present address: LPC, Caen, France.}, 
C.~Le~Brun$^{1}$\thanks{Present address: LPSC, Grenoble, France.}, 
M.~Lewitowicz$^{5}$,  
E.~Li\'egard$^{1}$,
F.~M.~Marqu\'es$^{1}$, 
P.~Roussel-Chomaz$^{5}$, 
M.-G.~Saint Laurent$^{5}$, 
M.~Shawcross$^{4,}$\thanks{Present address: Physics Department, 
University of 
Notre Dame, Indiana 46556, USA.}, 
 }

\bigskip

\address{ 
$^1$ Laboratoire de Physique Corpusculaire, IN2P3-CNRS, ISMRA et  
Universit\'e 
de
Caen, F-14050 Caen cedex, France \\ 
$^2$ IFIN-HH, P.O. Box MG-6, 76900 Bucharest-Magurele, Romania \\ 
$^3$ School of Physics and Astronomy, University of Birmingham, 
Birmingham B15 
2TT, 
United Kingdom \\ 
$^4$ School of Physics and Chemistry, University of Surrey, Guildford, 
Surrey, 
GU2 7XH, United Kingdom \\ 
$^5$ GANIL, CEA/DSM-CNRS/IN2P3, BP 5027, F-14021 Caen cedex, France \\ 
$^6$ Institut de Physique Nucl\'eaire, IN2P3-CNRS, F-91406 Orsay cedex, 
France 
}

\maketitle

\newpage

\begin{abstract}

A study of high energy (43--68~MeV/nucleon) one-neutron removal reactions 
on a 
range of
neutron-rich psd-shell nuclei (Z = 5--9, A = 12--25) has been undertaken. 
  The
inclusive longitudinal and transverse momentum distributions for the core 
fragments,  
together with
the cross sections have been measured for breakup on a carbon target. 
Momentum
distributions for reactions on tantalum were also measured for a subset of
nuclei.
  An extended version of the Glauber model incorporating second 
order noneikonal 
corrections
to the JLM parametrisation of the optical potential has been used to 
describe 
the
nuclear breakup, whilst the Coulomb dissociation is treated within first 
order 
perturbation
theory.  The projectile structure has been taken into account via shell 
model
calculations employing the psd-interaction of Warburton and Brown.  Both 
the 
longitudinal and transverse momentum distributions, together with 
the integrated cross sections were well 
reproduced by these calculations and spin-parity assignments are thus 
proposed 
for $^{15}$B, $^{17}$C,
$^{19-21}$N,  $^{21,23}$O, $^{23-25}$F.  
 In addition  to the large spectroscopic amplitudes for the $\nu2$s$_{1/2}$
intruder configuration in the $N=9$ isotones,$^{14}$B and $^{15}$C, 
significant $\nu2$s$_{1/2}^2$ admixtures appear to occur in the ground state of 
the
neighbouring $N=10$ nuclei $^{15}$B and $^{16}$C. 
 Similarly,  crossing  the N=14 subshell, the occupation of the
$\nu2$s$_{1/2}$ orbital is observed for $^{23}$O, $^{24,25}$F.  
 Analysis  of the longitudinal and transverse momentum 
distributions reveals that 
both carry spectroscopic information, often of a complementary nature.
The general utility of high energy nucleon removal reactions as a
spectroscopic tool is also examined.

\end{abstract}

\pacs{PACS number(s): 24.10.-i, 25.60.-t, 25.60.Dz, 25.60.Gc}

\newpage

%%%%%%%%%%%%%%%%%%%%%%%%%%%%%%%%%%%%%%%%%%%%%%%%%%%%%%%%%%%%%%%%%%%%%%%%%
%%%%%
%%

\section{Introduction}

High energy heavy-ion projectile fragmentation has been investigated now for 
some 
25 years
\cite{sco80,sto84}.  The initial emphasis centred on reactions employing 
stable beams  at
relativistic energies \cite{gre75,ols83} and the observed momentum 
distributions 
(typically Gaussian in form with FWHM $\sim$ 200~MeV/$c$) were 
interpreted 
within a
statistical description as reflecting the Fermi momentum of the removed 
nucleons
\cite{gol74}.  Further refinements lead to more sophisticated models 
incorporating the
peripheral nature  of the reaction process \cite{Hufner,fri83}.  Beyond 
relatively simple
considerations, such as the surface cluster model \cite{fri83}, the 
projectile 
structure 
played no role in the fragment distributions. 

More recently, the investigation of fragmentation reactions using 
radioactive  
beams has
lead to the recognition of narrow fragment momentum distributions (FWHM 
$\sim$
50~MeV/$c$) as a signature of the extended valence nucleon density
distributions in halo nuclei \cite{orr97}.  Originally these momentum 
distributions were
assumed, in terms of the transparent limit of the Serber model 
\cite{Serber}, 
to be a
direct mapping via the Fourier transform, of the valence nucleon(s) 
wave function. 
Detailed investigation of the role played by the reaction mechanisms, 
particularly in
the case of the single-neutron halo nucleus $^{11}$Be \cite{ann93,ann94} 
demonstrated that
such an interpretation was not generally applicable.  The longitudinal 
core fragment
momentum distributions have been suggested \cite{orr92,orr95,kel95} to be 
less 
influenced
by such effects and as such to represent a cleaner probe.  As foreseen in 
the 
original
work of Serber \cite{Serber} and more explicitly in the case of 
heavy ions by  
H\"ufner
and Nemes \cite{Hufner}, the requirement of core survival drives few 
nucleon 
removal to
probe essentially only that part of the valence nucleon(s) wave function 
residing outside
the core \cite{han96,Esbensen}.  Consequently, in the spirit of the 
spectator-core model
of Hussein and McVoy \cite{Hussein}, various Glauber-type approaches to 
modelling the
dissociation of energetic beams of nuclei far from stability have been 
developed 
\cite{Sag90,Sag94,han96,Esbensen,Hencken,oga97,Nego,Tostev,Yabana,par00}. 
The 
essential results
are that the momentum distributions, as first recognized by Bonaccorso and 
Brink
\cite{bonacc5} and Sagawa and Yazaki \cite{Sag90}, reflect 
the angular momentum ($l$) of the removed nucleon, 
whilst the
corresponding cross section may provide a measure of the associated
spectroscopic strength \cite{Han00}.

Recently, experimental measurements of single-nucleon removal reactions 
including
$\gamma$-ray detection have demonstrated that significant population of 
core 
fragment
excited states may occur 
\cite{Han00,nav98,aum00,gui00,nav00,mad01,Lola02,enders02}.  
The inclusion of 
the core
states within the theoretical framework \cite{Tostev,Tostev2}, as foreseen
in the original work of Sagawa and Yazaki \cite{Sag90}, has lead 
to the proposal
that such reactions (often termed ``knockout'' \cite{Han04}) may be used as 
a 
spectroscopic tool
\cite{Tostev,nav98}.  To date, however, this approach has been largely 
confined to
selected weakly bound halo and near dripline systems  
\cite{nav98,aum00,gui00,nav00,mad01,Lola02,enders03}.   

The objective of the present study was to undertake a systematic study of 
single-neutron
removal reactions over a range of neutron-rich nuclei.  For these 
purposes the 
light
psd-shell nuclei were selected as the nuclear  structure could be 
reliably 
calculated
within the shell model. The region encompassed a number of nuclei of 
current 
interest
 and the production rates were relatively high. 
 The 
use of high
energy nucleon-removal reactions as a spectroscopic tool could thus be 
verified on a
number of near stable nuclei with well known ground state structures.  
Additionally, the
structural evolution with isospin across the psd-shell, as expressed in 
the 
core fragment
observables, could be explored.  As will be demonstrated, even inclusive 
measurements of
the core fragments when executed using a broad range high acceptance 
spectrograph offer a
means to survey changes in structure over a wide range of isospin in a 
single 
measurement.

Earlier measurements of halo nuclei have suggested that the influence of  
nuclear and
Coulomb dissociation on the core fragment longitudinal momentum 
distributions 
is
relatively weak \cite{orr92,orr95,kel95,bau98}.  In order to explore 
further 
such
effects, both Carbon and Tantalum targets have been used in the present 
work. 
The results
obtained for the longitudinal momentum distributions and cross sections 
on the 
Carbon target have been briefly described elsewhere \cite{sau00}.  Here further 
details of the
experimental techniques are given along with a detailed account of the 
theoretical models.  In addition, the results obtained for the transverse momentum
distributions using the Carbon target are presented, as are the longitudinal
distributions from reactions on Tantalum.

The paper is organized as follows.  The experimental setup and techniques 
are 
described 
in Section II and the experimental results are presented in Section III.  
Sections IV to
VII are devoted to the description of eikonal based modelling of nuclear 
and Coulomb
dissociation. The results and comparison to calculations are discussed in 
Section VIII.  
A discussion, in the light of the present results, of the utility of 
single-nucleon removal as a spectroscopic tool is presented in section 
IX.  
The paper concludes (Section X) with a
summary and perspectives.  Explicit analytical formulae pertaining to 
the 
Coulomb dissociation calculations are presented in Appendix A.  
The results of the shell model and cross section calculations 
are tabulated in Appendix B.

%%%%%%%%%%%%%%%%%%%%%%%%%%%%%%%%%%%%%%%%%%%%%%%%%%%%%%%%%%%%%%%%%%%%%%%%%
%%%%%

\section{Experimental method} 

The experiment was performed at the GANIL coupled cyclotron facility. The 
secondary beams
were produced via the fragmentation of an intense ($\sim 1\mu$Ae) 70 
MeV/nucleon
$^{40}$Ar$^{17+}$ primary beam on a 490 mg/cm$^2$ thick carbon target.  
The 
reaction
products were collected and analysed (B$\rho$=2.880 Tm)  using the SISSI 
device
\cite{ann97} coupled to the beam analysis spectrometer.  The resulting 
secondary beam was
composed  of $^{12-15}$B, $^{14-18}$C, $^{17-21}$N, $^{19-23}$O and 
$^{22-25}$F nuclei
with energies between  43 and 68 MeV/nucleon and intensities ranging from 
$\sim$600~pps
($^{15}$C) to  $\sim$1~pps ($^{25}$F). The energy spread in the beam, as 
defined by the
acceptance of the beam analysis spectrometer, was $\Delta E/E$=1\%. 
Secondary 
reaction
targets of carbon (170 mg/cm$^2$) and tantalum (190 mg/cm$^2$) were used.

Owing to the large energy spread in the secondary beam, an energy-loss 
spectrometer was required to
undertake a high resolution measurement of the core fragment momentum  
distributions
\cite{orr97}.  In the present case, the SPEG spectrometer \cite{bia89} 
was 
employed and
operated at a central angle of zero degree in a dispersion matched mode for 
which an intrinsic
resolution of $\delta p/p$=4.5$\times$10$^{-4}$  (FWHM) was achieved.  
The final resolution, including target effects was  $\delta 
p/p$=3.5$\times$10$^{-3}$ 
(FWHM).  The
overall momentum acceptance of the spectrometer was 7\%, which permitted 
the 
momentum
distributions for the fragments resulting nuclei from one-neutron removal 
on all the
nuclei of interest to be obtained in a single 
setting for each
target (B$\rho_{SPEG}$=2.551 Tm for the carbon target and  
B$\rho_{SPEG}$=2.615 Tm for
tantalum).

Importantly, the broad angular acceptances of the spectrometer --- 
4$^{\circ}$ 
in the
horizontal (bending) and vertical planes --- provided, in the case of the 
carbon target, 
for almost complete collection (see below) of the core fragments, 
obviating 
any
ambiguities in the integrated cross sections and longitudinal momentum 
distributions that
may arise from limited transverse momentum acceptances 
\cite{rii93,riihab,kel97}. In the
case of the tantalum target (Section VIIC and ref. \cite{sauthese}), the  
effects of
multiple scattering and Coulomb deflection resulted in greatly 
reduced effective
transverse momentum acceptances that curtailed the extraction of any 
reliable 
transverse
momentum distributions or cross sections.  An investigation of the 
effects of 
incomplete
transverse acceptances on the longitudinal momentum distributions is 
presented 
elsewhere \cite{sauthese,car00}.   

Ion identification at the focal plane of SPEG was achieved using the  
energy 
loss derived
from a gas ionisation chamber and the time-of-flight between a thick 
plastic 
stopping
detector and the cyclotron radio-frequency.   Additional information was  
provided by the
residual energy measurement furnished by the plastic detector and the 
time-of-flight with
respect to a thin-foil microchannel plate detector located at the exit of 
the 
beam
analysis spectrometer. Two large area drift chambers straddling the focal 
plane of SPEG
were employed  to determine the angles of entry of each ion and, 
consequently, 
allowed the
focal plane position spectra to be  reconstructed.  The calibration in 
angle 
was performed using a tightly 
collimated
beam and a calibrated mask placed at the entrance to the spectrometer.  
The 
momentum of
each particle was derived from the reconstructed focal plane position.  
Calibration in
momentum was achieved  by removing the reaction target and stepping the 
mixed 
secondary
beam of  known rigidity along the focal plane.  This procedure also 
facilitated a
determination of the efficiency across the focal plane for the collection 
of 
the reaction
products ---  the range of angles accepted by the spectrograph being 
restricted 
at the
limits of the focal plane.  Where necessary, corrections were then applied 
to 
account for
any  reduction in efficiency.

The angles of incidence on target of the secondary beam particles were  
determined using
two beam tracking detectors (each comprising four 10$\times$10~cm$^2$ 
drift 
chambers)
located in the analysis line of the spectrometer.  The calibration in 
incident 
angle was
derived from the trajectories reconstructed at the focal plane of the 
spectrometer for a
measurement made with the target removed and the spectrometer set to the 
same 
rigidity as
the analysis line. Consequently, the transverse momentum distributions 
for the 
core
fragments could be reconstructed on an event-by-event basis from the 
incident 
projectile
angle and the core fragment outgoing angle.  The beam envelope was 
approximately Gaussian
in form and characterized by half-angles of $\Delta 
\theta_{1/2}$=0.35$^{\circ}
$ 
 and
$\Delta \phi_{1/2}$=0.5$^{\circ}$ (FWHM) in the horizontal and vertical 
planes
respectively.  Owing to the superior resolution in the  determination of 
the 
angles in the
bending (horizontal) plane of the spectrometer  
($\delta\theta$=0.1$^{\circ}$, 
$\delta
\phi$=0.4$^{\circ}$ (FWHM)), the transverse  momentum distributions 
presented here 
have been
reconstructed in this plane ($p_x$), with final resolutions including 
multiple
scattering of some 5\% 
being
achieved.   

The intensities of the various components of the secondary beam were 
derived 
from a
measurement of the primary beam current, which was recorded continuously 
during the
experiment using a non-interceptive monitor, with respect to runs taken 
with 
the
secondary reaction target removed and the spectrometer set to the same 
rigidity as the
beamline. A redundant check was also  provided by counting rates in two 
microchannel
plate detectors placed in the incident secondary beam: one, as noted 
above, at 
the exit
of the beam analysis spectrometer and another located on the upstream 
side of the 
secondary
reaction target. The final cross sections were determined using an 
average of 
these
three normalisations and the uncertainties quoted include contributions 
from 
both the
statistical uncertainty and that arising from the determination of the 
secondary beam
intensities (typically $\sim$ 7\%).

The core fragment angular distributions were inspected to ensure that no 
events were lost
due to the finite acceptance of the spectrometer ($\Delta \Omega$=5 msr).  
 For the reactions on
the C target only 
in the
cases of the broadest distributions did the losses exceed a few percent. 
The 
number of
events not detected was estimated based on extrapolations of Gaussian  
adjustments to the
measured angular distributions.  Corrections were also applied where 
necessary 
to those
nuclei falling near the limits of the focal plane (see above).  The final 
uncertainty in
the cross section includes an estimate ($\sim$ 5-10\%) of the 
uncertainties in 
these two
corrections.  As discussed in Section VIIC, for the reactions on the Ta 
target extremely 
broad angular distributions
were encountered, which precluded any reliable estimates of the transverse 
momentum distributions and
cross sections to be made. 

In many instances --- most notably $^{14}$B, $^{15,16}$C and $^{17,18}$N 
--- asymmetric longitudinal
momentum distributions
exhibiting low momentum tails 
were observed (Fig. \ref{fig:sysglau}).  
The origin of these events is discussed in Section VIIIA.
The cross sections reported here 
include these
events (typically $\sim$ 5\%).

To compare the measured distributions with the theoretical ones, all 
broadening effects
inherent in the measurements should be taken into account. These effects 
arise 
from
the differential energy losses of the projectile and the fragment in the 
target, energy
and angular straggling in the target and  the detector and spectrometer 
resolutions.  In
addition, for the longitudinal momentum  distributions the Lorentz 
transformation 
from the laboratory to projectile frame of reference must be taken into
account. In order to provide an estimate of the relative importance of 
these 
effects on
momentum distributions, a Monte Carlo  based simulation was developed.  
As an 
example,
an  evaluation of the effects for two nuclei with distributions 
representative 
of those encountered here are provided in Table 
\ref{tab1:widths}. 
The overall effect in the case of the longitudinal distributions 
is some 10--13\% and is dominated by the Lorentz 
contraction.  In the
case of the transverse momentum distributions, the 
broadening is relatively weak (at most some  5\%) and is dominated 
by the angular resolution of the spectrometer.

The widths of the momentum distributions were derived from Gaussian
adjustments to the central region ($\sim$FWHM) of each distribution, thus 
avoiding 
any bias introduced by low
momentum tails.  The use of other lineshapes (such as a Lorentzian), or a 
simple statistical
analysis \cite{kel95} produced essentially identical results.  The 
widths, noted 
FWHM$_{cm}$ in
Table \ref{tab2:res_c}, are quoted in the projectile frame and have been 
corrected for
the various broadening effects discussed above.

%%%%%%%%%%%%%%%%%%%%%%%%%%%%%%%%%%%%%%%%%%%%%%%%%%%%%%%%%%%%%%%%%%%%%%%%%
%%%%%
%%
\section{Results}

The core fragment longitudinal momentum distributions measured using a 
carbon 
target are displayed in Fig. \ref{fig:sysglau}.  In order to facilitate
their comparison,  each distribution is displayed over the same total 
momentum range of 
700~MeV/$c$.  The estimated widths, taking into account the various 
experimental effects
discussed above, are shown in Fig. \ref{fig:pxppar} and are listed in Table 
II 
(FWHM$^{cm}_{pz}$).
The corresponding single-neutron removal cross sections are tabulated in 
Table II 
and the evolution along the isotopic chains is presented in Fig.
\ref{fig:sefglau}.

The transverse momentum distributions $p_x$ from breakup on the carbon 
target 
are presented in Fig. \ref{fig:sysglaupx}.  The 
same 
total momentum range (here 600~MeV/$c$) has been used to display all the 
results to 
facilitate the comparison. The widths extracted from the measured 
distributions, 
taking into account all broadening effects (Section II), are listed in 
Table II (FWHM$^{cm}_{px}$).  A comparison of the widths of the 
longitudinal
and transverse momentum distributions is provided in Fig \ref{fig:pxppar}.  
The transverse 
distributions are systematically somewhat broader, a feature already 
observed
in reactions with stable beams \cite{silk88}.  More interestingly,
the transverse distributions exhibit the same trends as the longitudinal 
distributions,
suggesting that the sensitivity to the projectile structure expected for 
the latter is
also present to a similar degree in the transverse distributions.
As will be discussed further in Sections V-B and VIII, the transverse 
momentum 
distributions present
somewhat more complex forms than the longitudinal distributions.  It 
should be stressed that the widths
quoted here are only meant to serve as a comparative guide.

For a number of nuclei ($^{14}$B, $^{15-17}$C and  $^{17-19}$N) the 
core fragment longitudinal momentum distributions from
reactions on a Ta target were also measured (Fig. \ref{fig:systag}).
The momentum widths, which are seen to be almost identical with 
those
obtained on the C target, are listed in Table \ref{tab3:res_ta}.
As outlined in Section II, the corresponding transverse momentum 
distributions were observed to
be
much broader than the
acceptances of the spectrograph (see, for example, Fig. \ref{fig:px_ta}).  
As such no reliable cross sections could
be derived.

As noted in Tables \ref{tab2:res_c} and \ref{tab3:res_ta}, a number of the 
nuclei 
studied in the present 
work have 
been investigated elsewhere.
In the cases of 
$^{14}$B and $^{15,17,18}$C measurements made at similar energies on a Be 
target by 
Bazin {\em et al.} \cite{baz95,baz98} found longitudinal momentum 
distributions 
with widths in good agreement with those presented here.
The associated single-neutron removal cross sections are, however, 
significantly
smaller than those measured in the present work.  As pointed out in our 
earlier 
paper \cite{sau00}
the origin of this discrepancy lies in the rather limited
acceptances of the A1200 fragment separator.  This is clearly apparent 
from the 
transverse momentum distributions presented here (Fig. \ref{fig:sysglaupx}) 
\cite{sauthese}
and is also confirmed by more recent measurements of  
$^{14}$B \cite{gui00} and  
$^{16,17}$C  \cite{mad01} undertaken using the high acceptance
S800 spectrograph (Table \ref{tab2:res_c}).

Very recently a measurement for  
$^{16}$C (83~MeV/nucleon) on a C target has been reported \cite{Yam03}.  The 
results,
which were obtained using a new time-of-flight
technique to deduce the core momentum distribution \cite{Kan02}, are in
very good agreement with those reported here.

A measurement has also been carried out with the time-of-flight technique, 
at an energy somewhat higher (72~MeV/nucleon) 
than that employed 
here, of the breakup of $^{23}$O by carbon \cite{Kan02}.  The resulting 
$^{22}$O longitudinal momentum
distribution is in good agreement with the present work.  Moreover, as 
discussed
in Section VIII-D, the 
corresponding
one-neutron removal cross section (233$\pm$37~mb) is in good agreement with
that calculated here (Tables \ref{tab2:res_c} and 
\ref{tab4:res_o23})\footnote{As 
no reliable beam intensity normalization was 
available in the present study 
for $^{23}$O, no experimental cross section could be extracted}.   

At much higher energies ($\sim$900 MeV/nucleon), the single-neutron breakup 
of 
$^{17}$C by
carbon
has also been measured \cite{bau98,cor01}.  A width (FWHM) of 
141$\pm$6~MeV/$c$, 
slightly broader than
reported here, was extracted \cite{bau98} together with
a cross section of 129$\pm$22~mb \cite{cor01}.
Recently an experiment using the same setup as that of refs. 
\cite{bau98,cor01}
has been carried out to explore single-neutron removal on neutron-rich 
isotopes of 
N, O and F at some 900 MeV/nucleon. The preliminary results for the momentum
distributions --- in particular for $^{23}$O (Section VIII-D) --- are in 
very
good agreement with those reported here \cite{priv_lola}.  

In the case of reactions on the Tantalum target (Table \ref{tab3:res_ta}),
there is a relative paucity of work with high-Z targets with which
comparison can be made.  Indeed, of the nuclei measured here, the literature
reports results for only $^{14}$B and $^{15,17}$C.  In the case of the
former, very good agreement is found with the measurement made employing the
S800 spectrometer \cite{gui00}, whilst the earlier work of Bazin {\em et 
al.}
found a somewhat narrower core momentum distribution \cite{baz98}.
The core momentum distribution measured for $^{15}$C 
in the same experiment \cite{baz98}
is in reasonable agreement with that observed here.
In the case of the heavier isotope, $^{17}$C, a preliminary study 
\cite{orrrnbiii}
reported a momentum distribution consistent with that found in the present 
work.
Unfortunately the more complete study of Maddalena and
Shyam \cite{mad01c} does not quote any widths for the momentum 
distributions. 

In order to examine quantitatively the relationship between the 
projectile
structure and the measured distributions we now turn to a detailed 
development
of the necessary reaction theory.

%%%%%%%%%%%%%%%%%%%%%%%%%%%%%%%%%%%%%%%%%%%%%%%%%%%%%%%%%%%%%%%%%%%%%%%%%
%%%%%
%%
%%%%%%%%%%%%%%%%%%%%%%%%%%%%%%%%%%%%%%%%%%%%%%%%%%%%%%%%%%%%%%%%%%%%%%%%%

\section{Extended Glauber model for nuclear dissociation}

The ensemble of data presented in the previous section constitutes a 
test not only for the
structure models of neutron rich nuclei in p--sd shell but also for the 
description of the
reaction mechanisms involved. The aim of this section is to provide a 
model 
that, starting
from realistic projectile wave functions   and taking properly into 
account 
the reaction
mechanism, can explain the momentum distributions and cross sections. 
In the formal development, the principle features of which 
 are
similar to approaches developed  by  Esbensen \cite{Esbensen}, Bertsch 
{\em et al.}
\cite{Hencken,Bertsch}, Negoita {\em et al.} \cite{Nego} and Tostevin 
\cite
{Tostev,Tostev2}, we have attempted to 
retain the role 
played by the
wave function via its Wigner transform as this reveals  clearly the 
momentum content.
The formulae obtained for the  fragment momentum 
distributions and cross sections
explicitly display, in the spirit of the Glauber model, the distorting 
functions arising from
the reaction mechanism. The details concerning the  practical calculation 
of 
some basic
ingredients --- in particular the S-matrix elements --- are also 
provided.

We shall assume that the ground state of the projectile $(J^{\pi})$ could 
be  
approximated
by a superposition of configurations of the form  $[I_c^{\pi}\otimes 
nlj]^{J^{\pi}}$, where
$I_c^{\pi}$ denotes the core states and $nlj$ are the quantum numbers
specifying the  single particle
wave function of the last neutron, taken here as  Woods-Saxon wave 
functions 
evaluated using the 
effective
separation energy $S^{eff}_n=S_{n}+E_{ex}^c$ ($E_{ex}$ being the excitation 
energy of 
the core state).
We neglect coupling of core states to the final state and dynamical 
excitation 
of excited
core states in the reaction. In this approximation the reaction can 
populate a 
given core
state only to the extent that there is a nonzero spectroscopic factor 
$C^2S(I^\pi_c,nlj)$ in the
projectile ground state. When more than one configuration contributes to 
a 
given core state,
then the total cross section for one neutron removal is written, 
following
refs. \cite{Tostev,Tostev2}, as an 
incoherent
superposition of single particle cross sections:
\be
\sigma_{-1n}(I^\pi_c)=\sum_{nlj} C^2S(I^\pi_c,nlj) \; 
\sigma_{sp}(nlj,S^{eff}_n).
\label{eq1}
\ee
The total inclusive one-neutron removal cross section 
($\sigma_{-1n}^{Glauber}$) is then the sum over the cross sections to all 
core states. A similar relation holds for the momentum distributions. 
The coordinate system 
used in the calculations is sketched in Fig. \ref{fig:figglau}, whereby 
the 
impact parameters for the neutron and the core are given by,
\ba
\vec b_n=\vec R_{\perp}+\frac{A_c}{A_c+1}\vec s,\\
\vec b_c=\vec R_{\perp}-\frac{1}{A_c+1}\vec s.
\label{eq2}
\ea
We shall neglect recoil effects, so that ($A_c\gg 1$),
\ba
\vec b_n=\vec b+\vec s+O\left(\frac{1}{A_c}\right),\\
\vec b_c=\vec b +O\left(\frac{1}{A_c}\right).
\label{eq3}
\ea

Since the core states are not coupled by the interaction, the core plays 
a 
spectator role. Thus it is sufficient to consider only
the neutron degrees of freedom as described by the  wave 
function
$\Psi_{JM}$ corresponding to the coupling scheme $[[\vec
l\otimes\vec\frac{1}{2}]\vec j\otimes\vec I_c]\vec J$. We assume only one 
bound
state in the projectile and we need to consider
density matrix elements of the form:
\be
\rho^J_{M'M}(\vec r)=<\Psi_{JM'}\vert\Psi_{JM}>_{spins},
\label{eqfn1}
\ee
where $<>_{spins}$ means average over spin coordinates. A little angular
momentum algebra leads to,
\be
\rho^J_{M'M}(\vec
r)=\sum_{\lambda\mu}(-)^{J-2j-I_c-\frac{1}{2}}\frac{\hat l^2\hat j^2\hat 
J}
{\sqrt{4\pi}}C^{l l \lambda}_{0 0 0}C^{J ~\lambda~ J}_{M' \mu M}
W\left(jljl,\frac{1}{2}\lambda\right)W\left(JjJj,I_c\lambda\right)
R_{nlj}^2(r)Y_{\lambda\mu}(\hat r),   
\label{eqfn2}
\ee
with the property,
\be
Tr\rho=\frac{1}{\hat 
J^2}\sum_M\rho_{M'M}^J\delta_{M'M}=R_{nlj}^2(r)/4\pi\equiv\rho_{val}(r),
\label{eqfn3}
\ee
with $R_{nlj}(r)$ the radial part of the single particle wave function. 
It is useful to introduce also the projected density,
\ba
\tilde \rho_{M'M}^J(\vec s)&=&\int_{-\infty}^{\infty}dz\rho_{M'M}^J(\vec 
r),\\
\tilde\rho_{val}(s)&=&\frac{1}{\hat J^2}\sum_M\tilde\rho^J_{M M}(s).
\label{eqfn4}
\ea

If $S_c$ and $S_n$ are the S-matrices in impact parameter representation 
for 
the core 
and neutron-target interactions, the absorption cross section (or 
stripping) 
is  given by,
\be
\sigma_{abs}=\frac{1}{\hat J^2}\int d\vec b\sum_k\sum_M\vert 
<\phi_{k}\vert 
\Omega_{abs}\vert\Psi_{JM}>\vert^2,
\label{eq4}
\ee
where $\phi_k$ are scattering states and $\Omega_{abs}^2=(1-\vert 
S_n\vert^2)\vert S_c\vert^2$ is the transition operator for neutron 
absorption.  As the neutron is absorbed,  only the 
scattering states are available to the core. In this case the closure 
relation
  is,
\be
\sum_k\vert\phi_k><\phi_k\vert=1.
\label{eq5}
\ee
Combining Eqs. (\ref{eq4}) and (\ref{eq5}) we obtain,
\be
\sigma_{abs}=\frac{1}{\hat J^2}\sum_M\int d\vec b\int d\vec r \; 
\left(1-\vert 
S_n(\vec b+\vec s)\vert^2\right) \; \vert S_c(b)\vert^2 \; 
\rho_{MM}^{J}(\vec r),
\label{eq6}
\ee
which we rewrite in the form
\be
\sigma_{abs}=\int d\vec b d \vec s \; (1-\vert S_n(\vec b+\vec 
s)\vert^2)\vert 
S_c(b)\vert^2\tilde\rho_{val}( s)\equiv\int d\vec s 
D_{abs}(s)\tilde\rho_{val}(s),
\label{eq11}
\ee
where $D_{abs}$ is  the distortion kernel  \cite{Nego}. 
This kernel 
has a very intuitive physical interpretation: it is the convolution 
product of 
the survival 
probability for the core and the probability for the absorption of the 
neutron, 
as originally defined by H\H{u}fner and Nemes \cite{Hufner}. 
In fact the general absorption operator can be decomposed as follows,
\be
1-\vert S_c\vert ^2\vert S_n\vert ^2=\vert S_c\vert ^2(1-\vert S_n\vert 
^2)+\vert S_n\vert ^2(1-\vert S_c\vert ^2)+(1-\vert S_n\vert ^2)(1-\vert 
S_c\vert ^2).
\label{eq12}
\ee
where the first term corresponds to neutron absorption, the second to 
core 
absorption and the double scattering term to the absorption of both the core 
and the neutron. 
For the last two (inelastic) channels the total cross section is formally 
identical 
to Eq. (\ref{eq11}) with  appropriate redefinitions of the distorting 
kernels.

For diffraction a similar formula to Eq. (\ref{eq4}) holds, except that 
the transition 
operator is replaced in this case by $\Omega_{diff}=S_cS_n-1$. We shall 
again assume 
a structureless continuum and treat it via sum rules. In addition it is 
assumed that the projectile has only one 
bound state. Since the scattering states should be orthogonal 
to the ground state, 
the closure relation is in this case:
\be
\sum_M\vert\Psi_{JM}><\Psi_{JM}\vert+\sum_k\vert\phi_k><\phi_k\vert=1.
\label{eq13}
\ee 
Clearly, the ``one bound state assumption'' will lead to an overestimation 
of the
breakup cross section, since any additional bound state  will
subtract cross section \cite{JeffCDCC}.
 The orthogonality condition allows us to replace the ``$-1$'' 
in the 
definition 
of the diffraction transition operator by any function which does not 
couple 
the nucleon 
coordinates. For the purpose of convergence,  the most convenient form 
for this operator 
is $\Omega_{diff}=S_c(S_n-1)$. With this, the total cross section for 
diffraction is,
\be
\sigma_{diff}=\frac{1}{\hat J^2}\int d\vec b \; 
\left\{\sum_M<\Psi_{JM}\vert\vert\Omega_{diff}\vert^2\vert\Psi_{JM}>-\sum
_{M'M
}\vert<\Psi_{JM}\vert\Omega_{diff}\vert\Psi_{JM'}>\vert^2 \right\},
\label{eq14}
\ee
which demonstrates that the diffraction cross section is given by the 
fluctuation of 
the transition operator in the ground state. In terms of the density 
matrix 
(\ref{eqfn1}-\ref{eqfn4}) the nuclear diffraction cross section is,
\ba
\sigma_{diff}=\int d\vec bd \vec s \;\vert S_c(b)\vert^2\vert 1-S_n(\vec 
b+\vec s)\vert^2 \: \tilde\rho_{val}(s) \nonumber \\
-\frac{1}{\hat J^2}\sum_{M'M}\int d \vec b \; \left|\int d\vec 
s\Omega_{diff}(\vec b,\vec s)\tilde\rho_{M'M}^J(\vec s)\right|^2.
\label{eq15}
\ea
The first term is similar in structure to (\ref{eq11}) and provides the 
main 
contribution 
to the diffraction cross section. The second is a small correction 
arising 
from 
the orthogonality requirement. One can further simplify this correction
by observing that it arises essentially from the diagonal elements of the 
density matrix. 
Indeed, if one considers the following integral,
\be
I=\int d \vec s dz \; \Omega\left(\vec b,\vec s \right) 
\;\rho_{M'M}^J(\vec r),
\label{eq16}
\ee
and expands the  density in multipoles,  
\be
\rho_{M'M}^J(\vec 
r)=\sum_{\lambda\mu}\rho_{\lambda\mu}(r)P_{\lambda\mu}(\theta)e^{i\mu\phi
},
\label{eq17}
\ee
where $\cos\theta=z/r$ and $\phi=\phi(x,y)$ and assume that the 
dependence 
of 
$\Omega$ on angles of $\vec s$ is weak and may be neglected,  we are 
left with,
\be
I=\sum_{\lambda\mu}\int sdsd\phi \; \Omega(b,s)f_{\lambda\mu} \: 
e^{i\mu\phi}\sim\delta_{\mu 0}.
\label{eq18}
\ee
Therefore the main contribution to such integrals comes from multipoles 
with 
$\mu=0$ 
and only the diagonal elements of density matrix  contribute ($M'=M$, see 
(\ref{eqfn2})). 
Our final formula for diffraction cross section is then,
\ba
\sigma_{diff}=\int d\vec bd \vec s \; \vert S_c(b)\vert^2\vert 1-S_n(\vec 
b+\vec s)\vert^2 \; \tilde\rho_{val}(s) \nonumber\\
-\frac{1}{\hat J^2}\sum_{M}\int d \vec b \; \vert S_c(b)\vert^2 \left| 
\int 
d\vec s(1-S_n(\vec b+\vec s))\tilde\rho_{MM}^J(s)\right|^2.
\label{eq19}
\ea
As  mentioned earlier the first term is dominant and can be written in 
the 
form,
\be
\sigma_{diff}^{(1)}=\int d\vec s \; D_{diff}(s) \; \tilde\rho_{val}(s),
\label{eq20}
\ee
with $D_{diff}\simeq D_{abs}$.

Further simplifications arise if one observes that for spin independent 
transition operators one can neglect the intrinsic spin of the nucleon 
and the
coupling to core spin. In 
this case the 
total spin $J$ is replaced by the angular momentum $l$. Thus, for example 
the 
density matrix (\ref{eqfn2}) becomes,
\be
\rho_{m'm}^l(\vec r)=R_l^2(r)\sum_{\lambda\mu}\frac{\hat 
l^2}{\sqrt{4\pi}\hat 
\lambda} \; C^{l l \lambda}_{0 0 0} \; C^{~l ~l ~\lambda}_{-m' m \mu} \; 
Y_{\lambda\mu}(\hat r),
\label{eq21}
\ee
which has the same properties as in Eqs. (\ref{eqfn3}-\ref{eqfn4}).
The total cross sections for absorption and diffraction become,
\ba
\sigma_{abs}&=&\int d\vec b d \vec s \; \left(1-\vert S_n(\vec b+\vec 
s)\vert^2\right)\vert S_c(b)\vert^2 \; \tilde\rho_{val}( s),\\
\sigma_{diff}&=&\int d\vec bd \vec s \; \left| S_c(b)\right|^2 \left| 
1-S_n(\vec 
b+\vec s)\right|^2 \; \tilde\rho_{val}(s)\nonumber\\
&&-\frac{1}{\hat l^2}\sum_{m}\int d \vec b \; \vert S_c(b)\vert^2 \left| 
\int 
d\vec s \; \left(1-S_n(\vec b+\vec s)\right) \; 
\tilde\rho_{mm}^l(s)\right|^2.
\label{eq22}
\ea

Finally one may note that the absorption cross section is fully 
equivalent with 
the
corresponding cross section obtained by Hussein and McVoy in 
the core
spectator model \cite{Hussein}.

%%%%%%%%%%%%%%%%%%%%%%%%%%%%%%%%%%%%%%%%%%%%%%%%%%%%%%%%%%%%%%%%%%%%%%%%%
%%%%%
%%
\section{Momentum distributions}
\subsection{Longitudinal momentum}
As final state interactions are neglected the scattering states are taken 
as plane waves. 
The basic matrix element which gives the localization probability in 
momentum space is,

\be
\frac{dP}{d\vec k}=\frac{1}{(2\pi)^3\hat J^2}\sum_{MM_Im_s} \left| 
\langle\, 
e^{-i\vec 
k\vec r}\chi_{\frac{1}{2}m_s}(\sigma)\chi_{IM_I}(\xi)\vert\Omega(\vec b,
\vec s) \vert \Psi_{JM}(\vec 
r,\sigma,\xi) \,\rangle \right| ^2.
\label{eq23}
\ee
Integration is taken over nucleon ($\vec r,\sigma$) and core spin ($\xi$)
coordinates. The intrinsic momentum 
distribution ($W_0(\vec k)$) is obtained by choosing $\Omega\equiv 1$. 
For 
absorption and diffraction one uses the appropriate operators defined in 
the previous 
section. After applying some angular momentum algebra, one finds,
\be
\frac{dP}{d\vec k}=\frac{1}{(2\pi)^3\hat l^2}\sum_{m_l} \left|\int d\vec 
r 
e^{i\vec k\vec r} \; \Omega(\vec b,\vec s) \; R_l(r) \; Y_{lm_l}(\hat r) 
\right|^2.
\label{eq24} 
\ee
The longitudinal momentum distribution ($k_z$) is obtained by integration 
over 
the unobserved 
components ($k_x$ and $k_y$). To obtain closed formulae it is useful to 
introduce 
the partial Wigner transform of the wave function,
\be
w_m(\vec s,k_z)=\frac{1}{\sqrt{2\pi\hat l^2}}\int dz 
e^{ik_zz}R_l(r)Y_{lm}(\hat r),
\label{eq25}
\ee
in terms of which the total Wigner transform is,
\be
W(\vec s,k_z)=\sum_m\vert w_m(\vec s,k_z)\vert^2,
\ee
with the properties,
\ba
\int dk_zW(\vec s,k_z)=\tilde \rho_{val}(s),\\
\int d \vec s dk_z W(\vec s,k_z)=1,\\
\int d\vec s W(\vec s,k_z)=W_0(k_z).
\label{eq26}
\ea
 The longitudinal momentum distribution for absorption 
is thus calculated as,
\be
\left(\frac{d\sigma}{dk_z}\right)_{abs}=\int d\vec b d\vec s \; 
\left(1-\vert 
S_n(\vec b+\vec s)\vert^2\right)\vert S_c(b)\vert^2 \; W(\vec 
s,k_z)\equiv\int 
d\vec s \;D_{abs}(s)W(\vec s,k_z).
\label{eq27}
\ee
For diffraction the situation  is somewhat more complicated 
and  three terms must be considered,
\ba
\left(\frac{d\sigma}{dk_z}\right)_{diff}^{(1)}&=&\int d\vec b d\vec s \; 
\vert 
S_c(b)\vert^2\vert
S_n(\vec b+\vec s)-1\vert^2 \; W(\vec s,k_z),\\
\left(\frac{d\sigma}{dk_z}\right)_{diff}^{(2)}&=&\sum_{mm_1m_2}\int d\vec 
b 
\;\vert S_c(b)\vert^2\int d \vec s \; w_{m_1}(\vec s,k_z)w^{*}_{m_2}(\vec 
s,k_z)\nonumber\\
&&\times\int d \vec s_1 \; \tilde\rho_{m_1m}(s_1)\; \left(S_n(\vec b+\vec 
s_1)-1\right)\int d\vec s_2\tilde\rho^{*}_{m_2m}(s_2) \; 
\left(S^{*}_n(\vec 
b+\vec s_2)-1\right),\\
\left(\frac{d\sigma}{dk_z}\right)_{diff}^{(3)}&=&-2\Re\sum_{mm_1}\int 
d\vec b 
\vert S_c(b)\vert^2\int d\vec s_1 \; \tilde\rho_{m_1m} \; 
\left(S^*_n(\vec 
b+\vec s_1)-1\right)\nonumber\\ 
&&\times\int d\vec s_2 \; w_m(\vec s_2,k_z)w^*_{m_1}(\vec s_2,k_z)\; 
\left(S_n(\vec 
b+\vec s_2)-1\right).
\label{eq28}
\ea
To a good approximation one can use again the fact that the main 
contribution 
to diffraction arises 
from the diagonal part of the density matrix. In this case the second and 
third terms become,
\ba
\left(\frac{d\sigma}{dk_z}\right)_{diff}^{(2)}&=&\sum_m\int d\vec b\vert 
S_c(b)\vert^2\int d \vec s\vert w_m(\vec s,k_z)\vert^2 \left| \int d\vec 
s_1 
\left(S_n(\vec b+\vec 
s_1)-1\right)\tilde\rho_{mm}(s_1)\right|^2,\nonumber\\
\left(\frac{d\sigma}{dk_z}\right)_{diff}^{(3)}&=&-2\Re\sum_m\int d\vec 
b\vert 
S_c(b)\vert^2
\int d\vec s_1\tilde\rho_{mm}(s_1)\left(S^*_n(\vec b+\vec 
s_1)-1\right)\nonumber\\
&&\times\int d\vec s_2\vert w_m(\vec s_2,k_z)\vert^2 \left(S_n(\vec b+\vec 
s_2)-1\right).
\label{eq29}
\ea
It can be checked that integration over $k_z$, in the range
($-\infty,+\infty$) leads exactly to the integrated 
cross sections derived in the previous section. It has been shown by 
Bonaccorso and Bertsch \cite{bonacc3}
that the error introduced by these limits is less that 5 \% for the beam
energies considered in the present experiment.

%%%%%%%%%%%%%%%%%%%%%%%%%%%%%%%%%%%%%%%%%%%%%%%%%%%%%%%%%%%%%%%%%%%%%%%%%
%%%%%%%%%%%%%%%%%%%%%%%%%%%%%%%%%%%%%%%%%%%%%%%%%%%%%%%%%%%%%%%%%%%%%%%%%
%

\subsection{Transverse momentum}
Transverse momentum distributions ($k_x,k_y$) are obtained by projecting 
the
probability (\ref{eq24}) onto the axis of interest. For example we have,
\be
\frac{dP}{dk_x}=\frac{1}{2\pi \hat l^2}\sum_m \int_{-\infty}^{\infty}dy 
dz \;
\left| \int_{-\infty}^{\infty}dx e^{ik_x x}\Omega(\vec b, \vec 
s)R_l(r)Y_{lm}
(\hat
r)\right|^2,
\label{eqt1}
\ee
For the other transverse direction one simply exchange the indices
$x\longleftrightarrow y$.
The corresponding cross sections are given by
\be
\left(\frac{d\sigma}{dk_x}\right)_{abs}=\frac{1}{2\pi\hat l^2}\sum_m
\int d\vec b_n \left[1-\vert S_n(b_n)\vert^2\right]\int dy dz\left|\int 
dx 
e^{ik_x x}
S_c(\vert \vec b_n- \vec s \vert ) R_l(r) Y_{lm}(\hat r)\right|^2,
\label{eqt2}
\ee
and
\be
\left(\frac{d\sigma}{dk_x}\right)_{diff}=\frac{1}{2\pi\hat l^2}\sum_m
\int d\vec b \vert S_c(b)\vert^2\int dy dz\left|\int dx e^{ik_x x}
\left(S_n(\vert \vec b+ \lambda\vec s \vert )-1\right) R_l(r) Y_{lm}(\hat 
r)
\right|^2.
\label{eqt3}
\ee
with $\lambda=\frac{A_c}{A_c+1}$ as a measure of recoil effects. In the
derivation, we assume small excitation energies in a structureless 
continuum
(plane wave approximation) and neglect the orthogonality requirement
described in the preceding subsection. If $\lambda=1$, integration of 
equations
(\ref{eqt2}-\ref{eqt3}) over $k_x$ leads to the total cross sections 
(\ref{eq11}) and (\ref{eq20})
respectively. It should be noted  that the transverse momentum 
distributions 
are 
symmetric around $k_x=0$. There is no asymmetry in the nuclear breakup 
transverse
momentum distributions, principally due to the straight line 
approximation for
 the
trajectory and the inherent neglect of the conservation of the energy. This 
symmetry means that only 
half of the distributions need  to be computed. The necessity to evaluate a 
five 
dimensional
integral for each value of $k_x$, for two different transition operators 
and 
for
each core state renders the  calculations onerous. One may reduce the 
complexity  
by assuming that the angle dependence $(\vec b,\vec s)$ of the transition
operators is weak and can be replaced by an average value. For example, 
in the
case of diffraction one has,
\be
\bar{S_n}(b,s)=\frac{1}{2\pi}\int_0^{2\pi}d\phi
\left(S_n \left(\sqrt{b^2+\lambda^2s^2+2\lambda b 
s\cos\phi}\right)-1\right).
\label{eqt4}
\ee
The use of average transition operators allows a straightforward integration 
over
angles of $\vec b$ and reduces the integration domain to $(0,\infty)$ for
$x,y$ and $z$ variables. The computational time is thus reduced by one 
order of
magnitude and the precision in the calculation is significantly 
increased.
We have checked that averaged transition operators contain almost the 
same
transverse momentum components as the original ones, at least for impact 
parameters in the range of the strong absorption radius.

Equations (\ref{eqt2}) and (\ref{eqt3}) illustrate the
essential
difference between the longitudinal and transverse momentum 
distributions. 
The former are given essentially by the Wigner transform of the valence
nucleon wave function weighted by distortion kernels for absorption and
diffraction, whilst for transverse momentum distributions additional 
components
 appear due to nuclear interaction. The neutron-target interaction 
influences strongly the  transverse momentum distribution for diffraction 
and 
leads to a broader distribution as compared to that arising from 
absorption 
(see Fig.~\ref{fig:testpxpz}).

%%%%%%%%%%%%%%%%%%%%%%%%%%%%%%%%%%%%%%%%%%%%%%%%%%%%%%%%%%%%%%%%%%%%%%%%%
%%%%%
%%
\section{Coulomb dissociation}

In this section we describe the Coulomb dissociation 
within the framework of the eikonal approximation
which is formally equivalent to a first order perturbation theory. We use 
the
long wavelength approximation for the transition operator and obtain
a general formula for any electric multipolarity $(Elm)$. However, the 
calculations are done only for dipole and quadrupole transitions, since
these give rise to the principle contributions to the cross sections. 

The Coulomb excitation amplitude for a projectile in the  field of
 a target may be expressed in terms of the electric multipole matrix
elements characterizing the electromagnetic decay of nuclear states. If 
the
 charge distributions of the two nuclei do not overlap during the 
collision,
then the relative motion takes place along a classical Rutherford 
trajectory.
At very high energies, the trajectory is well approximated by a straight 
line
and the first order eikonal approximation should give reasonable results 
for
 momentum distributions and integrated cross sections.

Our starting point for the Coulomb amplitude ($f_C$), is the result 
of
Bertulani and Baur \cite{Bertul} in a first order eikonal approximation 
(formally
equivalent to the first order Born approximation),
\be
 f_C(\vec{Q},\vec{q})=i{Z_t \alpha \over \beta \gamma}k R^2
\sum_{lm} i^m \hat l \left({\omega \over c}\right)^l \; G_{Elm} \; 
\Phi_m(Q) 
\; {\cal M}(Elm),
\label{coul:eq1}
\ee
where ${\cal M}(Elm)$ are the matrix elements for electric transition of 
multipolarity $( lm )$
\be
 {\cal {M}}(Elm)=\sum_{j=1,2}Z_{j}\int\varphi_{f}^{*}(\vec{r})
r_{j}^{l} \; Y_{lm}(\hat{r}_{j}) \: \varphi_{i}(\vec{r})\: {\rm d}\vec{r},
\label{coul:eq2}
\ee
and $\varphi_{i(f)}$ are initial (final) state of the projectile. This is 
the 
long wavelength
limit of the transition operator written in terms of cluster coordinates 
$\vec{r}_j (j=1,2)$.
The two clusters are characterized by  masses ($m_j$), charges ($Z_j$), 
and 
 momenta
($\vec{k}_j$). The relative motion of the outgoing clusters is described 
in 
terms of the
relative momentum, 
$$\vec{q} =(m_2 \vec{k}_1-m_1 \vec{k}_2)/(m_1+m_2),$$ 	 
whilst the momentum
change in the scattering ($\vec{Q}$) is given by	 $Q = 
2k\sin{\theta\over 2}$, where
$\theta$ and $k$ are respectively the scattering angle and the incident 
momentum in 
the center of mass. The 
other notations
in formula (\ref{coul:eq1}) are:  the target charge, $Z_t$; the  velocity of 
the 
projectile in
units of the speed of light, $\beta$; the relativistic Lorentz factor, 
$\gamma$; 
the fine-structure
constant, $\alpha$; and the interaction radius, R. In addition  
$$\hbar\omega =\varepsilon + 
E_q
=\varepsilon+{\hbar^2q^2\over 2\mu},$$ 
is the excitation energy, i.e. the sum of the
(absolute) binding energy and the kinetic energy of the separated 
clusters. 
%The functions 
%$G_{Elm}=i^{l+m}\overline{G}(\beta,\gamma)$ are the relativistic 
%functions 
%defined by 
%Winther and Adler  \cite{Winth}, 
The  relativistic functions of Winther and Adler \cite{Winth} are used in 
the
form $G_{Elm}=i^{l+m}\overline{G}(\beta,\gamma)$ with the complex phase
factorized out, 
whilst the nondimensional functions 
$\Phi_m(Q)$, 
also defined in refs.
\cite{Bertul,Winth},  contain information on reaction mechanism whereby 
$$\Phi_m(Q) = 
\int_1^\infty
J_m (QRx) \: K_m\left({\omega R x\over c\gamma \beta}\right) \: x dx,$$  
and standard notation has been employed 
for the Bessel functions. The transition matrix elements (\ref{coul:eq2}) 
are 
calculated using a
simple shell model wave function for the ground state,
\be
\varphi_i(\vec{r}) = {\cal R}(r)Y_{LM}(\hat r).
\ee
Note the change in notation with respect to the previous sections for the 
quantum 
numbers of the
wave function: $L,M$ instead of $l,m$. If final state interactions are 
also 
neglected, 
the continuum
may be considered  to be structureless and the final state wave function 
may 
be treated as  
plane waves. 
This is a good approximation for small excitation energies. 

Under the above approximations, the matrix element in (\ref{coul:eq2}) is 
given by,
\ba
{\cal {M}}(Elm)=(Z_1\beta_1^l-(-)^lZ_2\beta_2^l)\int 
d\vec{r}~e^{-i\vec{q}\vec{
r}}~r^l~ Y_{lm}(\hat r)~ {\cal R}_L(r)~ Y_{LM}(\hat r)\nonumber\\ 
= \sqrt{4\pi} Z_l^{eff}~\hat l~\hat L~ \sum_{\lambda \nu}i^{-\lambda} 
\hat 
{\lambda}^{-1}
Y_{\lambda \nu}(\hat q)~ C^{L l \lambda}_{0~ 0~ 0} ~C^{L l \lambda}_{M m 
\nu} 
\int_0^\infty r^2
d r r^l j_{\lambda}(qr){\cal R}_{L}(r),
\label{coul:eq4}
\ea
 with an obvious notation for the 
effective charge $Z^{eff}$
and  $\beta_{1(2)} =m_{2(1)}/(m_1+m_2)$.  Since the spin 
orientations are not
specified, the  differential cross section for Coulomb excitation is 
obtained 
by averaging
 the square of the Coulomb amplitude over the magnetic projections,
\be
d^4\sigma = \hat L^{-1} \sum_{M}\left| f_C(\vec{Q}, \vec{q})\right|^2 \:
{d \vec{q} Q dQ \over (2\pi)^2 k^2}.
\label{coul:eq5}
\ee
The main contribution in (\ref{coul:eq4}) is given by dipole ($E1$) and 
quadrupole ($E2$) 
transitions, therefore only terms with $l=1$ or $l=2$ are included.
The complexity of Eq. (\ref{coul:eq5}) arise from the evaluation of 
integrals,

\be
f_{m_1m_2}(\xi) = 2 R^2 \xi^2 \int_0^\infty Q dQ 
\Phi_{m_1}^*(Q,\xi)\Phi_{m_2}
(Q,\xi), 
\label{coul:eq6}
\ee
where,
 $$ \xi = {\omega R \over \gamma v} = {\omega \over c} {R \over \beta 
\gamma},$$ is the
adiabaticity parameter, defined in terms of the excitation energy ($ 
\hbar\omega$) and the
minimum impact parameter $R$, given by  $$ R = R_p + R_t + {\pi Z_p Z_t 
e^2 
\over 4 E_{lab}
\gamma},$$ which includes a correction due to the deviation of the 
trajectory 
from a 
straight line \cite{Winth}. The function (\ref{coul:eq6}) is obtained in 
general by 
numerical integration except for the diagonal term which admits a simple 
analytical
expression. However, one can profitably perform first the integration 
over the 
azimuthal 
angle ($\phi$) of the relative momentum, $$ d \vec{q} = q^2 dq \sin 
\theta d 
\theta d \phi.$$
This integration automatically selects only the diagonal terms in 
(\ref{coul:eq6}) and,
$$ f_{mm} = \overline K_m (\xi),$$
with functions $\overline K$  given by:
$$  \overline K_0 (\xi) =  ( K_1^2 - K_0^2 ),$$
$$  \overline K_1 (\xi) = \frac{2}{\xi}  K_0 K_1 -  ( K_1^2 - K_0^2 ),$$
$$  \overline K_2 (\xi) = \frac{4}{\xi^2} K_1^2 +  ( K_1^2 - K_0^2 ),$$
where $K_m$  are modified Bessel functions of the first kind and 
$\overline K_{-m}= \overline K_m$.
After some simple calculation we obtain the following expression for the 
differential cross section:
\be
{d^2 \sigma \over q^2 dq \sin \theta d \theta } = 
{ Z_t^2 \alpha^2 \over \sqrt {4 \pi }}\sum (-)^{L+m} 
(-)^{{\lambda_2-\lambda_1+
l_1-l_2 \over 2}} \: \hat l_1^2 \hat l_2^2 \hat \lambda_1 \hat \lambda_2 
\hat S^{-1} \left({\omega \over c} \right)^{l_1-1} \left({\omega \over c} 
\right)^{l_2-1}
\label{coul:eq7}
\ee
$$\times~
Z_{l_1}^{eff}~ Z_{l_2}^{eff}~ \overline G_{l_1m}~\overline G_{l_2m}~
\overline K_m(\xi)$$
$$
\times~I_{Ll_1\lambda_1}(q)~ I_{Ll_2\lambda_2}(q)~ C^{L l_1 \lambda_1}_{0 
~0~ 0}
~C^{L l_2 \lambda_2}_{0~ 0~ 0}~ C^{\lambda_1 \lambda_2 S}_{0~ 0~ 0}
~C^{l_1 l_2 S}_{m -m 0} \left\{^{S l_1 l_2}_{L \lambda_2 
\lambda_1}\right\}
\overline Y_{S0}(\hat q),
$$
where $I_{Ll\lambda}(q)$ is a shorthand notation for the radial integrals
 appearing in
(\ref{coul:eq4}).
The summation runs over all quantum numbers except $L$. The angular 
dependence 
of the cross
section is given by functions $\overline Y_{S0}$ which are the standard 
spherical harmonics
defined without the phase factor $e^{im \phi}$ (where $\phi$ has already 
been 
integrated). The
energy dependence of the reaction mechanism is governed by the functions 
$\overline G$ and
$\overline K$. However, the magnitude of the cross section for a given  
multipolarity
transition is mainly determined by the effective charge. If the  clusters 
have 
equal charge
to mass ratio, then the dipole transition has  vanishing cross section in 
this 
approximation.
This is readily understood from classical arguments since in this case 
the 
dipole field
acts on the two clusters with the same force in the same direction and 
does not lead to
 dissociation. This is a consequence of the assumption of a well defined 
cluster
structure for the projectile. Experimentally an 
appreciable, but not complete
suppression of the $E1$ transition occurs. The interference term $E1E2$ does
not contribute to the total cross section. 

Various observables may be readily obtained from (\ref{coul:eq7}) by 
appropriate 
integration or change of variables. 
The total Coulomb dissociation 
cross section $\sigma_C = \sigma_{E1} + \sigma_{E2}$ can be obtained by 
numerical 
integration of equation (\ref{coul:eq7}). In practice we have used the 
closed 
formulae given in  Appendix A.
For momentum distributions a change of variables is made such that,
$$d\vec q~=~dq_x~dq_y~dq_z~=~2\pi q^2~dq~\sin\theta~d\theta~=
~2\pi~q_r~dq_r~dq_z.$$
The radial momentum distribution is obtained by integrating over the
longitudinal
momenta,
$${d\sigma\over q_r~dq_r}~=~\int_{-\infty}^{\infty} dq_z~\sum_{\alpha,S}
A_S^\alpha(q)\overline Y_{S0}\left({q_z\over 
q}\right)~=~2\sum_{\alpha,S=even}
\int_0^\infty A_s^\alpha (q)\overline Y_{S0}\left({q_z\over 
q}\right)dq_z,$$
where we have used a shorthand notation for the general cross section 
(\ref{coul:eq7}). 
In addition, $q^2=q_r^2+q_z^2$, $\cos\theta=q_z/q$,
$\sin\theta=q_r/q$ and ($\alpha$) denotes all summation indices appearing 
in 
(\ref{coul:eq7}), different from $S$. There is no $E1E2$ asymmetry in the 
radial momentum 
distribution since the interference term in (\ref{coul:eq7}) contains 
only odd 
S-values.
Similarly, one can demonstrate that the longitudinal momentum 
distribution 
takes the 
form,
$${d\sigma\over dq_z~}(q_z>0)~=~\sum_{\alpha,S}\int_{q_z}^\infty 
A_S^\alpha (q)\overline Y_{S0}\left({q_z\over q}\right)q~dq,$$
and
$${d\sigma\over dq_z~}(q_z<0)~=~\sum_{\alpha,S}\int_{\vert 
q_z\vert}^\infty 
(-)^S~A_s^\alpha (q)\overline Y_{S0}\left({q_z\over q}\right)q~dq.$$
This leads imediately to,
$${d\sigma_{E1,E2}\over dq_z~}(q_z<0)~=~{d\sigma_{E1,E2}\over 
dq_z~}(q_z>0),$$
$${d\sigma_{E1E2}\over dq_z~}(q_z<0)~=-~{d\sigma_{E1E2}\over 
dq_z~}(q_z>0).$$
From the above relations it can be seen that the loss of information
--- concerning e.g. the $E1E2$ interference term --- inherent when an 
integration 
over
all variables is performed, may be partially compensated for  by 
measuring the
longitudinal momentum distribution. 
 In an inclusive measurement, such as that performed here, only
the core-like particle momentum is measured. To compare with the data, we 
have
 to 
transform the theoretical momentum distribution which is given as a 
function
of $\it {relative}$ $\it {momentum}$ to a function of $\it {fragment}$ 
$\it
{momentum}$. 
This is done most easily in the projectile rest frame, taking into 
account 
momentum conservation,
$$\vec q =\beta_1\vec q_1-\beta_2\vec q_2$$
$$\vec q_1+\vec q_2=0,$$
and
$${d\sigma\over dq_{z1}}=\int 
dq_z~dq_{z2}~f(q_z)~\delta(q_z-\beta_1q_{z1}
+\beta_2q_{z2})~\delta(q_{z1}+q_{z2})~=~f(q_{z1}).$$
where $f$ is a generic notation for the theoretical momentum 
distribution. A 
similar formula holds for the other fragment.

%%%%%%%%%%%%%%%%%%%%%%%%%%%%%%%%%%%%%%%%%%%%%%%%%%%%%%%%%%%%%%%%%%%%%%%%%
%
%%%%%%%%%%%%%%%%%%%%%%%%%%%%%%%%%%%%%%%%%%%%%%%%%%%%%%%%%%%%%%%%%%%%%%%%%
%
\section{S-matrices and optical model potentials}\label{sec:smatrix}

The remaining physics is to describe interaction of the core and the 
removed nucleon with the
target. These enter through the associated S-matrices, $S_{c,n}$, 
expressed  as a 
function of impact
parameter. Previously, S-matrix calculations have been based on the 
optical 
limit 
of the eikonal
model \cite{Hencken,Nego,Tostev,nav98}. In this approach the 
nucleus-nucleus 
phase shifts are
entirely determined by nucleon-nucleon collisions in the density overlap 
volume.  The energy
dependence is dictated by the total nucleon-nucleon  cross sections 
$\sigma_{NN}$. However,
diffraction dissociation is sensitive to the refractive power of the 
optical 
potential and
these effects  are not very well controlled in the above approximation. 

Recently, a more
fundamental approach has been introduced by Bonaccorso and Carstoiu 
\cite{Bonacc} and 
Tostevin \cite{Tostev2}, whereby the G-matrix interaction of 
Jeukenne, Lejeune and
Mahaux (JLM) \cite{Jeuken}, which is obtained in a Brueckner-Hartree-Fock 
approximation from
the Reid soft core nucleon-nucleon potential, has been employed. 
This interaction is complex, 
density and energy
dependent and, therefore, provides simultaneously both the real and 
imaginary 
parts of the
optical potential. The optical potential calculation with this 
interaction is 
described in
detail in refs. \cite{Bonacc,Trache}. The  
single particle
densities for the core and target were generated in a spherical 
Hartree-Fock + 
BCS calculation
using the density functional of Beiner and Lombard\cite{Beiner}. The 
strength 
of the surface
term in the functional was  slightly adjusted  in order to reproduce the 
experimental
binding energy for each nucleus. The matter radii resulting from these 
calculations are
in good agreement with the existing experimental values of Liatard {\em 
et al.} 
\cite{liatard}
and with results of relativistic mean field calculations (RMF) by Ren 
{\em et al.}
\cite{ren95,ren96a,ren96b}, as shown in Fig. \ref{fig:mat-radii}. In Fig.
\ref{fig:mat-radii}, other experimental radii values derived from reaction 
cross sections at higher energy are also presented \cite{oza01,tan88}. 
Except for the case of oxygen isotopes, these values are systematically 
slightly smaller than those obtained by Liatard {\em et al.}.

The resulting optical potentials were renormalised in order to reproduce 
the 
total cross
section for neutron target interaction in an eikonal calculation 
including 
noneikonal
corrections up to second order. For the core target potentials the 
renormalisation constants
have been taken from ref. \cite{Trache}. It should  be noted that the 
potentials 
are strong and
the eikonal expansion (as defined by Wallace \cite{Wallace}) does not 
converge 
at low
energies. 

The
resulting S-matrix elements and transmission coefficients are displayed 
in Fig.
\ref{fig:smatrix} for n+$^{12}$C and $^{14}$C+$^{12}$C at energies of 30 
MeV 
and 30 MeV/nucleon
respectively. One sees clearly substantial changes in the shape and 
magnitude 
of all matrix
elements for the neutron if higher order noneikonal corrections are taken 
into 
account. These
effects are most pronounced at low impact parameters where neutron 
absorption
profile shows an important contribution from trajectories reflected inside 
the barrier
superimposed on a characteristic strong absorption at the nuclear surface. 
The
eikonal approximation in lowest order underestimates the interaction range 
and
the absorption in the nuclear interior. Given the surface dominance of the
breakup reactions, this approximation will lead to an underestimation of the
stripping and dissociation cross sections. 
The distortion 
kernels 
obtained from
these S-matrix elements are displayed in Fig. \ref{fig:dfunc}. It is  
clear that
diffraction dissociation is most sensitive to the noneikonal corrections. 
The 
asymptotic
behaviour of the distorting kernels is most affected and this has 
important 
consequences for the calculation of the 
momentum distributions,  since large impact parameters probe the 
low momentum
content of the projectile wave function. 

Finally, the neutron S-matrix  
 has been
checked against  known experimental total cross sections  for four 
targets \cite{Web}. The 
results are displayed in Fig. \ref{fig:jlm}. The comparison is reasonably 
good for 
all targets
except  tantalum  where the absorptive potential is too strong to 
be treated in
the eikonal approximation. Nevertheless, we have been able to find 
normalization constants
which reproduce, at least qualitatively the experimental cross sections. 
The second order eikonal calculation for n+$^{12}$C match reasonably well
recently evaluated data for elastic and reaction cross sections in the 
range
20--100 MeV\cite{chad} (Fig. \ref{fig:jlm}b).

%%%%%%%%%%%%%%%%%%%%%%%%%%%%%%%%%%%%%%%%%%%%%%%%%%%%%%%%%%%%%%%%%%%%%%%%%
%%%%%
%%
\section{Discussion}

A number of features are
apparent from the systematics of the core fragment  momentum 
distributions and 
associated single-neutron removal cross sections for reactions on Carbon 
(Figs. 1-3).  Firstly, the crossing 
of the N=8 shell 
and N=14 subshell closures are associated with a significant reduction in 
the widths of both
the longitudinal and transverse momentum  
distributions for $^{14,15}$B, $^{15}$C, $^{23}$O and $^{24,25}$F 
compared to
the neighbouring less exotic nuclei. 
Such behaviour is a clear
indication of the role played by the structure of the projectile owing
to the large valence $\nu$2s$_{1/2}$ admixtures expected
in the ground state wave functions (as discussed in detail below).  
Secondly, as a result of this and the weak binding of the valence neutron
$^{14}$B (S$_n$=0.97~MeV) and $^{15}$C (S$_n$=1.22~MeV) exhibit enhanced 
cross sections 
in comparison to the neighbouring isotopes, suggesting a  spatial
delocalisation of the valence neutron orbital\cite{sau00}.

In order to analyse quantitatively the measurements presented here, we 
now proceed to 
make a comparison with the results of calculations using the extended 
Glauber model described above coupled with the results of shell model 
calculations.

%%%%%%%%%%%%%%%%%%%%%%%%%%%%%%%%%%%%%%%%%%%%%%%%%%%%%%%%%%%%%%%%%%%%%%%%%
%%%%%%%%%%%%%%%%%%%%%%%%%%%%%%%%%%%%%%%%%%%%%%%%%%%%%%%%%%%%%%%%%%%%%%%%%

\subsection{Cross sections and longitudinal momentum distributions: 
Carbon target}

The spectroscopic factors $C^2S(I^\pi_c,nlj)$ employed here 
were calculated with the shell-model code OXBASH \cite{oxbash} using the 
WBP \cite{war92a} interaction within the 1$p$-2$s$1$d$ model space. 
In order to avoid energy-shift 
effects \cite{war92b} only pure 0, 1 or 2 $\hbar\omega$ excitations were 
considered. 
The older Millener-Kurath \cite{mil75}
interaction
(PSDMK) was also investigated and the results were found, for the nuclei 
examined here, 
to be comparable to those obtained using the WBP interaction.
Where known, the experimentally established spin-parity ($J^\pi$) 
assignments 
and core excitation energies \cite{as82,as86,as87,end90} have been used
for calculation of cross sections and momentum distributions.
 In all other cases, the shell-model 
predictions 
were employed.  It was found that the cross sections were relatively
insensitive to the excitation energies of the core states.
A detailed listing of the spectroscopic factors and calculated cross 
sections
are given in Appendix B\footnote{Since our original publication
\cite{sau00,sauthese} the $^{22,24}$F 
and $^{23}$O calculations 
have been revised.}.  

Aside from the spectroscopic factors the principle parameters entering
into the calculations were the S-matrices and the geometry of the 
Woods-Saxon
potential used to define the single-particle wave function.
As described in section VII, the JLM calculations of the S-matrices were
verified through comparison with measured total, elastic and reaction 
cross section data. The strength parameter of the single-particle potential 
was fixed by
fitting the known experimental one-neutron separation energy.
The radius and  
diffusivity of the Woods-Saxon
potential were fixed at $r_{ws}$=1.15~fm and $a_{ws}$=0.5~fm for the 
isotopes of B and C, and $r_{ws}$=1.2~fm,
and $a_{ws}$=0.6~fm for N, O and F.  These values were 
chosen to
provide a good global agreement with the measured cross sections. Better
agreement is obtained if for example radius and  diffusivity parameters are 
tuned locally.
The sensitivity to the choice of these parameters may be illustrated by 
the example of
$^{16}$C breakup at 50 MeV/nucleon, for which a change in geometry  to 
$r_{ws}$=1.20 fm and $a_{ws}$=0.65 fm   leads to an 
increase of 
about 27 \%  in the total one-neutron removal cross section.
As might be expected the shape and  width of the momentum distributions 
were found to 
be rather insensitive to the 
radius and diffusivity since these parameters mainly affect the 
single-particle
asymptotic normalization coefficient of the wave function.
It should also be noted that in the present calculations the ground 
and excited states of the core 
were assumed to have the same density distributions. As such the same 
Woods-Saxon geometry was employed
for all core states of each projectile.

In order to facilitate the comparison of the calculated and measured 
momentum distributions,
the former were filtered through a Monte Carlo simulation (Section II) to 
take 
into account the experimental 
broadening effects.
% vino aici
As may be seen in 
Figs. \ref{fig:sysglau} and \ref{fig:sefglau},  the 
measured distributions and cross sections for all the nuclei included in 
the present
study, including those nuclei with well established structure, are well 
reproduced with 
the exception of
$^{22}$F, where the cross section is somewhat underestimated.

It is, however, apparent that for a number of nuclei the calculated 
momentum 
distributions are slightly broader than the experimental ones\footnote{The
asymmetric nature of some of the distributions is discussed in the following
section.}, 
in particular the low and high momentum wings are somewhat more pronounced.
This effect  appears to arise from the specific shapes of the distorting 
functions ($D(s)$, Section V) at low
impact parameters, as shown in
Fig. \ref{fig:dfunc}. The use of distortion kernels calculated with less 
realistic
black disk S-matrix elements leads to a strong suppression of the high 
momentum
components in the wave function, and momentum distributions consequently 
become much narrower. 

To illustrate the contributions from the different mechanisms leading
to the removal of the neutron, the various contributions --- absorption, 
diffraction and Coulomb dissociation --- to the
breakup of $^{14}$B, $^{15}$C, $^{17}$C and $^{21}$O are displayed in
Fig. \ref{fig:exglau}. The Coulomb dissociation cross section is typically 
less 
than 1~mb in all but the
most favourable cases --- $^{14}$B and $^{15}$C --- for which the Coulomb 
induced
breakup was estimated to amount to some 7~mb.
As expected \cite{han96}, absorption and diffraction 
result in distributions with very similar lineshapes. The diffraction 
cross section is, however,
smaller than that for 
absorption for well bound states, whilst the two are essentially 
identical for weakly
bound states (Fig. \ref{fig:sefglau}). 
This evolution with binding is also illustrated in
Fig. \ref{fig:testeiko} for single-neutron removal at 50~MeV/nucleon from 
an $A=17$ system 
comprising a core  and a single-valence neutron. 
In the case where the neutron occupies an $s$-wave configuration (left 
panel) the absorption and diffraction 
cross sections are almost equal independent of the binding energy. 
If the beam energy is increased, absorption becomes the 
dominant process,
whilst at lower energies  diffraction is favoured. 
For a $d$-wave valence neutron the cross section is dominated for all 
energies by absorption 
and the contribution 
from diffraction decreases as $S_n$ increases.

The effects of applying the noneikonal corrections described in Section 
VII are also 
displayed in Fig. \ref{fig:testeiko}. 
These corrections lead to an increase in the total (absorption and 
diffraction) 
cross section: for example, for a valence neutron binding energy of 
1~MeV, an
increase of some 12\% occurs. As expected \cite{esb01} the effect is even 
more pronounced at a lower energies (e.g., some 19\% at 30~MeV/nucleon).  

Following Brown \etal \cite{brown} we introduce a quenching factor
$R_s=\sigma_{-1n}/\sigma_{-1n}^{Glauber}$ in analogy with the $z$ factor 
of Pandharipande
\etal \cite{panda}. Individual values are plotted in Fig. \ref{fig:rs_flo} as 
a
function of projectile mass number.  Averaging over the 22 one-neutron 
removal 
reactions for which cross sections were measured here, one obtains 
$R_s=0.98\pm 
0.16$.
  Note that our data base includes both loosely ($S_n\approx 1$ MeV) and
well bound nuclei ($S_n\approx 2-8$ MeV ).  For the loosely bound systems
($^{14}$B and $^{15}$C ) $R_s=0.8\pm
0.1$ in agreement with the values deduced by Brown \etal \cite{brown} and 
Enders 
\etal
\cite{enders03} for
$^8$B and $^9$C.
 
\subsection{Transverse momentum distributions: Carbon target}

The transverse momentum distributions ($p_x$) 
were calculated as described in Sections V and VI.
In order to make a comparison with the experimentally measured 
distributions, the 
calculated distributions were filtered through a simulation which took 
into account the
various broadening effects --- straggling, the resolution of the tracking 
detectors and 
the spectrometer (Section II and Table \ref{tab1:widths}) --- together with the 
finite angular acceptances of the spectrograph.
The most significant effect was that arising from the acceptances, 
whereby the
high momentum wings of the distributions were reduced in 
intensity\footnote{As the emittance of the
beam was relatively large (Section II), the angular acceptances of the 
spectrometer do not introduce a
sharp cut-off in the transverse momentum distributions.}.
Good agreement, as may be seen in Fig. \ref{fig:sysglaupx}, was found 
between the
measured and calculated distributions.

As mentioned in Section V-B, the transverse momenta 
are strongly influenced by the
interaction with the target and the distributions arising from 
absorption and diffraction exhibit different lineshapes. 
This effect is explored in Fig.~\ref{fig:testpxpz}, where the 
longitudinal
and transverse distributions have been calculated for $s$, $p$ and $d$-wave 
single-particle
states with binding energies of 1~MeV in an A=14 system.   In the case
of the longitudinal momenta both the $s$ and (to a lesser extent) $p$-wave 
states result in relatively
narrow distributions, whilst the $d$-wave state may be identified with a 
broad distribution
exhibiting two symmetric peaks.

In the case of the transverse momenta, the contribution arising from 
diffraction
is systematically much broader than that from absorption and as such 
dominates the high momentum
components of the total distribution.  In addition, only the $s$-wave
configuration gives rise to a relatively narrow  distribution.  The 
$p$ and $d$-wave
states are relatively broad with the former presenting a flat topped 
distribution with a small
central dip.
These features, combined with the intrinsic structure of the projectile 
results in
transverse distributions significantly different in form from the 
longitudinal distributions.
In particular, the presence in a mixed configuration of  a non negligible 
$s$-wave component
will manifest itself in the transverse momentum distribution as a narrow 
feature 
superimposed on a much broader
component.  This is particulary well illustrated by the results for  
$^{23}$O
and
$^{24,25}$F  (Fig.~\ref{fig:sysglaupx}). 
Thus, whilst the lineshapes of the transverse momentum distributions may 
be more complex, 
they remain
sensitive to the nature of the projectile ground state and may furnish 
spectroscopic information
in a complementary manner.

As noted in the preceding section a number of the longitudinal momentum
distributions exhibit low momentum tails (Fig. \ref{fig:sysglau}).
Such asymmetric distributions may arise from dissipative core-target 
collisions, such as observed 
in stable beam
fragmentation \cite{sco80,sto84,mou81} or, more likely in the case of weakly
bound systems as a result of 
diffractive/elastic  
breakup \cite{JeffCDCC}. 
Experimentally a correlation exists between the longitudinal and 
transverse
momenta for events in the low momentum tail, as displayed in 
Fig. \ref{fig:trainepx}. 
Here the data for $^{15}$C, where the yield is 
dominated by breakup to the $^{14}$C ground state, was analysed so as to
minimise any momentum shifts arising from core excited states.
When events situated in the tail are selected 
the corresponding 
transverse momentum distribution is broad.  In contrast,
for events with $p_z$ greater than the mean momentum\footnote{We 
note 
that such events should reflect most directly the intrinsic momentum 
of the removed neutron.} the 
transverse momentum distribution is much narrower, with a width
identical to that of the total distribution (Fig. \ref{fig:sysglaupx}).  The 
low momentum events constituting the tail thus, on average,
exhibit a much larger scattering angle.  This result is consistent
with the observation by Tostevin {\em et al.} \cite{JeffCDCC} of 
asymmetric longitudinal momentum distributions at scattering angles 
away
from zero degrees in the breakup of $^{15}$C.  
Furthermore, the
average momenta of the distributions were observed to be downshifted 
with increasing scattering angle.  Such energy nonconservation 
effects
cannot be described within the framework of the eikonal 
approximation
employed here.  As described in ref. \cite{JeffCDCC}, fully 
dynamical
coupled discretised continuum channels calculations are capable of
reproducing these effects, suggesting that the origin is 
diffractive/elastic
breakup.  A model independent confirmation could be furnished
by fully exclusive measurements in which the beam velocity neutrons 
from breakup (a signature of diffractive dissociation) are measured
in coincidence with the core fragments and deexcitation gamma rays.

%%%%%%%%%%%%%%%%%%%%%%%%%%%%%%%%%%%%%%%%%%%%%%%%%%%%%%%%%%%%%%%%%%%%%%%%%
%%%%%%%%%%%%%%%%%%%%%%%%%%%%%%%%%%%%%%%%%%%%%%%%%%%%%%%%%%%%%%%%%%%%%%%%%

\subsection{Longitudinal and transverse momentum distributions: Ta target}

As noted earlier, the longitudinal momentum distributions were measured
for the breakup of $^{14}$B, $^{15-17}$C and  $^{17-19}$N on Ta (Fig.
\ref{fig:systag}), and distributions almost identical in width and
form to those obtained for reactions on C were observed. 
Owing to the Coulomb deflection of the projectile and core fragment in
the field of the target nucleus very broad transverse momentum 
distributions were encountered
experimentally.
As an example the  distribution obtained for 
$^{15}$C is displayed in Figure \ref{fig:px_ta}.  It is clearly evident 
that
the acceptances ($p_x$ $\simeq$ $\pm$200~MeV/$c$) of the spectrograph were 
too limited
to allow either transverse momentum distribution or the single-neutron
removal cross section to be determined.

An estimate of the effects of Coulomb orbital deflection on the 
dissociation of $^{15}$C  
is shown in Fig. \ref{fig:px_ta}.
For simplicity the transverse distribution has been calculated assuming 
pure Coulomb
breakup (which is expected to dominate for the breakup of $^{15}$C) and 
considering only the
dominant $s$-wave component in the ground state wave function.
The solid line in Fig. \ref{fig:px_ta} corresponds to the 
assumption that
the core fragment is deflected, following dissociation, along a
classical Rutherford 
trajectory. In this
case the angle of deflection depends only on the impact 
parameter $b$. The final transverse distribution was simulated assuming 
a distribution of impact parameters 
given by the calculations described in Section VI for 
$b > b_{min}=R_{core}+R_{target}$ joined smoothly with a diffuse shape  
for $b < b_{min}$ mocked up by a Woods-Saxon form factor. The calculation in
Fig. \ref{fig:px_ta} shows that the  broadening effect in the transverse
momentum distribution is largely explained by core  deflection in the target
Coulomb field and suggests the breakdown of the straight line trajectory
assumption at intermediate energies for heavy targets.

We can now turn to the  longitudinal core fragment momentum 
distributions on Ta target. 
Calculations including nuclear and Coulomb components compare well with the
data as displayed in Fig. \ref{fig:systag}. As for the results obtained with
the carbon target the theoretical predictions have been filtered through a
Monte Carlo simulation (Section II) to take account the experimental 
effects.  
Details of the calculation are shown in Fig. \ref{fig:excoul} for 
$^{15,17}$C. 
In the case of $^{15}$C, with a ground state dominated by an $s$-wave  
 valence neutron and low
$S_n$, the nuclear and Coulomb  distributions are almost identical, an 
example
 of the ``numerical coincidence'' pointed out 
by Hansen \cite{han96}. 
In the case of $^{17}$C, which is dominated by a $d$-wave valence neutron 
the Coulomb and nuclear distributions are not identical. The Coulomb 
interaction
samples large impact parameters and selects small momentum components and 
the
corresponding momentum distribution is narrow.
 For the cases detailed here, the laboratory grazing angle
is about  1.4$^\circ$, whilst the measured angular range is $\pm 2^\circ$.
Clearly a nuclear component must be present in these data and the 
calculations
presented 
in  Fig. \ref{fig:excoul} suggest that the absolute  nuclear and Coulomb
contributions predicted by
the Glauber model coupled to first order perturbation theory for Coulomb
dissociation seem to be realistic. 

%%%%%%%%%%%%%%%%%%%%%%%%%%%%%%%%%%%%%%%%%%%%%%%%%%%%%%%%%%%%%%%%%%%%%%%%%
%%%%%%%%%%%%%%%%%%%%%%%%%%%%%%%%%%%%%%%%%%%%%%%%%%%%%%%%%%%%%%%%%%%%%%%%%

\subsection{Momentum distributions as a spectroscopic tool}

On the basis of the preceding comparisons, the reaction 
mechanism on the carbon target appears to be understood and, except for the 
low
momentum tails, 
well described by a 
Glauber type approach within the  eikonal approximation. 
When the nuclear structure is well known, as is the case for the 
 nuclei closest to stability, the data are well 
reproduced by the model.
In this section the manner by which the spin-parity 
assignments given in our earlier paper \cite{sau00} to nuclei
with poorly established ground state structure will be outlined.
This will also be instructive in illustrating the sensitivity
of the inclusive core fragment momentum distributions to the 
structure
of the projectile. 

A comparison of the measured core fragment longitudinal momentum
distributions from reactions on the carbon target and those 
calculated 
for the possible ground-state spin-parities is provided in   
Fig. \ref{fig:spectro}. Here both the calculated and experimental 
distributions are displayed on an absolute scale (mb/[MeV/$c$])
without any normalisation, except for  
$^{23}$O where no cross section could be extracted experimentally
(Section III).  In this case the 
calculated momentum distributions have been 
normalised so as to best reproduce the 
experimental distribution. 
In most of the cases the choice between the 
various $J^\pi_{gs}$ is clear and the favoured spin-parity 
assignments, represented by the solid lines in 
Fig. \ref{fig:spectro}, are listed in 
Table \ref{tab2:res_c}\footnote{Owing to an error in the compilation
of Ref. \cite{sauthese}, the likely spin-parity assignments for 
$^{24}$F 
were given 
as 1$^+$, 3$^+$ rather than 2$^+$, 3$^+$ in our original paper
\cite{sau00}.}.  
Interestingly, in all the cases
presented here, the favoured spin-parity assignments correspond to 
those
suggested by the shell model calculations.  

In two cases ($^{17}$C and $^{23}$O) the spin-parity
assignments are not directly evident from inspection of 
Fig. \ref{fig:spectro}.  
As it has been the object of recent attention
\cite{Kan02,Bro03,Kan03} we will first turn our attention to 
$^{23}$O.
The form of the single-neutron removal longitudinal momentum 
distribution 
obtained here is well 
reproduced by both $J^\pi_{gs}$=1/2$^+$ and 5/2$^+$ assignments (similar 
results hold for the transverse 
momentum
distribution).  The former, however, leads to a predicted
cross section of some 224~mb (Appendix B, 
Table \ref{tab4:res_o23})\footnote{The calculated cross section
listed
in Table 1 of our original paper \cite{sau00} omitted the strong 
yield to the 
3$^+$ state predicted at around 4.8~MeV in $^{22}$O 
(Table \ref{tab4:res_o23}).}, a factor of 
around 4 higher
than for $J^\pi_{gs}$=5/2$^+$.  Based on the systematics of the 
cross sections
obtained here for the Oxygen isotopes (Fig. \ref{fig:sefglau}), a 
1/2$^+$ assignment 
was favoured \cite{sau00}.
This conclusion is, contrary to the arguments made recently 
by the the authors of
Ref. \cite{Kan02}, confirmed by their measured cross section 
for
single-neutron removal --- 233$\pm$37~mb as compared to a value of 
185~mb
which we predict (Table \ref{tab4:res_o23}), using the ground state 
structure 
given in Appendix B,
at their beam energy of 72~MeV/nucleon (a very similar result was 
also found by Brown {\em et al.} \cite{Bro03}).  Further support for this
ground state structure for $^{23}$O may be found in the preliminary 
results for single-neutron removal at very high energy
(936~MeV/nucleon) 
obtained using the FRS \cite{priv_lola}.  Not only can we
reproduce well the measured longitudinal momentum distribution
assuming a 1/2$^+$ ground state, 
but a cross section of
82~mb is predicted (Table \ref{tab4:res_o23} ) as compared to the 
experimental 
value of 84$\pm$11~mb.

In the case of $^{17}$C the three possible ground state spin-parity
assignments are shown in Fig.
\ref{fig:spectro}. It is clear that the
1/2$^+$ assignment grossly overestimates the cross section. 
The 3/2$^+$ and 5/2$^+$ assignments reproduce the data quite well,
with the former providing a marginally better description of the
central part of the distribution.
As noted in Appendix B, a $J^\pi_{gs}$=3/2$^+$ results in a 
large yield to the $^{16}$C 2$^+_1$ state in single-neutron removal
from $^{17}$C.  The observation by 
Maddalena \etal \cite{mad01} and Datta Pramanik \etal \cite{Dat03} 
of the corresponding 1.76~MeV gamma-ray
transition thus confirms directly the 3/2$^+$ assignment.  

Interestingly, the transverse momentum distribution measured here 
for 
$^{17}$C also supports this assignment.  This may be seen in 
Fig.~\ref{fig:pxtool} where the predictions (obtained with the same
spectroscopic factors as above and without adjusting  the parameters
of the reaction
calculation) for the
three possible spin-parity assignments are compared to the measured 
distribution.  This illustrates, as noted earlier, that the 
transverse
momentum distributions can carry spectroscopic 
information complementary to that provided by the longitudinal
momenta.

In the light of the results described here, the evolution of the 
core fragment momentum distributions with $T_z$ may be understood, 
in particular through the competing contributions of the 
valence neutron $\nu$2s$_{1/2}$ and 
$\nu$1d$_{5/2}$ admixtures.  In summary, following the crossing of 
N=8, the 
ground states of the N=9 -- $^{14}$B, $^{15}$C -- and N=10 isotones 
-- $^{15}$B, $^{16}$C -- are significantly influenced by the intruder 
$\nu$2s$_{1/2}$ and 
$\nu$2s$_{1/2}^2$ configurations, respectively.  As the neutron 
number 
increases 
so too does the contribution from $\nu$1d$_{5/2}$ configurations.  
This reaches
a maximum at N=14 and then, as expected from the naive shell model, 
the
the $\nu$2s$_{1/2}$ orbital is occupied for N=15 and 16.   
Similar conclusions may also be drawn from the 
interaction
cross section measurements of Ozawa \etal, which exhibit
enhancements for  $^{23}$O and $^{24,25}$F \cite{oza01,oza00}.

%%%%%%%%%%%%%%%%%%%%%%%%%%%%%%%%%%%%%%%%%%%%%%%%%%%%%%%%%%%%%%%%%%%%%%%%%
%%%%%

\section{Conclusions}

An investigation of high-energy one-neutron removal reactions on
23 neutron-rich p-sd shell nuclei has been presented.  By studying
isotopic chains extending from strongly bound near stable systems
to weakly bound near dripline nuclei, the evolution of structure with
isospin, as expressed by the core fragment observables
(longitudinal and transverse momentum distributions and 
inclusive cross sections) has been
explored.  Experimentally, the measurements were carried out
using a broad range high resolution, high acceptance spectrometer which
permitted the data to be collected at a single magnetic 
field setting.
Data were recorded using both Carbon and Tantalum targets
in order to explore nuclear and Coulomb induced breakup.
In the case of the Carbon target data, the large angular
acceptances of the spectrometer were sufficient to encompass the full
range of transverse momenta, thus permitting unambiguous measurements of
the longitudinal and transverse momentum distributions and associated
cross sections to be made.  Owing to the large Coulomb 
deflection present in the
reactions on the Tantalum target, the effective transverse momentum
acceptances were very limited.  As such only the longitudinal momentum
distributions could be deduced.

From the theoretical standpoint, an extended version of the Glauber 
model 
which incorporates effective NN interactions and second order
noneikonal corrections to the JLM parameterisation of the optical potential
has been developed.  The treatment of Coulomb dissociation using
first order perturbation theory has also been described.
Particular emphasis has been devoted to retain the r\^ole played by  
the valence-neutron wave function via its Wigner transform in mapping the 
intrinsic
momentum components onto the measured distributions. Despite a number of
simplifying assumptions the
model predictions agree very well with the experimental data, in particular
those obtained for nuclei near stability with relatively well known 
structure.

In the case of nuclear induced breakup the model 
suggests that for the longitudinal momentum distribution the 
reaction mechanism factorises in a manner such that  
only the low or surface momentum
components in the wave function are selected.   As a consequence 
only the asymptotic 
part of the 
wave function is probed. In the transverse momentum 
distributions
this factorization does not occur and additional momenta arising 
from interactions with the target come into play. 
The principle drawback of the present approach, which is inherent
to all Glauber-type models, is the neglect of energy and 
momentum conservation in describing diffractive/elastic breakup.  
The predicted 
momentum distributions
are thus always symmetric and the low momentum tails observed here
for some of the weakly bound nuclei cannot be reproduced.
As noted in Section VIII-B, the description of such asymmetries requires the
implementation of fully dynamical calculations (see, for example,
\cite{JeffCDCC}).

Shell model spectroscopic factors calculated using the Warburton-Brown
effective interaction  
formed the structural
input 
for the calculations.  The resulting momentum distributions and 
cross sections were found to be in very good agreement (except for the 
cross section for $^{22}$F which was underpredicted by some 30\%)
with the measurements.
This agreement, especially for those nuclei with well established structure,
suggests that the 
longitudinal momentum  distributions and associated inclusive cross sections  
constitute a spectroscopic tool and ground state spin-parity assignments
were proposed 
for $^{15}$B, $^{17}$C,
$^{19-21}$N,  $^{21,23}$O, $^{23-25}$F.
In addition to the dominance of the $\nu2$s$_{1/2}$
intruder configuration in the $N=9$ isotones,$^{14}$B and $^{15}$C, 
significant $\nu2$s$_{1/2}^2$ admixtures were found to occur in the 
ground states of the
neighbouring $N=10$ nuclei $^{15}$B and $^{16}$C. 
Similarly, following the crossing  the N=14 subshell, the occupation of the
$\nu2$s$_{1/2}$ orbital is clearly observed for $^{23}$O, $^{24,25}$F.

The calculations of the transverse momentum distributions were also seen to
agree well with the measurements.  Thus, whilst being systematically 
somewhat
broader than the longitudinal distribution the
transverse distribution also carries structural information.   Interestingly, 
due to the interplay of the projectile structure and reaction mechanism the 
transverse momenta were seen to often carry information in a complementary 
manner.  In particular, the competition between $s$ and $d-$wave valence 
neutron configurations can exhibit itself directly in the transverse momentum
distribution.  Such complementary information may be of utility when
conducting experiments with weak beams.

Ultimately, the experimental determination of the core excited states
populated in the reaction is required if detailed, unambiguous spectroscopic 
information is to be deduced.  As seen elsewhere 
\cite{nav98,aum00,gui00,nav00,mad01,Lola02,enders02} this may be obtained 
for 
bound core states using
large scale NaI or Germanium-detector arrays.  As the neutron dripline is
approached and the core itself becomes weakly bound, coincident neutron
detection will also become necessary. 

Finally, in terms of perspectives, we conclude with some more general
observations concerning the
use of high-energy single-nucleon removal reactions as a probe of structure.  
As noted above, the reaction probes only the surface content of the 
projectile wavefunction.  As such comparatively simple wavefunctions have
been employed to 
describe the valence nucleon.
These wavefunctions, weighted by spectroscopic factors derived from
large scale shell model calculations are, as described here, coupled to 
relatively
sophisticated reaction model calculations.  As is evident from the present
work and has already been  
pointed out by others \cite{Han00,Han04}, remarkably good agreement has been
achieved to date in describing the measurements.

A few caveats should, however, be added.  First, given the
uncertainties inherent in the calculations, such as those described in 
Section VIII-A,
uncertainties of order $\pm$10\% should be ascribed to the predicted cross
sections\footnote{As noted in Section VIII-A, the widths of the 
momentum distributions are far less sensitive.}.  
Coupled with the experimental uncertainties --- typically 
of a similar
order --- it would appear that deficiencies in our modelling 
of $\sim$10-20\% 
in cross section could easily be overlooked.
High precision data obtained employing beams with very well established
structure would, therefore, provide a means to help validate the accuracy to
which the present approaches can be employed.  A recent 
reanalysis of very high-energy inclusive 
measurements of
single-nucleon removal from beams of $^{12}$C and $^{16}$O \cite{brown}
is an encouraging step in the this direction and dedicated 
experiments employing
coincident gamma-ray detection are to be expected.

Second, modelling employing ``realistic'' wavefunctions should be explored.
In the case of single-nucleon transfer reactions it has long been known
that despite their surface nature, the extraction of absolute 
spectroscopic factors can depend strongly on the description of the valence
nucleon wavefunction \cite{Sat80}. 
In this context, it is instructive to recall a recent reanalysis of (d,$^3$He) 
measurements by 
Kramer {\em et al.} \cite{Kra01}.  In this study it was demonstrated 
that whilst only the tail of the
bound-state wavefunction is sampled, it is very 
sensitive to the exact shape of the potential, thus introducing a significant 
model dependence in the calculated cross sections. 
In terms of weakly bound nuclei, the need to employ realistic wavefunctions 
was also found in the analysis of the p($^{11}$Be,$^{10}$Be(2$^+$))d 
reaction \cite{Win01}. 
Similar effects must almost certainly be addressed in the analysis of high-energy,
single-nucleon removal, and the levels to which they occur may provide a
limit to the relatively simple analyses employed to date.

\acknowledgements 

The support provided by the staffs of LPC and
GANIL during the experiment is gratefully acknowledged. Discussions with 
B.~A.~Brown and J.~A.~Tostevin related to theoretical aspects of the work are also
acknowledged as is the assistance of G.~Mart\'{\i}nez in preparing
and executing the experiment. This work was funded under the auspices of the 
IN2P3-CNRS 
(France) and
EPSRC (United Kingdom). Additional support from ALLIANCE programme 
(Minist\`ere des
Affaires Etrang\`eres and British Council), the Human Capital and 
Mobility 
Programme of
the European Community (contract n$^\circ$ CHGE-CT94-0056) and the GDR 
Exotic 
Nuclei
(CNRS) was received.  One of the authors (F.C.) acknowledges the support 
furnished by LPC-Caen and the 
IN2P3, including that provided within the framework 
of the IFIN-HH$-$IN2P3 convention.

%%%%%%%%%%%%%%%%%%%%%%%%%%%%%%%%%%%%%%%%%%%
%%%%%%%%%%%%%%%%%%%%%%% bibliography %%%%%%
%%%%%%%%%%%%%%%%%%%%%%%%%%%%%%%%%%%%%%%%%%%

\newpage

%%%%%%%%%%%%%%%%%%%%%%%%%%%%%%%%%%%%%%%%%%%%%%%%%%%%%%%%%%%%%%%%%%%%%%%%%

\appendix
\section{}
In this appendix explicit analytical expressions for the Coulomb 
dissociation 
cross section Eq. (\ref{coul:eq7}) are given for $s$, $p$ and $d$-wave 
functions 
for $E_1$, $E_2$ transitions and $E_1E_2$ interference. 
The short hand notations $FX_x$ are defined below.  $P_l$ denote 
Legendre polynomials where the argument 
$\cos \theta$ has been omitted. The functions $\overline{K}_\mu$ are defined 
in 
Section VI.

\begin{equation}
F1_{E1} = \overline{K}_0(\xi) + \overline{K}_1(\xi) \gamma^2 
\end{equation}

\begin{equation}
F2_{E1} = -2 \overline{K}_0(\xi) + \overline{K}_1(\xi) \gamma^2 
\end{equation}

\begin{equation}
F1_{E2} = 3\overline{K}_0(\xi) + \overline{K}_2(\xi) + 
\overline{K}_1(\xi) 
\gamma^2 (2-\beta^2)^2
\end{equation}

\begin{equation}
F2_{E2} = 3\overline{K}_0(\xi) -\overline{K}_2(\xi) + \frac{1}{2} 
\overline{K}_1(\xi) \gamma^2 (2-\beta^2)^2
\end{equation}

\begin{equation}
F3_{E2} = 9\overline{K}_0(\xi) + \frac{1}{2} \overline{K}_2(\xi) - 2 
\overline{K}_1(\xi) \gamma^2 (2-\beta^2)^2
\end{equation}

\begin{equation}
F1_{E1E2} = 2 \overline{K}_0(\xi) + \overline{K}_1(\xi) \gamma^2 (2 - 
\beta^2)
\end{equation}

\begin{equation}
F2_{E1E2} = 3 \overline{K}_0(\xi) - \overline{K}_1(\xi) \gamma^2 (2 - 
\beta^2)
\end{equation}

\vspace{1cm}

{\bf L=0 }

\begin{eqnarray}
\frac{d^2 \sigma_{E1}}{q^2dq\sin{\theta}d\theta} & = & \frac{4} {3} \; 
\frac{Z^2_t (Z^{eff}_1)^2 \alpha^2}{\gamma^2 \beta^2}  I_{011}^2 \times 
\left( 
\right. \nonumber\\
  & + &  F1_{E1} \nonumber\\
  & - &  P_{2} \; F2_{E1} \left. \right) \nonumber\\
\end{eqnarray}

\begin{eqnarray}
\frac{d^2 \sigma_{E2}}{q^2dq\sin{\theta}d\theta} & = & \frac{1} {105} \; 
\frac{Z^2_t (Z^{eff}_2)^2 \alpha^2}{\gamma^2 \beta^4}  (\omega/c)^2 
I_{022}^2 
\times \left( \right. \nonumber\\
  & + & 7 \; F1_{E2} \nonumber\\
  & + & 10 \; P_{2} \; F2_{E2} \nonumber\\
  & + & 6 \; P_{4} \; F3_{E2} \left. \right) \nonumber \\
\end{eqnarray}

\begin{eqnarray}
\frac{d^2 \sigma_{E1E2}}{q^2dq\sin{\theta}d\theta} & = & \frac{4}{5} \; 
\frac{Z^{eff}_1 Z^{eff}_2 Z^2_t \alpha^2} {\gamma^2 \beta^3} \; 
(\omega/c) 
I_{011}I_{022} \times \left( \right. \nonumber\\
  & + & P_{1} \; F1_{E1E2} \nonumber \\
  & + & P_{3} \; F2_{E1E2} \left. \right)\nonumber \\
\end{eqnarray}

{\bf L=1}

\begin{eqnarray}
\frac{d^2 \sigma_{E1}}{q^2dq\sin{\theta}d\theta} & = & \frac{4}{3} \; 
\frac{Z^2_t (Z^{eff}_1)^2 \alpha^2}{\gamma^2 \beta^2} \times \left( 
\right. 
\nonumber\\
    & + &  F1_{E1} \; (I_{110}^2 + 2 I_{112}^2) \nonumber\\
    & + & P_{2} \; F2_{E1} \; (2 I_{110}I_{112} - I_{112}^2) \left. 
\right) 
\end{eqnarray}

\begin{eqnarray}
\frac{d^2 \sigma_{E2}}{q^2dq\sin{\theta}d\theta} & = & \frac{1} {525} \; 
\frac{Z^2_t (Z^{eff}_2)^2 \alpha^2}{\gamma^2 \beta^4}  (\omega/c)^2\times 
\left( \right. \nonumber\\
  & + & 7  \; F1_{E2} \; (2 I_{121}^2 + 3 I_{123)^2} \nonumber\\
  & + & 2 \; P_{2} \; F2_{E2} \; (7 I_{121}^2 + 12 I_{123}^2 - 6 I_{121} 
I_{123}) \nonumber\\
  & + & 6 \; P_{4} \; F3_{E2} \; (I_{123}^2 - 4 I_{121} I_{123}) \left. 
\right) 
\end{eqnarray}

\begin{eqnarray}
\frac{d^2 \sigma_{E1E2}}{q^2dq\sin{\theta}d\theta} & = & \frac{4}{75} \; 
\frac{Z^{eff}_1 Z^{eff}_2 Z^2_t \alpha^2}{\gamma^2 \beta^3} \; 
(\omega/c)\times \left( \right.\nonumber\\
  & + & P_{1} \; F1_{E1E2} \; (5 I_{110}I_{121}  - 7 I_{112} I_{121} + 9 
I_{112} I_{123}) \nonumber \\
  & + & P_{3} \; F2_{E1E2} \; (- 5 I_{110} I_{123} - 6 I_{112}I_{121} + 4 
I_{112}I_{123}) \left. \right) \nonumber \\
\end{eqnarray}

%\vspace{1cm}

{\bf L=2}

\begin{eqnarray}
\frac{d^2 \sigma_{E1}}{q^2dq\sin{\theta}d\theta} & = & \frac{4} {75} \; 
\frac{(Z^{eff}_1)^2 Z^2_t \alpha^2}{\gamma^2 \beta^2}\times \left( 
\right. 
\nonumber\\
    & + & 5 \; F1_{E1} \; (2 I_{211}^2 + 3 I_{213}^2) \nonumber\\
    & + &2 \; P_{2} \; F2_{E1} \; (-\frac{1}{2} I_{211}^2 + 9 I_{211}  
I_{213} 
- 6 I_{213}^2) \left. \right) \nonumber\\
\end{eqnarray}

\begin{eqnarray}
\frac{d^2 \sigma_{E2}}{q^2dq\sin{\theta}d\theta} & = & \frac{1}{25725} \; 
\frac{(Z^{eff}_2)^2 Z^2_t \alpha^2}{\gamma^2 \beta^4} \; 
(\omega/c)^2\times 
\left( \right.\nonumber\\
  & + & 49 \; F1_{E2} \; (18 I_{224}^2 + 7 I_{220}^2 + 10 I_{222}^2) 
\nonumber 
\\
  & + & 10  \; P_{2} \; F2_{E2} \; (- 15 I_{222}^2 + 90 I_{224}^2 - 98 
I_{220} 
I_{222}- 72 I_{222} I_{224}) \nonumber \\
  & + & 6 \; P_{4} \; F3_{E2} \; (20 I_{222}^2 + 27 I_{224}^2 + 98 
I_{220} 
I_{224}- 100 I_{222} I_{224}) \left. \right)\nonumber \\
\end{eqnarray}

\begin{eqnarray}
\frac{d^2 \sigma_{E1E2}}{q^2dq\sin{\theta}d\theta} & = & \frac{4}{175} \; 
\frac{Z^{eff}_1 Z^{eff}_2 Z^2_t \alpha^2} {\gamma^2 \beta^3} \; 
(\omega/c)\times \left( \right.\nonumber\\
  & + & P_{1} \; F1_{E1E2} \; (7 I_{211}I_{222}  + 18 I_{213} I_{224} - 3 
I_{213} I_{222} - 7 I_{211} I_{220} ) \nonumber \\
  & + &  P_{3} \; F2_{E1E2} \; (- 12 I_{211} I_{224} - 8 I_{213}I_{222} + 
2 
I_{211}I_{222} +  \nonumber \\
  &    &  \; \; \;  \; 7 I_{213} I_{220} + 6 I_{213} I_{224}) \left. 
\right) 
\nonumber \\
\end{eqnarray}

\newpage

%\appendix

%%%%%%%%%%%%%%%%%%%%%%%%%%%%%%%%%%%%%%%%%%%%%%%%%%%%%%%%%%%%%%%%%%%%%%%%%%%
%%%%%%%%%%%%%%%% tables for oxsbash and cross section calcs. %%%%%%%%%%%%%%
%%%%%%%%%%%%%%%%%%%%%%%%%%%%%%%%%%%%%%%%%%%%%%%%%%%%%%%%%%%%%%%%%%%%%%%%%%%

\section{}\label{spec_fact}

In the following, listings are provided of the calculated spectroscopic
factors ($C^2S$) and cross sections ($\sigma(I^\pi_c)$)
to the core excited states ($E_{ex}^c$, $I^\pi_c$)  populated in 
single-neutron 
removal
from the projectile nucleus ($^A$Z, $J^\pi$) by the Carbon target.
The contributions arising from absorption ($\sigma_{abs}$) and diffractive
dissociation ($\sigma_{diff}$) are detailed (note: the latter includes the 
very 
small contributions
arising from Coulomb breakup) and the total inclusive cross section 
($\sigma_{-1n}^{Glauber}$)
to all core states is given.  Projectile $J^\pi_{gs}$ and core excited state 
energies
taken from the shell model predictions are marked in parentheses. 
Only the results for the preferred ground structure are given here: the 
listings 
for
other $J^\pi$ may be found in ref.~\cite{sauthese}.

%%%%%%%%%%%%%%%%%%%%%%%%%%%%%%%%%%%%%%%%%%%%%%%%%%%%%%%%%%%%%%%%%%%%%%%%%
%%%%%
%% new table updated by sauvan. Coulomb contribution added to
%% diff.
%%%%%%%%%%%%%%%%%%%%%%%%%%%%%%%%%%%%%%%%%%%%%%%%%%%%%%%%%%%%%%%%%%%%%%%%%

\vspace*{1.5cm}

\noindent \begin{tabular}{c c c c c c c c c}
\hline
\hline
 $^A$Z & $J^\pi$ & $E_{ex}^c$  & $I^\pi_c$& $nlj$  & $C^2S$ & $\sigma_{abs}$ & 
$\sigma_{diff}$ & $\sigma(I^\pi_c)$\\
  & & [MeV]  &  &   &  & [mb] & [mb] & [mb]\\
\hline
\hline
 $^{12}$B & $1^+$ & g.s.  & $3/2^-$ & 1p$_{1/2}$ & 0.71 & 19.9 & 14.7 & 34.6 
\\
 & & 2.124 & $1/2^-$ & 1p$_{3/2}$ & 0.27 & 6.6  & 4.4  & 11.0  \\
 & & 4.444 & $5/2^-$ & 1p$_{3/2}$ & 0.2  & 4.14 & 2.66  & 6.8  \\
 & & 5.02  & $3/2^-$ & 1p$_{3/2}$ & 0.36 & 7.2  & 4.5  & 11.7 \\
 & & 8.92  & $5/2^-$ & 1p$_{3/2}$ & 1.02 & 17.2 & 10.0  & 27.2 \\
\hline
\multicolumn{9}{r}{$\sigma_{-1n}^{Glauber}$=91 mb}\\
\hline
 $^{13}$B & $3/2^-$ & g.s.   & $1^+$ & 1p$_{3/2}$ &  0.61 & 13.0& 9.9 & 22.9 
\\
 & & 0.953  & $2^+$ & 1p$_{1/2}$ &  1.17 & 22.6& 16.7& 39.3 \\
\hline
\multicolumn{9}{r}{$\sigma_{-1n}^{Glauber}$=62 mb}\\
\hline
 $^{14}$B & $2^-$     & g.s. & $3/2^-$ & 1d$_{5/2}$ & 0.31 & 9.6 & 8.7 & 
18.3 \\
 & &      &         & 2s$_{1/2}$ & 0.64 & 57.0& 64.7& 121.7 \\
 & &3.483 & $3/2^+$ & 1p$_{1/2}$ & 0.41 & 8.5 & 7.1 & 15.6 \\
 & &3.68  & $5/2^+$ & 1p$_{1/2}$ & 0.8  & 16.2& 13.3& 29.5 \\
\hline
\multicolumn{9}{r}{$\sigma_{-1n}^{Glauber}$=185 mb}\\
\hline
 $^{15}$B & ($3/2^-$) & g.s.  & $2^-$ & 1d$_{5/2}$ & 0.28 & 5.8 & 5.3 & 11.1 
\\
 & &       &       & 2s$_{1/2}$ & 0.48 & 20.3& 22.3& 42.6 \\
 & &(0.89) & $1^-$ & 2s$_{1/2}$ & 0.27 & 9.4 & 9.9 & 19.3 \\
 & &(0.73) & $1^+$ & 1p$_{1/2}$ & 0.58 & 11.7 & 11.3 & 23 \\
 & &(0.96) & $3^-$ & 1d$_{5/2}$ & 0.47 & 8.7 & 7.6 & 16.3 \\
\hline
\multicolumn{9}{r}{$\sigma_{-1n}^{Glauber}$=112 mb}\\
\hline
\end{tabular}

\newpage

\noindent\begin{tabular}{c c c c c c c c c}
\hline
\hline
 $^A$Z & $J^\pi$ & $E_{ex}^c$  & $I^\pi_c$ & $nlj$  & $C^2S$ & 
$\sigma_{abs}$ & 
$\sigma_{diff}$ & $\sigma(I^\pi_c)$\\
  & & [MeV]  &  &   &  & [mb] & [mb] & [mb]\\
\hline
\hline
 $^{14}$C & $0^+$ & g.s.  & $1/2^-$ & 1p$_{1/2}$ &  1.67 & 26.4 & 15.9 & 
42.3 \\
 & & 3.089 & $1/2^+$ &            &       & & & \\
 & & 3.684 & $3/2^-$ & 1p$_{3/2}$ & 2.05  & 29.7 & 16.9 & 46.6 \\
\hline
\multicolumn{9}{r}{$\sigma_{-1n}^{Glauber}$=89 mb}\\
\hline
 $^{15}$C & $1/2^+$ & g.s. & $0^+$ & 2s$_{1/2}$ & 0.83  & 62.1 & 62.1 & 
124.2\\
 & & 6.094 & $1^-$ & 1p$_{3/2}$ & 0.16  & 2.8  & 1.9  & 4.7  \\
 & &       &       & 1p$_{1/2}$ & 1.03  & 16.3 & 10.9 & 27.2 \\
 & & 6.903 & $0^-$ & 1p$_{1/2}$ & 0.46  & 6.9  & 4.6  & 11.5 \\
\hline
\multicolumn{9}{r}{$\sigma_{-1n}^{Glauber}$=168 mb}\\
\hline
 $^{16}$C & $0^+$ & g.s.  & $1/2^+$ & 2s$_{1/2}$ & 0.6  & 19.3 & 17.5 & 36.8 
\\
 & &  0.740 & $5/2^+$ & 1d$_{5/2}$ & 1.23 & 21.9 & 16.2 & 38.1 \\
\hline
\multicolumn{9}{r}{$\sigma_{-1n}^{Glauber}$=75 mb}\\
\hline
 $^{17}$C & ($3/2^+$) & g.s. & $0^+$ & 1d$_{3/2}$  &  0.035 & 0.9 & 0.8 & 
1.7 \\
 & &  1.762 & $2^+$ & 1d$_{5/2}$ &  1.41  & 29.3& 25.5& 54.8 \\
 & &        &       & 2s$_{1/2}$ &  0.16  & 6.9 & 7.2 & 14.1 \\
 & &   4.1  & $2,3,4^+$ & 1d$_{5/2}$ &  0.76  & 12.5& 10.0& 22.5 \\
  & &       &       & 2s$_{1/2}$ &  0.22  & 6.1 &5.7 & 11.8 \\
\hline
\multicolumn{9}{r}{$\sigma_{-1n}^{Glauber}$=105 mb}\\
\hline
 $^{18}$C & $0^+$ &  g.s.   & $3/2^+$ & 1d$_{3/2}$ & 0.1  & 1.4 & 1.2 & 2.6 
\\
 & & (0.04)  & $5/2^+$ & 1d$_{5/2}$ & 2.8  & 43.3& 38.0& 81.3 \\
 & & (0.3)   & $1/2^+$ & 2s$_{1/2}$ & 0.65 & 17.2& 17.7& 34.9 \\
\hline
\multicolumn{9}{r}{$\sigma_{-1n}^{Glauber}$=119 mb}\\
\hline
\end{tabular}

\vspace*{0.7cm}

\noindent\begin{tabular}{c c c c c c c c c}
\hline
\hline
 $^A$Z & $J^\pi$ & $E_{ex}^c$  & $I^\pi_c$ & $nlj$  & $C^2S$ & 
$\sigma_{abs}$ & 
$\sigma_{diff}$ & $\sigma(I^\pi_c)$\\
  & & [MeV]  &  &   &  & [mb] & [mb] & [mb]\\
\hline
\hline
 $^{17}$N & $1/2^-$ & g.s.  & $2^-$ & 1d$_{5/2}$ & 0.59  & 11.1 & 7.6 & 18.7 
\\
 & &  0.120 & $0^-$ & 2s$_{1/2}$ & 0.12  & 3.4  & 2.7 & 6.1  \\
 & &  0.298 & $3^-$ & 1d$_{5/2}$ & 0.784 & 14.5 & 9.8 & 24.3 \\
 & &  0.397 & $1^-$ & 1d$_{5/2}$ & 0.36  & 9.9  & 7.8 & 17.7 \\
\hline
\multicolumn{9}{r}{$\sigma_{-1n}^{Glauber}$=67 mb}\\
\hline
 $^{18}$N & $1^-$ & g.s.  & $ 1/2^-$ &            &      & & & \\
 & &  1.374 & $ 3/2^-$ & 1d$_{5/2}$ & 0.65 & 13.4 & 10.1 & 23.5 \\
 & &	    &	       & 2s$_{1/2}$ & 0.195& 6.7  & 5.9 & 12.6 \\
 & & 1.850  & $1/2^+$  &	    &	   & & & \\
 & & 1.907  & $5/2^-$  & 1d$_{5/2}$ & 0.89 & 15.4 & 11.4 & 26.8 \\
 & & 3.129  & $7/2^-$  & 1d$_{5/2}$ & 0.42 & 7.6  & 5.4  & 13 \\
 & & 3.2    & $3/2^-$  & 2s$_{1/2}$ & 0.15 & 4.1  & 3.4  & 7.5  \\
 & &	    &	       & 1d$_{3/2}$ & 0.29 & 4.6  & 3.2  & 7.8 \\
\hline
\multicolumn{9}{r}{$\sigma_{-1n}^{Glauber}$=91 mb}\\
\hline
 $^{19}$N & ($1/2^-$) & g.s.  & $1^-$   & 1d$_{3/2}$ & 0.02  & 0.3 & 0.24 & 
0.54 \\
 & &	    &	      & 2s$_{1/2}$ & 0.005 & 0.14& 0.16 & 0.3  \\
 & &  0.115 & $(2^-)$ & 1d$_{5/2}$ & 1.26  & 21.6& 16.9 & 38.5 \\
 & &  0.747 & $(3^-)$ & 1d$_{5/2}$ & 1.71  & 24.9& 18.7 & 43.6 \\
 & &  (0.936) & $(1^-)$ & 2s$_{1/2}$ & 0.35  & 8.6  & 7.7  & 16.3  \\
\hline
\multicolumn{9}{r}{$\sigma_{-1n}^{Glauber}$=99 mb}\\
\hline
 $^{20}$N & ($2^-$) & g.s.   & $1/2^-$ & 1d$_{5/2}$ & 0.36  & 8.5 & 7.9 & 
16.4 \\
 & &  (1.68) & $3/2^-$ & 1d$_{5/2}$ & 0.66  & 12.4& 10.7& 23.1 \\
 & &  (2.17) & $5/2^-$ & 1d$_{5/2}$ & 0.38  & 6.8 & 5.8 & 12.6 \\
 & &  (3.9)  & $7/2^-$ & 1d$_{5/2}$ & 1.73  & 26.8& 21.6& 48.4 \\
\hline
\multicolumn{9}{r}{$\sigma_{-1n}^{Glauber}$=101 mb}\\
\hline
$^{21}$N & ($1/2^-$) & g.s.  &  $2^-$ & 1d$_{5/2}$ & 1.744 & 28.3 & 25.1 & 
53.4 \\
 & &  (0.6) &  $3^-$ & 1d$_{5/2}$ & 2.61  & 40.1 & 35.1 & 75.2 \\
 & &  (0.74)&  $1^-$ & 2s$_{1/2}$ & 0.45  & 11.1 & 11.1 & 22.2 \\
\hline
\multicolumn{9}{r}{$\sigma_{-1n}^{Glauber}$=151 mb}\\
\hline
\end{tabular}

\newpage

\noindent\begin{tabular}{c c c c c c c c c}
\hline
\hline
 $^A$Z & $J^\pi$ & $E_{ex}^c$ & $I^\pi_c$ & $nlj$  & $C^2S$ & $\sigma_{abs}$ & 
$\sigma_{diff}$ & $\sigma(I^\pi_c)$\\
  &  & [MeV]  &  &   &  & [mb] & [mb] & [mb]\\
\hline
\hline
$^{19}$O & $5/2^+$ &  g.s.  & $0^+$ & 1d$_{5/2}$ & 0.685 & 14.1 & 9.8 & 23.9 
\\
 & & 1.982 & $2^+$ & 1d$_{5/2}$ & 0.48  & 8.4  & 5.6 & 14.0  \\
 & &	   &	   & 1d$_{3/2}$ & 0.019 & 0.3  & 0.2 & 0.5  \\
 & &	   &	   & 2s$_{1/2}$ & 0.009 & 0.25 & 0.2 & 0.45 \\
 & & 3.555 & $4^+$ & 1d$_{5/2}$ & 1.24  & 19.7 & 12.6& 32.3 \\
 & & 3.92  & $2^+$ & 1d$_{5/2}$ & 0.22  & 3.4  & 2.2 & 5.6  \\
 & &	   &	   & 2s$_{1/2}$ & 0.06  & 1.34 & 0.96 & 2.3  \\
 & & 5.25  & $2^+$ & 1d$_{3/2}$ & 0.023 & 0.3  & 0.18 & 0.48  \\
 & &	   &	   & 1d$_{5/2}$ & 0.016 & 0.2 & 0.18 & 0.38  \\
 & & 5.38  & $3^+$ & 2s$_{1/2}$ & 0.11  & 2.2  & 1.5 & 3.7 \\  
\hline
\multicolumn{9}{r}{$\sigma_{-1n}^{Glauber}$=84 mb}\\
\hline
$^{20}$O & $0^+$ & g.s.  & $5/2^+$ & 1d$_{5/2}$ & 3.43 & 51.3 & 34.5 & 85.8 
\\
 & & 0.096 & $3/2^+$ & 1d$_{3/2}$ & 0.05 & 0.66 & 0.44 & 1.1  \\
 & & 1.471 & $1/2^+$ & 2s$_{1/2}$ & 0.28 & 5.3  & 4.0  & 9.3  \\
\hline
\multicolumn{9}{r}{$\sigma_{-1n}^{Glauber}$=96 mb}\\
\hline
$^{21}$O & ($5/2^+$) & g.s.   & $0^+ $ & 1d$_{5/2}$ & 0.345 & 6.7 & 5.4 & 
12.1 \\
 & & 1.67   & $2^+ $ & 1d$_{5/2}$ & 1.3   & 21.8& 16.4& 38.2 \\
 & &	    &	     & 2s$_{1/2}$ & 0.004 & 0.1 & 0.1& 0.2  \\
 & & 3.57   & $4^+ $ & 1d$_{5/2}$ & 2.59  & 38.0 & 27.3& 65.3 \\
 & & 4.072  & $2^+ $ & 1d$_{5/2}$ & 0.09  & 1.3 & 0.9 & 2.2  \\
 & &	    &	     & 2s$_{1/2}$ & 0.05  & 1.0  & 0.9 & 1.9  \\
 & & 4.456  & $0^+ $ &  	  &	  & & & \\
 & & 4.85   & $4^+ $ &  	  &	  & & & \\
 & & 5.23   & $2^+ $ & 1d$_{5/2}$ & 0.12  & 1.6 & 1.1 & 2.7  \\
\hline
\multicolumn{9}{r}{$\sigma_{-1n}^{Glauber}$=123 mb}\\
\hline
$^{22}$O & $0^+$ & g.s.   & $5/2^+$ & 1d$_{5/2}$  & 5.22 & 74.9 & 56.9& 
131.8 \\
 & & 1.33   & $1/2^+$ & 2s$_{1/2}$  & 0.23 & 4.3  & 3.7 & 8.0	\\
 & & 2.2    & $3/2^+$ & 1d$_{3/2}$  & 0.03 & 0.33 & 0.27& 0.6	\\
 & & 3.08   & $5/2^+$ & 1d$_{3/2}$  & 0.14 & 1.7 & 1.2 & 2.9	\\ 
\hline
\multicolumn{9}{r}{$\sigma_{-1n}^{Glauber}$=143 mb}\\
\hline
$^{23}$O & ($1/2^+$) & g.s.  & $0^+$ & 2s$_{1/2}$  & 0.8   & 31.7 & 33.8 & 
65.5 \\
 & &  3.38 & $2^+$ & 1d$_{3/2}$  & 0.053 & 0.7  & 0.5  & 1.2  \\
 & &	   &	   & 1d$_{5/2}$  & 2.1   & 30.5 & 24.5 & 55.0 \\
 & & (4.62) & $0^+$ & 2s$_{1/2}$  & 0.11 & 2.2  & 1.9  & 4.1  \\ 
 & & (4.83) & $3^+$ & 1d$_{5/2}$  & 3.08 & 40.5  & 31.7  & 72.2  \\ 
 & & (6.5)  & $2^+$ & 1d$_{5/2}$  & 0.24 & 2.9  & 2.1  & 5.0  \\ 
 & & (6.64) & $0^-$ & 1p$_{1/2}$  & 0.36 & 3.4  & 2.6  &  6.0 \\ 
 & & (6.9)  & $1^-$ & 1p$_{1/2}$  & 0.94 & 8.6  & 6.5  &  15.1 \\ 
\hline
\multicolumn{9}{r}{$\sigma_{-1n}^{Glauber}$=224 mb}\\
\hline
\end{tabular}

\newpage

\noindent\begin{tabular}{c c c c c c c c c}
\hline
\hline
 $^A$Z & $J^\pi$ & $E_{ex}^c$  & $I^\pi_c$ & $nlj$  & $C^2S$ & 
$\sigma_{abs}$ & 
$\sigma_{diff}$ & $\sigma(I^\pi_c)$\\
  & & [MeV]  &  &   &  & [mb] & [mb] & [mb]\\
\hline
\hline
$^{22}$F & $4^+$ & g.s.   & $5/2^+$ & 1d$_{5/2}$ & 0.56 & 9.1 & 6.4 & 15.5 
\\
 & & 0.2799 & $1/2^+$ & 	   &	  & & & \\
 & & 1.73  & $3/2^+$ & 1d$_{5/2}$ & 0.24 & 3.6 & 2.4 & 6.0  \\
 & & (1.9)    & $9/2^+$ & 1d$_{5/2}$ & 0.96 & 13.6& 9.1 & 22.7 \\
 & & (3.56)   & $3/2^+$ & 1d$_{5/2}$ & 0.09 & 1.19& 0.81 & 2.0  \\
 & & (3.64)   & $7/2^+$ & 1d$_{5/2}$ & 0.087 & 1.15& 0.75 & 1.9  \\
 & &	    &	      & 2s$_{1/2}$ & 0.015& 0.3& 0.2 & 0.5   \\
 & & (3.7)    & $9/2^+$ & 1d$_{5/2}$ & 0.58 & 7.4& 4.8 & 12.2  \\
 & & (4.02) & $1/2^+$ & 	   &	  & & & \\
 & & (4.45)   & $7/2^+$ & 1d$_{5/2}$ & 0.03 & 0.4& 0.2 & 0.6  \\
 & &	    &	      & 2s$_{1/2}$ & 0.02& 0.3& 0.2 & 0.5   \\ 
 & & (4.84)   & $11/2^+$ & 1d$_{5/2}$ & 0.54 & 6.5& 4.1 & 10.6  \\
 & &	    &	      & 1d$_{3/2}$ & 0.04& 0.4& 0.3 & 0.7   \\
 & & (4.91)   & $13/2^+$ & 1d$_{5/2}$ & 0.52 & 6.3& 3.9 & 10.2  \\
 & & (5.31)   & $7/2^+$ & 1d$_{5/2}$ & 0.07 & 0.8& 0.5 & 1.3  \\
 & &	    &	      & 2s$_{1/2}$ & 0.03& 0.5& 0.3 & 0.8   \\
 & & (5.5)   & $7/2^+$ & 1d$_{5/2}$ & 0.012 & 0.2& 0.1 & 0.3  \\
 & & (6.95)   & $13/2^+$ & 1d$_{5/2}$ & 0.075 & 0.8& 0.5 & 1.3  \\     
\hline
\multicolumn{9}{r}{$\sigma_{-1n}^{Glauber}$=87 mb}\\
\hline
$^{23}$F & ($5/2^+$) & g.s.  & $4^+$ & 1d$_{5/2}$ & 1.2  & 15.9 & 11.1 & 
27.0\\
 & & (0.2) & $3^+$ & 1d$_{5/2}$ & 0.76 & 10.0 & 6.8  & 16.8\\
 & &	   &	   & 2s$_{1/2}$ & 0.06 & 1.1  & 0.9  & 2.0 \\
 & & (0.7) & $2^+$ & 1d$_{5/2}$ & 0.64 & 8.1  & 5.6  & 13.7\\
 & &	   &	   & 2s$_{1/2}$ & 0.03 & 0.63 & 0.49 & 1.12\\
 & & 1.41  & $5^+$ & 1d$_{5/2}$ & 1.4  & 17.0 & 11.4 & 28.4\\
 & & (1.6) & $1^+$ & 1d$_{5/2}$ & 0.06 & 0.7  & 0.5  & 1.2  \\
 & &(1.65) & $3^+$ & 2s$_{1/2}$ & 0.07 & 1.2  & 0.8  &  2.0\\
 & &(1.67) & $2^+$ & 1d$_{5/2}$ & 0.31 & 3.7  & 2.5  & 6.2 \\
 & &(2.3)  & $1^+$ & 1d$_{5/2}$ & 0.27 & 3.1 & 2.1& 5.2 \\
 & &(3.5)  & $5^+$ & 1d$_{5/2}$ & 0.11 & 1.2 & 0.8  &  2.0 \\
\hline
\multicolumn{9}{r}{$\sigma_{-1n}^{Glauber}$=106 mb}\\
\hline
$^{24}$F & ($3^+$) & g.s. & $5/2^+$ & 1d$_{5/2}$ & 0.09  & 1.6 & 1.3 & 2.9 
\\
 & &	   &	     & 2s$_{1/2}$ & 0.74  & 22.3& 21.1& 43.4\\
 & & (1.8) & $1/2^+$ & 1d$_{5/2}$ & 0.073 & 1.1 & 0.8 & 1.9 \\
 & & (2.9) & $7/2^+$ & 1d$_{5/2}$ & 0.44  & 6.0 & 4.4 & 10.4\\
 & & (3.2) & $5/2^+$ & 1d$_{5/2}$ & 0.37  & 4.9 & 3.6 & 8.5 \\
 & & (3.7) & $9/2^+$ & 1d$_{5/2}$ & 0.96  & 12.3& 9.0 & 21.3\\
 & & (4.2) & $7/2^+$ & 1d$_{5/2}$ & 0.38  & 4.7 & 3.4 & 8.1 \\
 & & (4.4) & $3/2^+$ & 1d$_{5/2}$ & 0.2   & 2.5 & 1.7 & 4.2 \\
 & &(4.61) & $5/2^+$ & 1d$_{5/2}$ & 0.09   & 1.1  & 0.8 & 1.9 \\ 
 & &       &         & 2s$_{1/2}$ & 0.04   & 0.7 & 0.5 & 1.2 \\ 
 & &(4.65) & $9/2^+$ & 1d$_{5/2}$ & 0.2   & 2.4 & 1.7 & 4.1 \\
 & &(4.78) & $5/2^+$ & 2s$_{1/2}$ & 0.052   & 0.9 & 0.7 & 1.6 \\
 & &(5.6) & $5/2^+$ & 1d$_{5/2}$ & 0.12   & 1.4 & 0.9 & 2.3 \\
 & &(6.77) & $3/2^+$ & 1d$_{5/2}$ & 0.16   & 1.7 & 1.2 & 2.9 \\ 
\hline
\multicolumn{9}{r}{$\sigma_{-1n}^{Glauber}$=115 mb}\\
\hline
\end{tabular}

\noindent\begin{tabular}{c c c c c c c c c}
\hline
\hline
 $^A$Z & $J^\pi$ & $E_{ex}^c$  & $I^\pi_c$ & $nlj$  & $C^2S$ & 
$\sigma_{abs}$ & 
$\sigma_{diff}$ & $\sigma(I^\pi_c)$\\
  & & [MeV]  &  &   &  & [mb] & [mb] & [mb]\\
\hline
\hline
$^{25}$F & ($5/2^+$) & g.s.  & $3^+$ & 2s$_{1/2}$ & 0.82 & 21.6 & 21.0 & 
42.6 \\
 & &	   &	   & 1d$_{5/2}$ & 0.08 & 1.3  & 1.0    & 2.3  \\
 & & (0.1) & $2^+$ & 2s$_{1/2}$ & 0.64 & 16.6 & 16.0 & 32.6 \\
 & & (0.8) & $1^+$ & 1d$_{3/2}$ & 0.06 & 0.7  & 0.6  & 1.3 \\
 & & (2.2) & $4^+$ & 1d$_{5/2}$ & 1.0  & 13  & 10.0  & 23  \\
 & & (2.5) & $3^+$ & 1d$_{5/2}$ & 0.7  & 8.4  & 6.3  & 14.7 \\
 & & (2.8) & $1^+$ & 1d$_{5/2}$ & 0.18 & 2.1  & 1.6  & 3.7  \\
 & & (2.9) & $1^+$ & 1d$_{5/2}$ & 0.1  & 1.2  & 0.9  & 2.1  \\
 & & (3.5) & $5^+$ & 1d$_{5/2}$ & 1.6  & 18.3 & 13.5 & 31.8 \\
\hline
\multicolumn{9}{r}{$\sigma_{-1n}^{Glauber}$=154 mb}\\

\end{tabular}

\newpage

%\section{TABLES}

%%%%%%%%%%%%%%%%%%%%%%%%%%%%%%%%%%%%%%%%%%%%%%%%%%%%%%%%%%%%%%%%%%%%%%%%%%%%
%%
%%  Tables
%%
%%%%%%%%%%%%%%%%%%%%%%%%%%%%%%%%%%%%%%%%%%%%%%%%%%%%%%%%%%%%%%%%%%%%%%%%%%%%
%%
\newpage
\onecolumn{

%%%%%%%%%%%%%%%%%%%%%%%%%%%%%%%%%%%%%%%%%%%%%%%%%%%%%%%%
%%%%%%    table 1 : experimental effects %%%%%%%%%%%%%%%
%%%%%%%%%%%%%%%%%%%%%%%%%%%%%%%%%%%%%%%%%%%%%%%%%%%%%%%%

\begin{table}[!htbp]
\caption{Examples of the contributions of different experimental broadening 
effects
 on the widths (FWHM) of the core fragment momentum distributions (see 
text). }

\begin{center}
\begin{tabular}{ c | c c c c c c c c c}
\hline
   & $^A$Z   &            & Lorentz&   & target&   & spectrometer &                 
 & broadening \\
 &  & intrinsic       & $\Rightarrow$ &   &$\Rightarrow$&   &$\Rightarrow$&   
measured       & [\%] \\
\hline
FWHM$_{pz}$ & $^{14}$B & 56.3& & 59 & &  63 & & 63.6 & 11  \\
(C target)              & $^{19}$N & 166& & 175.5 & & 187.6 & & 188.4 & 11.6 
\\
\hline
FWHM$_{pz}$ &  $^{14}$B & 57& & 60 & &  61.8 & & 62.2 & 8.4  \\
(Ta target)             &  $^{19}$N & 176& & 186 & &  193 & & 194 & 9.3 \\
\hline
\multicolumn{10}{l}{}\\
%\multicolumn{10}{l}{}\\
\hline
 & $^A$Z      &  & target   &&  tracking  &    & spectrometer &  & 
broadening \\
 &  & intrinsic &$\Rightarrow$&    &$\Rightarrow$&    &$\Rightarrow$&   
measured    
& [\%] \\
\hline
FWHM$_{px}$  & $^{14}$B & 75& & 77 & &  76.8 & & 79 & 5  \\
(C target)   & $^{19}$N & 226& & 227 & &  228 & & 229 & 1.3 \\
\hline
\end{tabular}
\label{tab1:widths}
\end{center}
\end{table}

%%%%%%%%%%%%%%%%%%%%%%%%%%%%%%%%%%%%%%%%%%%%%%%%%%%%%%
%%%%%%%%%%%% Table 2 : results on C target, summary
%%%%%%%%%%%%%%%%%%%%%%%%%%%%%%%%%%%%%%%%%%%%%%%%%%%%%%

\begin{table}
\caption{Summary of the results obtained with the Carbon target.  Where 
available
the results of other experiments are also listed.}
\begin{center}
\begin{tabular}{ccccccccc}
\hline
$^A$Z & Energy  &  FWHM$_{pz}^{lab}$ & FWHM$_{pz}^{cm}$ & FWHM$_{px}^{lab}$ 
&  
FWHM$_{px}^{cm}$ &  $\sigma_{-1n}$ & 
$\sigma_{-1n}^{Glauber}$ & $J^\pi$\\
      & [MeV/nucleon] &   [MeV/$c$] &  [MeV/$c$] &  [MeV/$c$] &  [MeV/$c$]          
 & [mb]            &
[mb] & \\
\hline
$^{12}$B & 64 & 158 $\pm$ 3.5 & 142 $\pm$  3.5 & 175 $\pm$ 3 & 173 $\pm$ 3 & 
81 
$\pm$ 5 & 91  & 1$^+$    \\
$^{13}$B & 54 &  150 $\pm$ 7  &  135 $\pm$ 7 &  178 $\pm$ 2 & 176 $\pm$ 2 & 
59 
$\pm$ 4 & 62  & 3/2$^-$  \\
$^{14}$B & 50 &  63.6 $\pm$ 0.5  &  56.5 $\pm$ 0.5 &  79 $\pm$ 2 & 75 $\pm$ 
2 & 153 
$\pm$ 15 & 185 & 2$^-$    \\
         & 86 & & 57 $\pm$ 2 $^a$& & &  48 $\pm$ 5 $^a$   &     &     \\
         & 59 & & 55 $\pm$ 2 $^b$ & & &  176 $\pm$ 16 $^b$&     &     \\
$^{15}$B & 43 & 82 $\pm$ 2.5     & 73 $\pm$ 2.5 &  84 $\pm$ 9 & 80 $\pm$ 9 & 
108 
$\pm$ 13 & 112  & 3/2$^-$ $^c$\\
\hline
$^{14}$C & 67 & 200 $\pm$ 5 & 180  $\pm$  5 &  222 $\pm$ 3 & 220 $\pm$ 3 & 
65 $\pm$ 
4 & 89  & 0$^+$\\
$^{15}$C  & 62 & 71 $\pm$ 0.7 & 63.5  $\pm$  0.7 & 81 $\pm$ 1 & 86 $\pm$ 1 & 
 159 
$\pm$ 15& 168  & 1/2$^+$\\
         & 54 & &                   & & &  137 $\pm$ 16 $^d$&  &  \\
         & 85 & & 67   $\pm$ 3 $^a$ & & &  33 $\pm$ 3 $^{a,q}$ &   & \\
$^{16}$C & 55 & 121 $\pm$ 2 & 108 $\pm$ 2 & 143 $\pm$ 3 & 140 $\pm$ 3 &65 
$\pm$ 6& 
75  & 0$^+$\\
	 & 62 & &      & & &  77 $\pm$ 9 $^e$ &   & \\
	 & 83 & & 90 $\pm$ 9  $^r$    & & &  65$^{+15}_{-10}$ $^r$ &   & \\         
$^{17}$C & 49 & 125 $\pm$ 3 & 111  $\pm$ 3 & 169 $\pm$ 9 & 166 $\pm$ 9 &84 
$\pm$ 9 
& 105  & 3/2$^+$ $^{c,a,e,o,t}$\\
         & 84  & & 145 $\pm$ 5 $^a$  & & &  26 $\pm$ 3 $^{a,q}$  &   & \\
	 & 62  & &  & & &  115 $\pm$ 14 $^e$  &   & \\
         & 96.8  & & 94 $\pm$ 19 $^f$ & & &  41 $\pm$ 4 $^{f,q}$  &   & \\
         & 904  & & 141 $\pm$ 6 $^g$ & & &  129$\pm$22 $^h$   &   & \\
$^{18}$C & 43 & 143 $\pm$ 5 & 126  $\pm$ 5 & 159 $\pm$ 16 & 156 $\pm$ 16 & 
115 
$\pm$ 18& 119  & 0$^+$ \\
         & 86.2  & & 110 $\pm$ 12 $^f$   & & &  35 $\pm$ 2 $^{f,q}$ &   & \\
\hline
$^{17}$N & 65 & 158 $\pm$ 4 &141 $\pm$ 4 &217 $\pm$ 4 & 214 $\pm$ 4 &  55 
$\pm$ 5 & 
67  & 1/2$^-$ \\
$^{18}$N & 59 &  188 $\pm$ 3 &  168 $\pm$ 3 &  219 $\pm$ 3 & 216 $\pm$ 3 
&109 $\pm$ 
11 & 91  & 1$^-$ \\
$^{19}$N  & 53 &  199 $\pm$ 3 & 177 $\pm$ 3 &  229 $\pm$ 5 & 226 $\pm$ 5 & 
86 
$\pm$ 9 & 99  & 1/2$^-$ $^{c,g}$\\
 $^{20}$N & 48 & 184 $\pm$ 4 & 162 $\pm$ 4 & 220 $\pm$ 16 & 217 $\pm$ 16 & 
98 $\pm$ 
13 & 101 & 2$^-$ $^{c}$ \\
$^{21}$N & 43 & 173$\pm$ 7 & 149 $\pm$ 7 &  &  & 140 $\pm$ 44 & 151 & 
1/2$^-$ $^c$ 
\\
\hline
$^{19}$O & 68 & 214 $\pm$ 8 & 190 $\pm$ 8 & 253 $\pm$ 13 & 250 $\pm$ 13 &104 
$\pm$ 
12 & 84   & 5/2$^+$ \\
$^{20}$O & 62 &  247 $\pm$ 5 & 219 $\pm$ 5 & 254 $\pm$ 7  & 251 $\pm$ 7  & 
112 
$\pm$ 11 & 96   & 0$^+$ \\
$^{21}$O  &  56  & 237 $\pm$ 6 & 210 $\pm$ 6 & 246 $\pm$ 7 & 243 $\pm$ 7 
& 134 $\pm$ 14 & 123  & 5/2$^+$ $^{c,g}$ \\
$^{22}$O & 51 & 235 $\pm$ 4 & 206 $\pm$ 4 & 240 $\pm$ 16 & 237 $\pm$ 16 &120 
$\pm$ 
14 & 143  & 0$^+$ \\
 $^{23}$O & 47 & 135 $\pm$ 9 &  114 $\pm$ 9 & 162 $\pm$ 28.5 & 157 $\pm$ 
28.5 
 & -- $^n$          & 224 $^s$ & 1/2$^+$ $^c$ \\
         & 72  &  94 $\pm$ 12 $^j$ &  175 $\pm$ 14 $^j$ &  &  & 233 $\pm$ 37 
$^j$  
& 185 $^s$ & \\
         & 936 &       &  123 $\pm$ 10 $^p$ &  &  & 85 $\pm$ 15 $^p$  & 82 
$^s$ & \\
\hline
$^{22}$F & 64 & 212 $\pm$ 14 & 185 $\pm$ 14 &278 $\pm$ 28 & 274 $\pm$ 28 & 
121 
$\pm$ 16 & 87   & 4$^+$ \\
 $^{23}$F & 59 &  267 $\pm$ 4 &  235 $\pm$ 4 &  236 $\pm$ 10 &  232 $\pm$ 10 
& 114 $\pm$ 12& 106  & 5/2$^+$ $^{c,k,l}$\\
 $^{24}$F  & 54 & 151 $\pm$ 4 & 129 $\pm$ 4 & 203 $\pm$ 18 & 198 $\pm$ 18 
& 124 $\pm$ 16 & 115  & 3$^+$ 
$^{c,m}$ \\
$^{25}$F & 50 & 128 $\pm$ 8 & 106 $\pm$ 8 & 173 $\pm$ 45 & 168 $\pm$ 45 & 
173 $\pm$ 
46& 154  & 5/2$^+$ $^c$ \\
\hline \\
\multicolumn{9}{l}{\footnotesize
$a$ ref. \protect\cite{baz98} (Be target), $b$ ref. \protect\cite{gui00} 
(Be target),} \\
\multicolumn{9}{l}{\footnotesize
$c$ assignment from present experiment,  $d$ ref. \protect\cite{JeffCDCC}
$e$ ref. \protect\cite{mad01} (Be target), $f$ ref. \protect\cite{baz95} 
(Be target), } \\
\multicolumn{9}{l}{\footnotesize $g$ ref. \protect\cite{bau98}, 
(C target), $h$ ref. \protect\cite{cor01} 
(C target),} \\
\multicolumn{9}{l}{\footnotesize $i$ ref. \protect\cite{cat89}, 
$j$ ref. \protect\cite{Kan02} (C target), $k$ ref. \protect\cite{orr89}, $l$ 
ref. 
\protect\cite{goo74},
$m$ ref. \protect\cite{ree99},}\\
\multicolumn{9}{l}{\footnotesize $n$ no beam intensity normalization 
available,
$o$ ref. \protect\cite{Oga02}, $p$ ref. \protect\cite{priv_lola} (C 
target),} \\
\multicolumn{9}{l}{\footnotesize $q$ affected by limited transverse momentum 
acceptances,
$r$ ref. \protect\cite{Yam03}, $s$ see Table IV, $t$ ref. \protect\cite{Dat03} } \\
\end{tabular}

\label{tab2:res_c}
\end{center}
\end{table}

%%%%%%%%%%%%%%%%%%%%%%%%%%%%%%%%%%%%%%%%%%%%%%%%%%%%%%%%%%%%%%%%%%%%%%%%%%%% 
%%%       Table results Ta
%%%%%%%%%%%%%%%%%%%%%%%%%%%%%%%%%%%%%%%%%%%%%%%%%%%%%%%%%%%%%%%%%%%%%%%%%%%% 
\begin{table}[!htbp]
\caption{Summary of the results obtained with the Tantalum target.  Where 
available
the results of other experiments are also listed.}
\begin{tabular}{c c c c c c}
\hline

$^A$Z & Energy & FWHM$_{pz}^{lab}$ & FWHM$_{pz}^{cm}$ & $\sigma_{-1n}$ 
&$\sigma_{-1n}^{Glauber}$\\
      & [MeV/nucleon]   &  [MeV/$c$]           &   [MeV/$c$] & [mb] & [mb]\\
\hline
$^{14}$B & 50 & 62 $\pm$ 2 & 57 $\pm$ 2 & -- $^d$ &  864  \\
         & 59 &            & 59 $\pm$ 3 $^{b}$ & 638 $\pm$ 45 $^b$ &  \\
         & 86 &            & 48 $\pm$ 3 $^{a}$ & 157 $^{a,e}$ &   \\
\hline
$^{15}$C & 62 & 69 $\pm$ 0.5 & 63  $\pm$  0.5 & -- $^d$ & 978 \\
         & 85 &            & 67 $\pm$ 1 $^{a}$ & 75 $^{a,e}$ &   \\
$^{16}$C & 55 &106 $\pm$ 3 & 97 $\pm$ 3 & -- $^d$ & 193\\
$^{17}$C & 49 &131 $\pm$ 7 & 121  $\pm$ 7   & -- $^d$ & 280 \\
         & 61 &            &  & 350 $\pm$ 67 $^{c}$&  \\
         & 62 &            & $\sim$110 $^f$ &  &  \\        
\hline
$^{17}$N & 65 & 147 $\pm$ 5 & 134 $\pm$ 5 & -- $^d$ & 173 \\
$^{18}$N & 59 & 176 $\pm$ 5 & 159 $\pm$ 5 & -- $^d$ & 238 \\
$^{19}$N & 53 & 194 $\pm$ 11 & 176 $\pm$ 11 & -- $^d$ & 216 \\
\hline
\multicolumn{6}{l}{\footnotesize
$a$ ref. \protect\cite{baz98} (Ta target), $b$ ref. \protect\cite{gui00} 
(Au target),  $c$ ref. \protect\cite{mad01c} (Au target),} \\
\multicolumn{6}{l}{\footnotesize $d$ no reliable cross section could be 
estimated 
owing to 
very broad transverse momentum distributions (see text),} \\
\multicolumn{6}{l}{\footnotesize $e$ affected by restricted transverse 
momentum 
acceptances,
$^f$ ref. \protect\cite{orrrnbiii} ($^{238}$U target) } \\
\end{tabular}
\label{tab3:res_ta}
\end{table}

\bigskip
\bigskip
\bigskip

%%%%%%%%%%%%%%%%%%%%%%%%%%%%%%%%%%%%%%%%%%%%%%%%%%%%%%%%%%%%%
%% table o23 comparison glauber calcs at 47, 72 and 900 MeV/u
%%%%%%%%%%%%%%%%%%%%%%%%%%%%%%%%%%%%%%%%%%%%%%%%%%%%%%%%%%%%%
%%%%%%%%%%%%%%%%%%%%%%%%%%%%%%%%%%%%%%%%%%%%%%%%%%%%%%%%%%%
% tabel nou
\begin{table}
\caption{Predicted cross sections for one-neutron removal
 from $^{23}$O ($J^\pi_{gs}$=1/2$^+$) by a Carbon
target at 47 (present experiment), 72$^a$ and 900$^b$ MeV/nucleon. The core 
excited 
states 
and spectroscopic
factors are those listed in Appendix B.}
\begin{center}
\begin{tabular}{ccccc}
\hline
$E^c_x$ [MeV] &$I^{\pi}_c$& \multicolumn{3}{c}{ $E_{lab}$ [MeV/nucleon] } \\ 
  &    
&    47 & 72 &  900 \\
\hline
  g.s.  & $0^+$     & 65.5                    &53.0   &27.2  \\
 3.38   & $2^+$     & 1.2                     &0.99   &0.4  \\
 	&           &  55                     &45.4   &19.3  \\
 4.62   & $0^+$     &  4.1                    &3.33   &1.6  \\ 
 4.83   & $3^+$     & 72.2                    &59.8   &24.6  \\ 
 6.5    & $2^+$     &  5.0                    &4.2    &1.7  \\ 
 6.64   & $0^-$     &  6.0                    &4.9    &1.9  \\ 
 6.9    & $1^-$     & 15.1                    &13.2   &5.2  \\ 
\hline
        & $\sigma_{-1n}^{Glauber}$&224        &185&82 \\
\hline
        &$\sigma_{-1n}^{exp}$     &--         &233$\pm$37$^a$&84$\pm$11$^b$\\
\hline
{\footnotesize $^a$ \protect ref. \cite{Kan02}~$^b$ \protect ref. 
\cite{priv_lola} 
}\\
\end{tabular}
\label{tab4:res_o23}
\end{center}
\end{table}

%%%%%%%%%%%%%%%%%%%%%%%%%%%%%%%%%%%%%%%%%%%%%%%%%%%%%%%%%%%%%%%%%%%%%%%%%%%%
%% figures and captions %%%%%%%%%%%%%%%%%%%%%%%%%%%%%%%%%%%%%%%%%%%%%%%%%%%%
%%%%%%%%%%%%%%%%%%%%%%%%%%%%%%%%%%%%%%%%%%%%%%%%%%%%%%%%%%%%%%%%%%%%%%%%%%%%
%%

\newpage

%%%%%%%%%%%%%%%%%%%%%%%%%%%%%%%%%%%%%%%%%%%%%%%%%%%%%%%%%%%%%%%%%%%%%%%%
%%%%%%%%%%%  pz distribtions 
%%%%%%%%%%%%%%%%%%%%%%%%%%%%%%%%%%%%%%%%%%%%%%%%%%%%%%%%%%%%%%%%%%%%%%%%
%fig 1
\begin{figure}[!htbp]
\begin{center}
\mbox{\epsfig{file=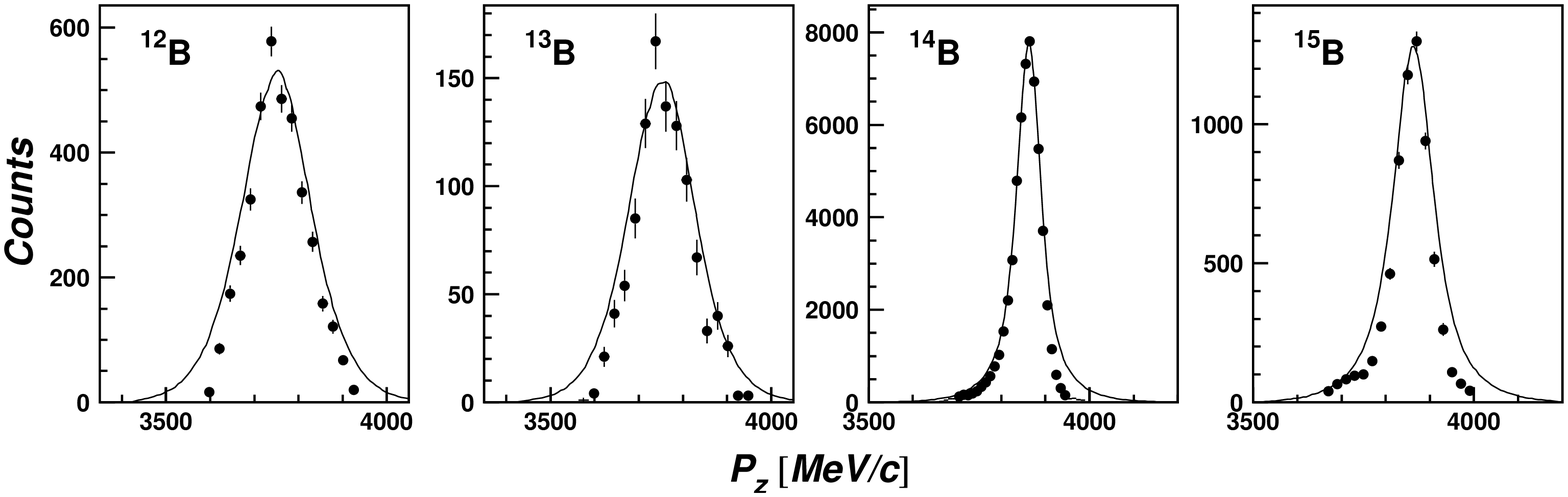,height=4cm}}
\mbox{\epsfig{file=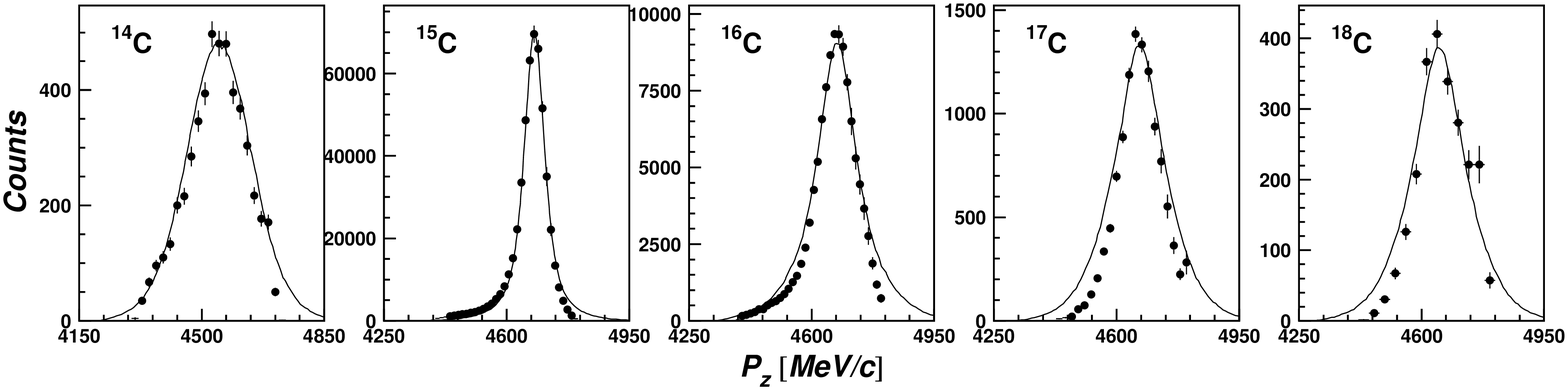,height=4cm}}
\mbox{\epsfig{file=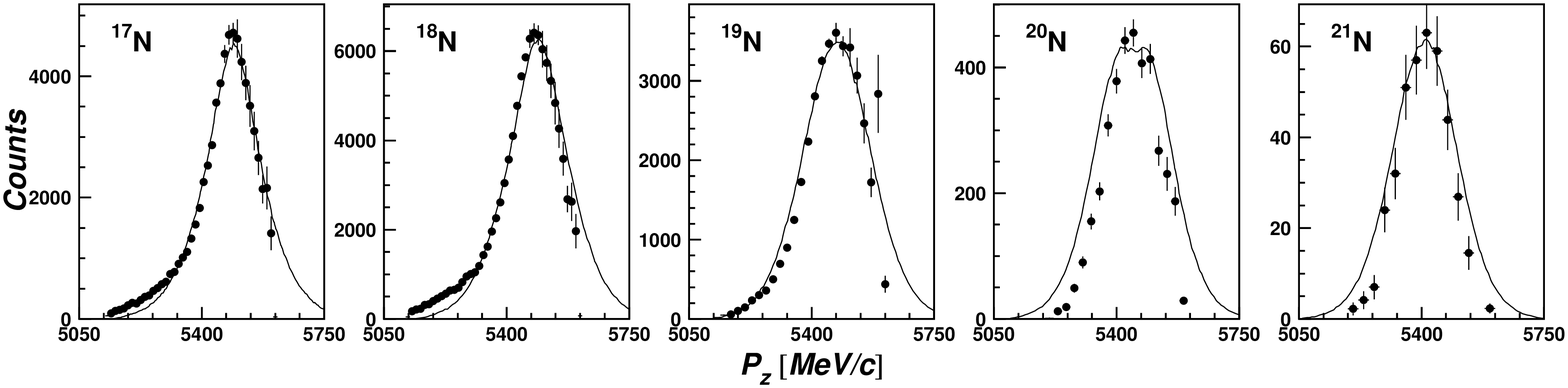,height=4cm}}
\mbox{\epsfig{file=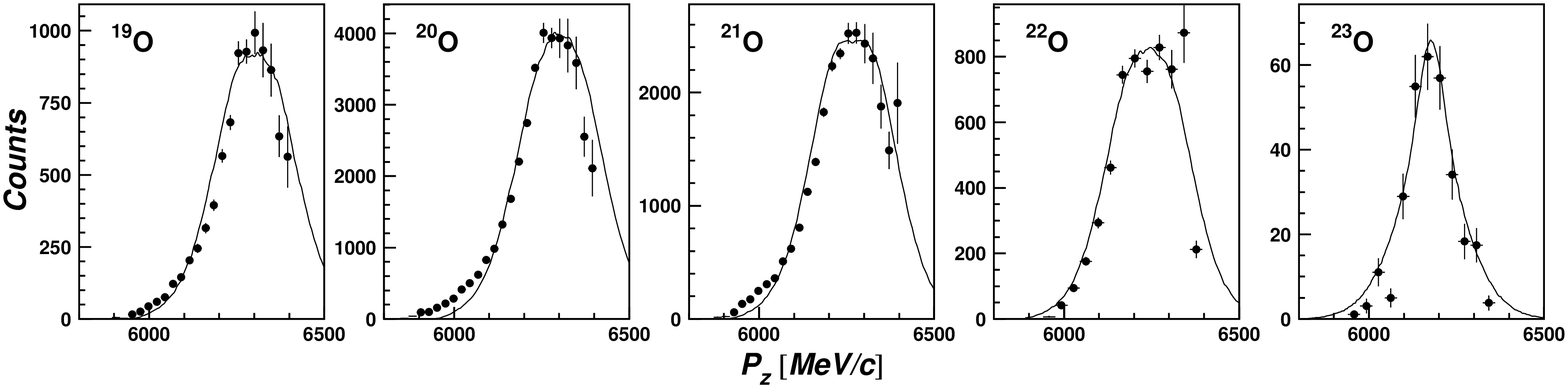,height=4cm}}
\mbox{\epsfig{file=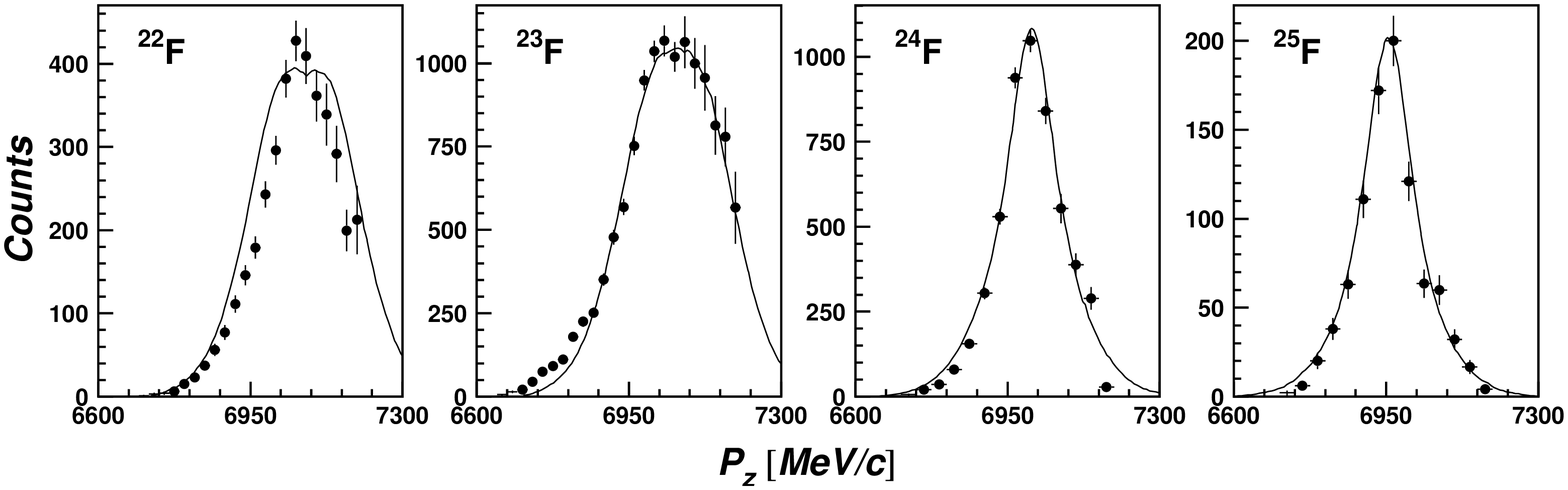,height=4cm}}
\end{center}
\caption{Comparison of the core fragment longitudinal momentum ($p_z$) 
distributions obtained using a 
Carbon target and the Glauber model calculations (solid line). }
\label{fig:sysglau}
\end{figure}

\newpage

%fig 2 

\begin{figure}[h]
\begin{center}
\mbox{\epsfig{file=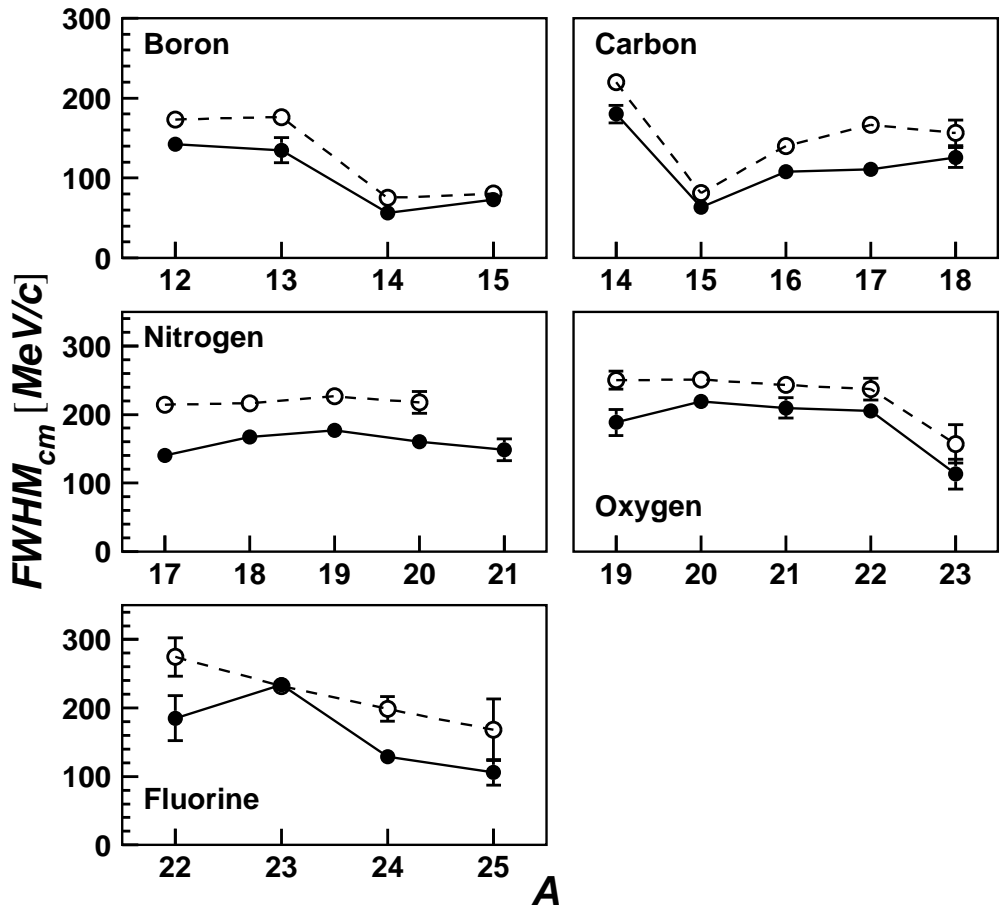,width=15cm}}
\end{center}
\caption{Comparison of the widths (FWHM) of the longitudinal (filled 
circles) 
and 
transverse (open circles) core fragment momentum distributions for 
reactions on the Carbon target.}
\label{fig:pxppar}
\end{figure}

%%%%%%%%%%%%%%%%%%%%%%%%%%%%%%%%%%%%%%%%%%%%%%%%%%%%%%%%%%%%%%%%%%%%%%%%
%%%%%%%%%%%  cross sections 
%%%%%%%%%%%%%%%%%%%%%%%%%%%%%%%%%%%%%%%%%%%%%%%%%%%%%%%%%%%%%%%%%%%%%%%%

\newpage

%fig 3 

\begin{figure}[h]
\begin{center}
\mbox{\epsfig{file=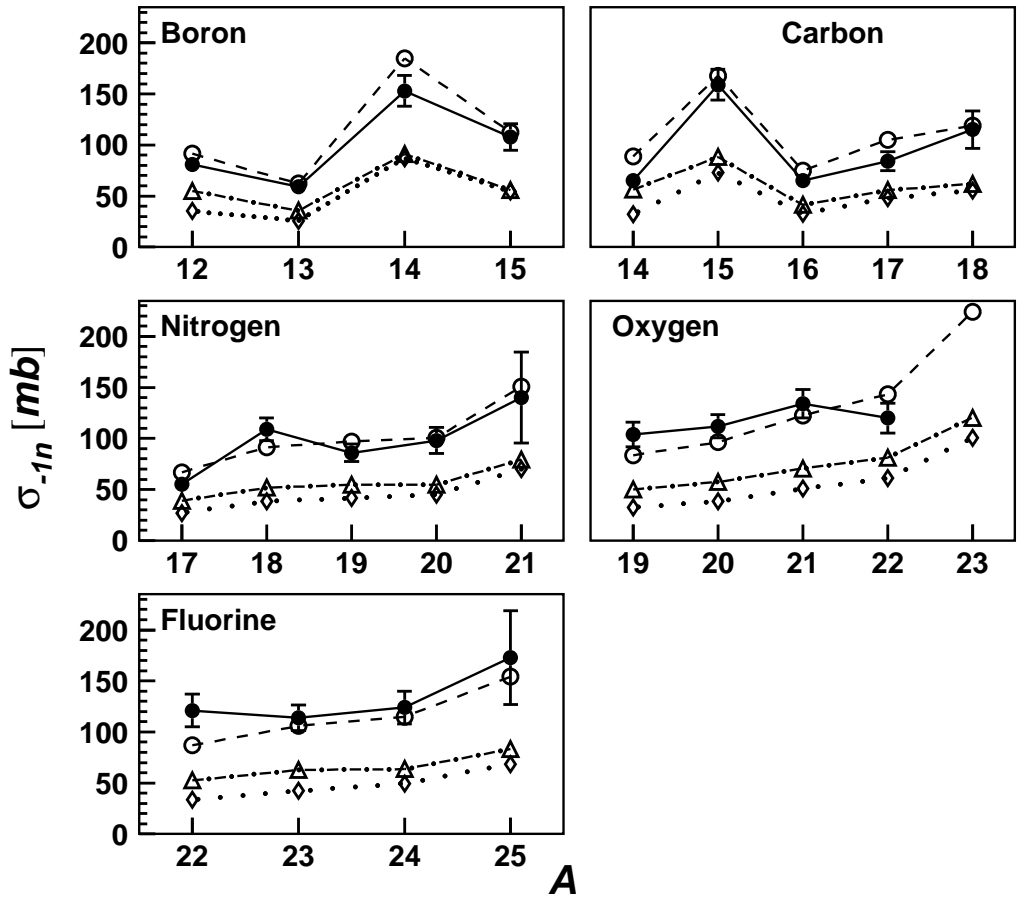,width=15cm}}
\end{center}
\caption{Experimental one-neutron removal cross sections (filled circles) 
compared 
to the results of the Glauber calculations -- open circle: total 
cros section, open 
triangle: absorption and open diamond: diffraction (see text for details). 
The 
points are 
connected by
lines to guide the eye.}
\label{fig:sefglau}
\end{figure}

\newpage

%%%%%%%%%%%%%%%%%%%%%%%%%%%%%%%%%%%%%%%%%%%%%%%%%%%%%%%%%%%%%%%%%%%%%%%%
%%%%%%%%%%%  px distribtions 
%%%%%%%%%%%%%%%%%%%%%%%%%%%%%%%%%%%%%%%%%%%%%%%%%%%%%%%%%%%%%%%%%%%%%%%%

%fig 4 

\begin{figure}[!htbp]
\begin{center}
\mbox{\epsfig{file=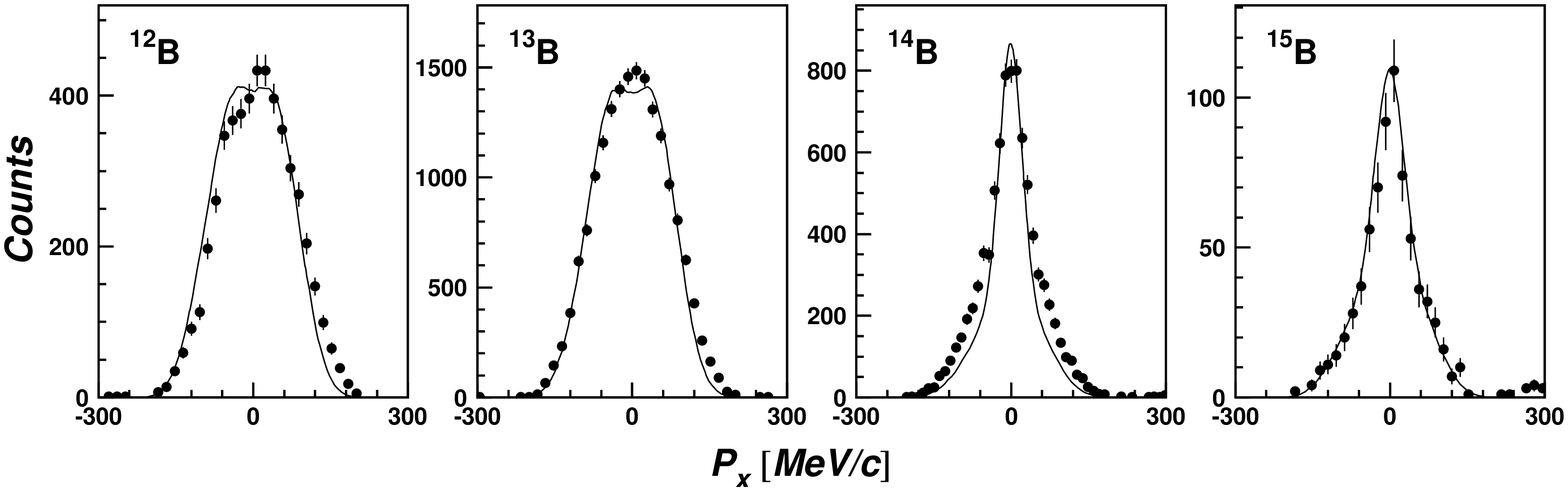,height=4cm}}
\mbox{\epsfig{file=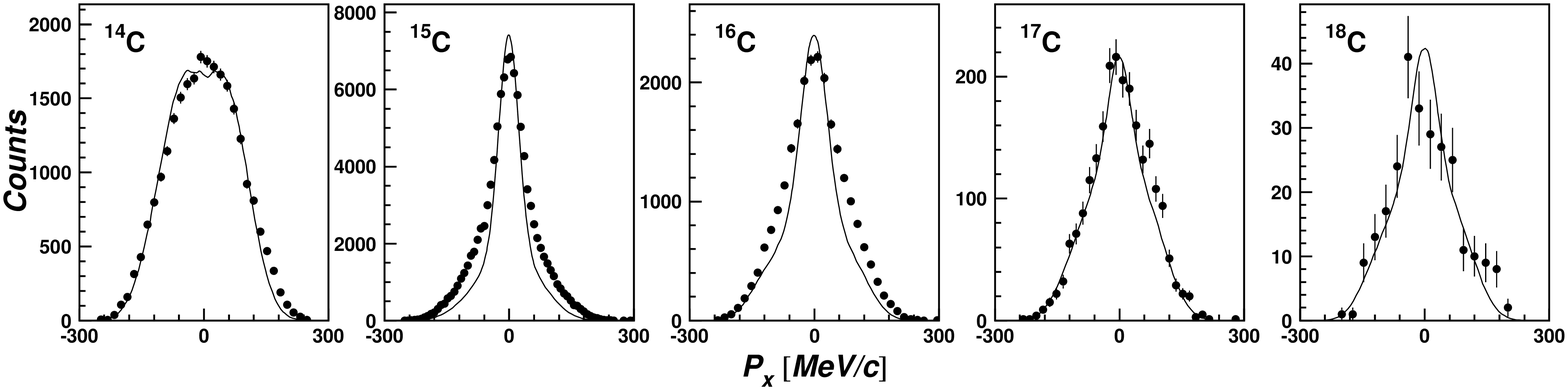,height=4cm}}
\mbox{\epsfig{file=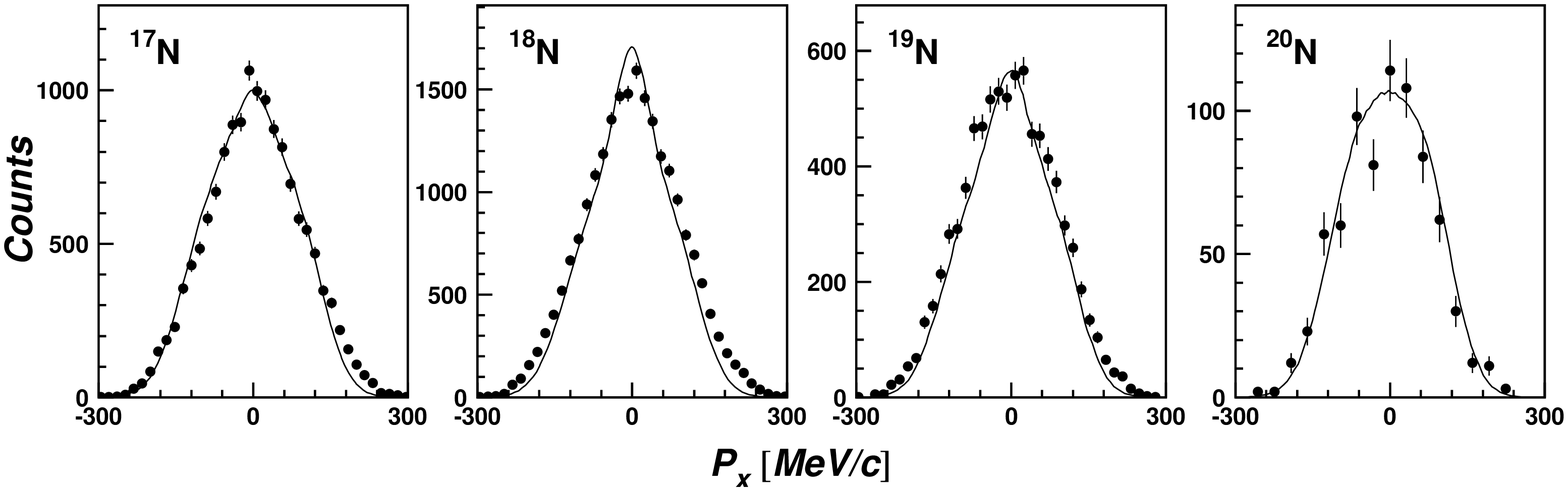,height=4cm}}
\mbox{\epsfig{file=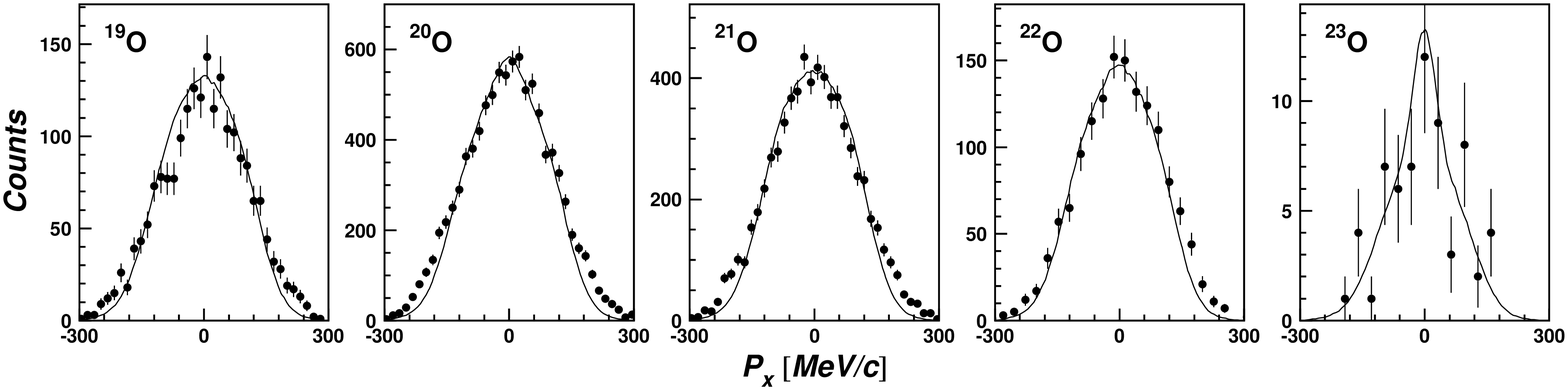,height=4cm}}
\mbox{\epsfig{file=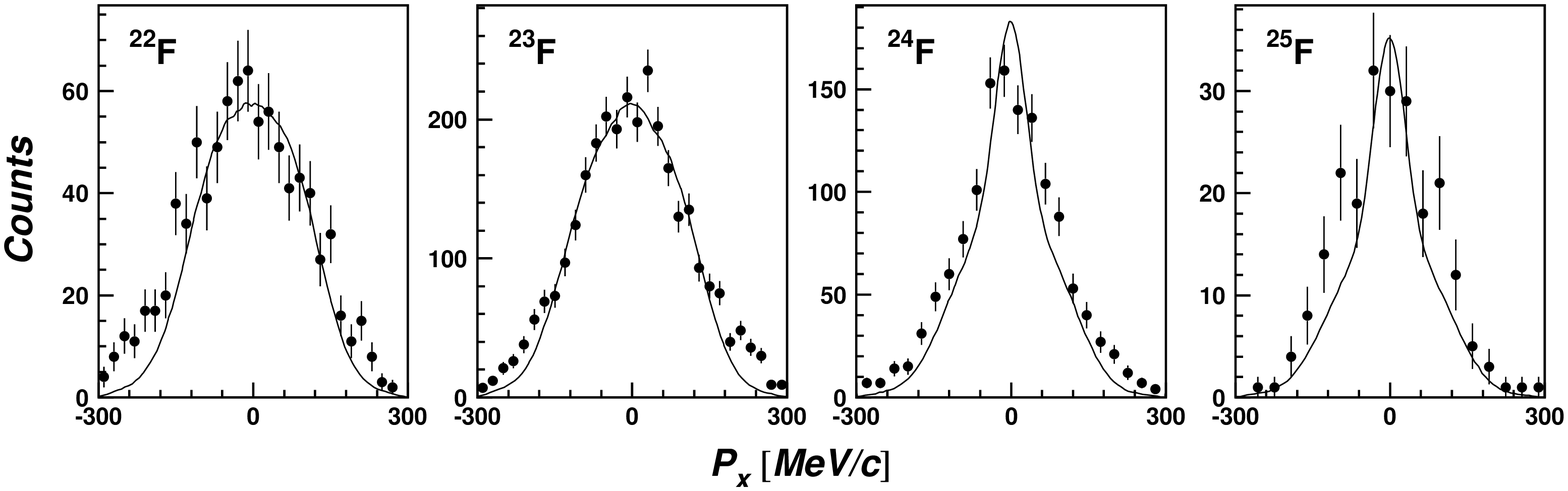,height=4cm}}
\end{center}
\caption{Comparison of the core fragment transverse momentum ($p_x$) 
distributions obtained using a 
Carbon target and the Glauber model calculations (solid line).}
\label{fig:sysglaupx}
\end{figure}

\newpage

%%%%%%%%%%%%%%%%%%%%%%%%%%%%%%%%%%%%%%%%%%%%%%%%%%%%%%%%%%%%%%%%%%%%%%%%
%%%%%%%%%%%  pz distribtions  Ta
%%%%%%%%%%%%%%%%%%%%%%%%%%%%%%%%%%%%%%%%%%%%%%%%%%%%%%%%%%%%%%%%%%%%%%%%

% fig 5

\begin{figure}[!htbp]
\begin{center}
\mbox{\epsfig{file=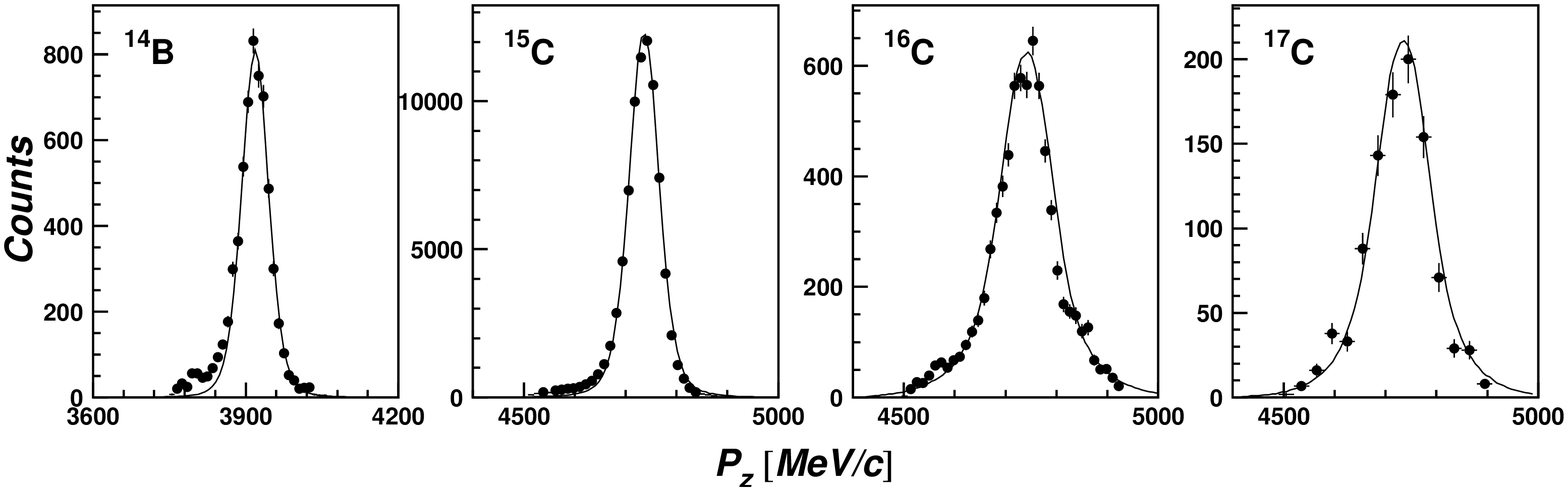,height=4cm}}
\mbox{\epsfig{file=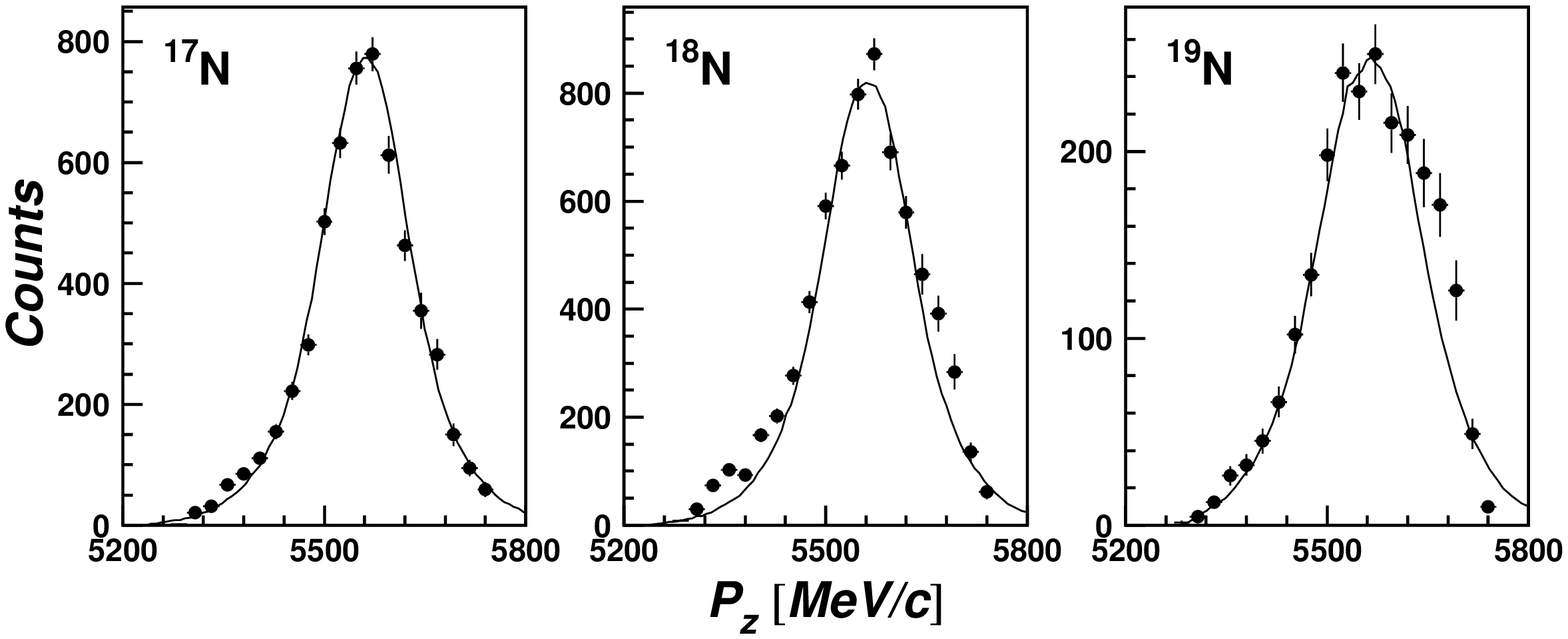,height=4cm}}
\end{center}
\caption{Comparison of the core fragment longitudinal momentum ($p_z$) 
distributions obtained using a 
Tantalum target and the Glauber model calculations (solid line).}
\label{fig:systag}
\end{figure}

\bigskip
\bigskip

%%%%%%%%%%%%%%%%%%%%%%%%%%%%%%%%%%%%%%%%%%%%%%%%%%%%%%%%%%%%%%%%%%%%%%%%
%%%%%%%%%%%  other figures
%%%%%%%%%%%%%%%%%%%%%%%%%%%%%%%%%%%%%%%%%%%%%%%%%%%%%%%%%%%%%%%%%%%%%%%%

%fig 6

\begin{figure}[h]
\begin{center}
\mbox{\epsfig{file=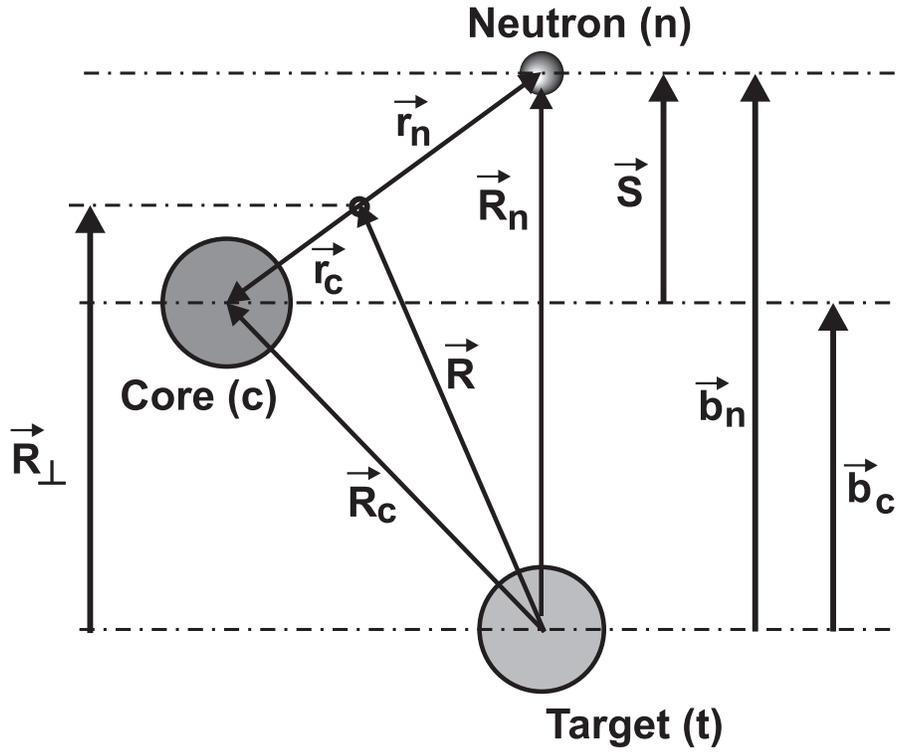,width=12cm}}
\end{center}
\caption{Coordinate system used in the Glauber model calculations.}
\label{fig:figglau}
\end{figure}

\newpage

%fig 7

\begin{figure}[h]
\begin{center}
\mbox{\epsfig{file=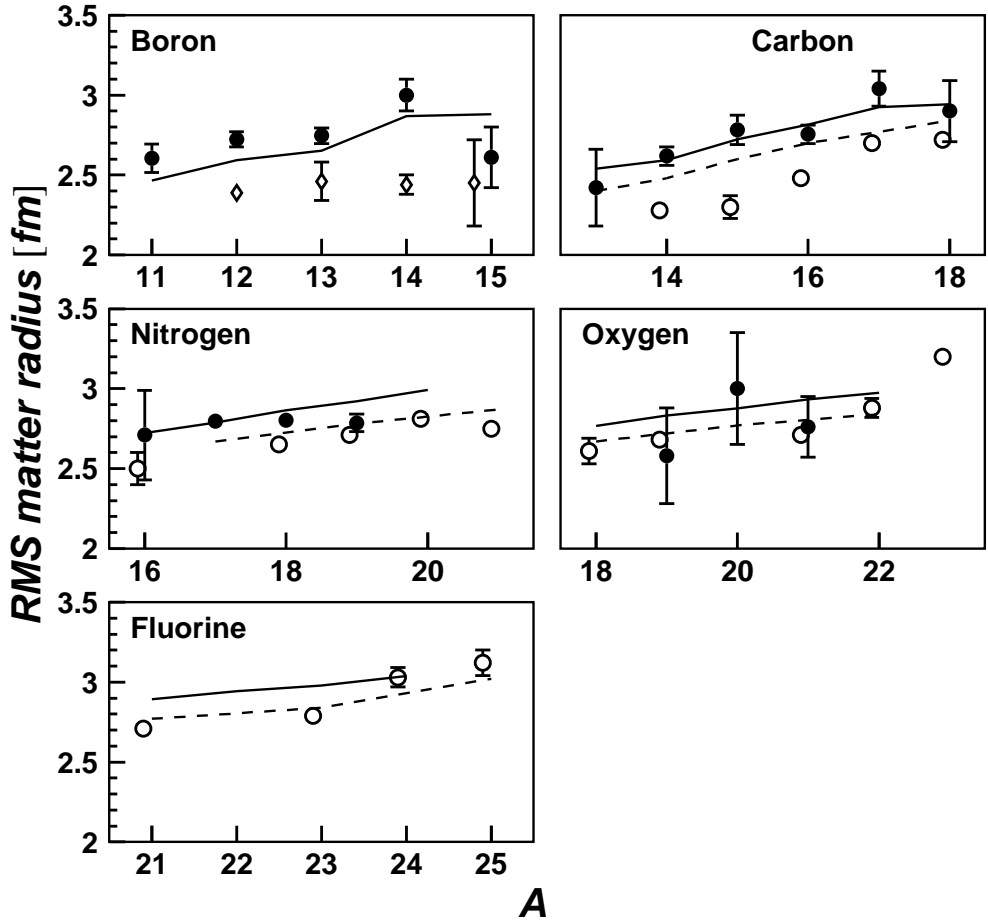,width=15cm}}
\end{center}
\caption{$RMS$ matter radii extracted from Hartree-Fock + BCS calculations 
of 
single particle densities (black line), compared with experimental values 
from Liatard 
\etal \protect\cite{liatard} (filled circles), Ozawa \etal 
\protect\cite{oza01} 
(open circles) and Tanihata \etal 
\protect\cite{tan88} (open diamonds). The relativistic mean field 
calculations 
 of Ren \etal \protect\cite{ren95,ren96a,ren96b} are indicated by the dashed 
lines.}
\label{fig:mat-radii}
\end{figure}

\newpage

%fig 8

\begin{figure}[h]
\begin{center}
\mbox{\epsfig{file=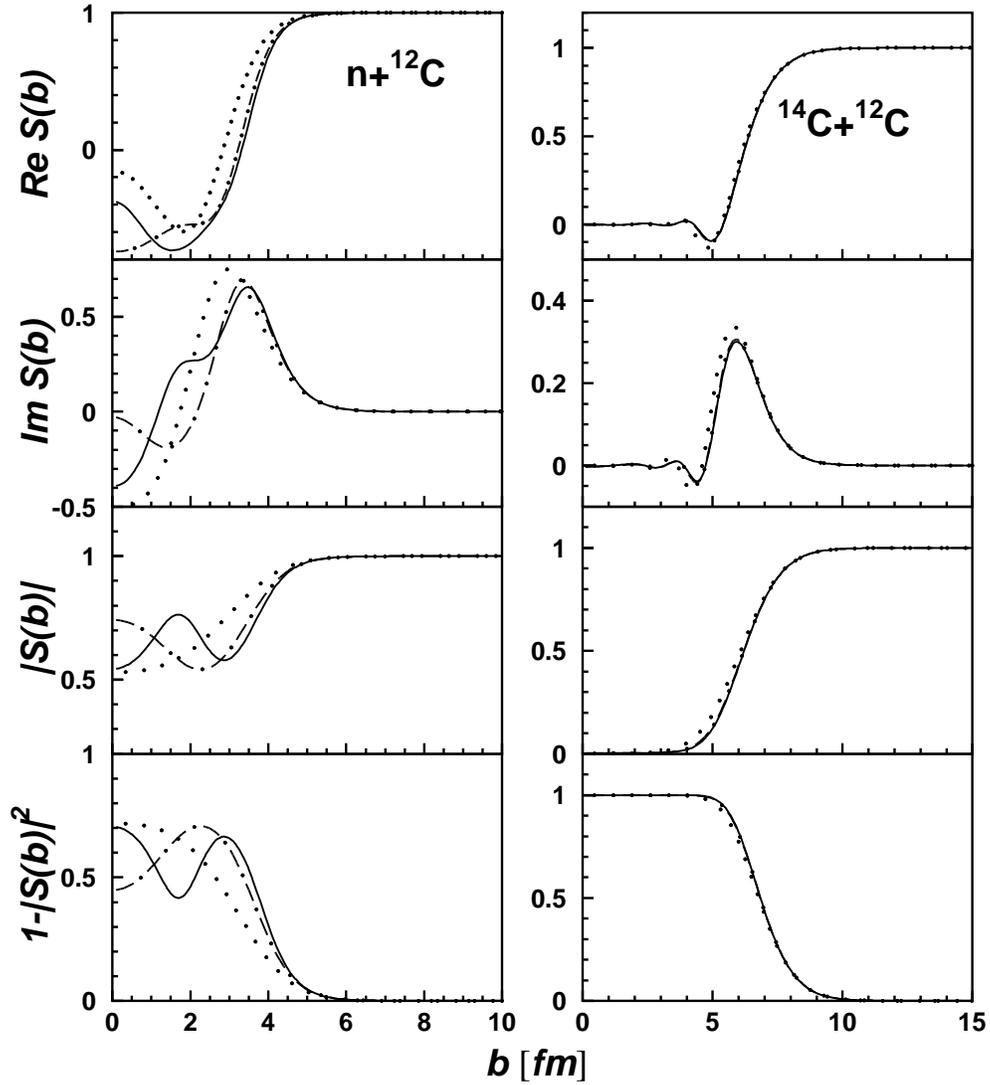,width=15cm}}
\end{center}
\caption{S-matrix, absorption profiles and transmission coefficients for 
n--$^{12}$C (left panels) 
and $^{14}$C--$^{12}$C  (right panels) interactions at 30 MeV/nucleon 
obtained from 
the JLM folding
model calculations. Results using the 
lowest order of the eikonal approximation are shown as a dotted line, those using the 
first 
order
corrections by the dot-dashed line, and those using second order
corrections by the solid line. For $^{14}$C+$^{12}$C the calculations 
converge 
with the use of first order corrections. }
\label{fig:smatrix}
\end{figure}

\newpage

%fig 9

\begin{figure}[h]
\begin{center}
\mbox{\epsfig{file=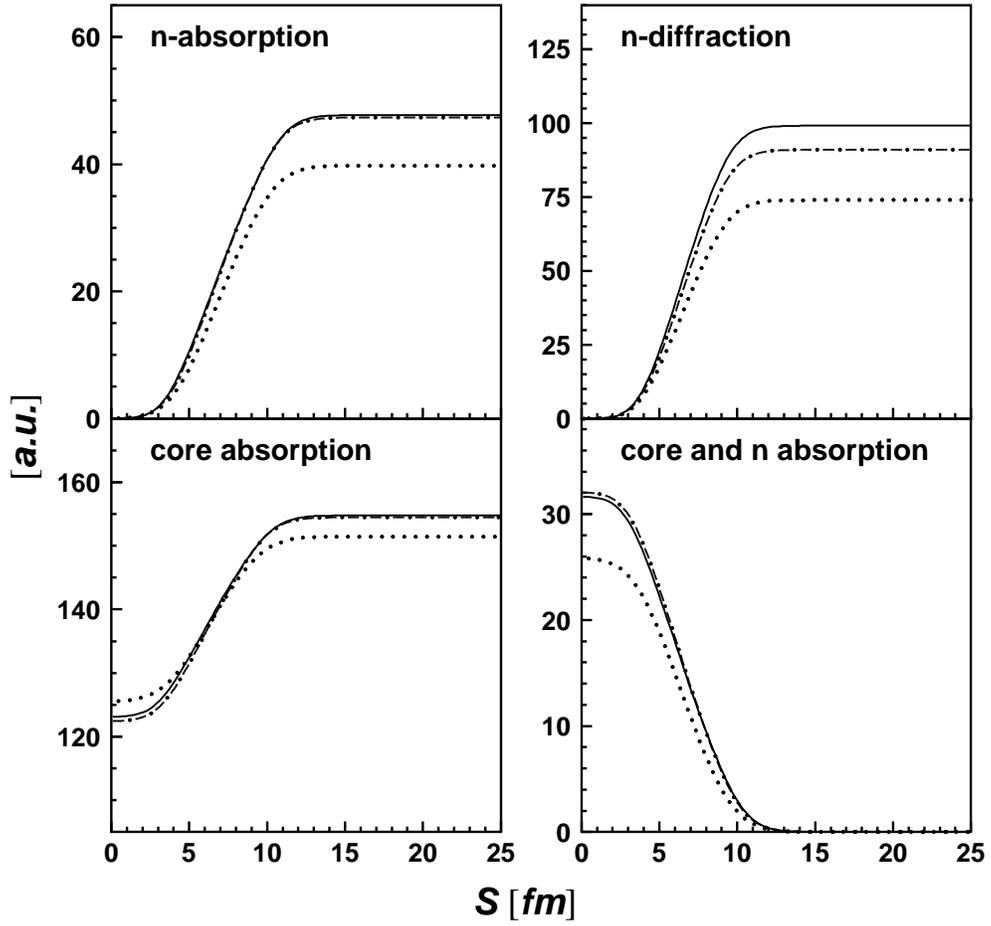,width=15cm}}
\end{center}
\caption{The effect of noneikonal corrections on the distortion kernels 
(see the text for definitions) as a function of impact parameter $s$, 
for 
neutron 
absorption and diffraction, core ($^{14}$C) absorption and 
for absorption of both the core and neutron by $^{12}$C at 30~MeV/nucleon. 
Dotted lines: lowest order eikonal approximation.  Dashed-dotted
lines: first order noneikonal corrections included.  Solid line: second 
order 
corrections
included. }
\label{fig:dfunc}
\end{figure}

\newpage

%fig 10
\begin{figure}[h]
\begin{center}
\mbox{\epsfig{file=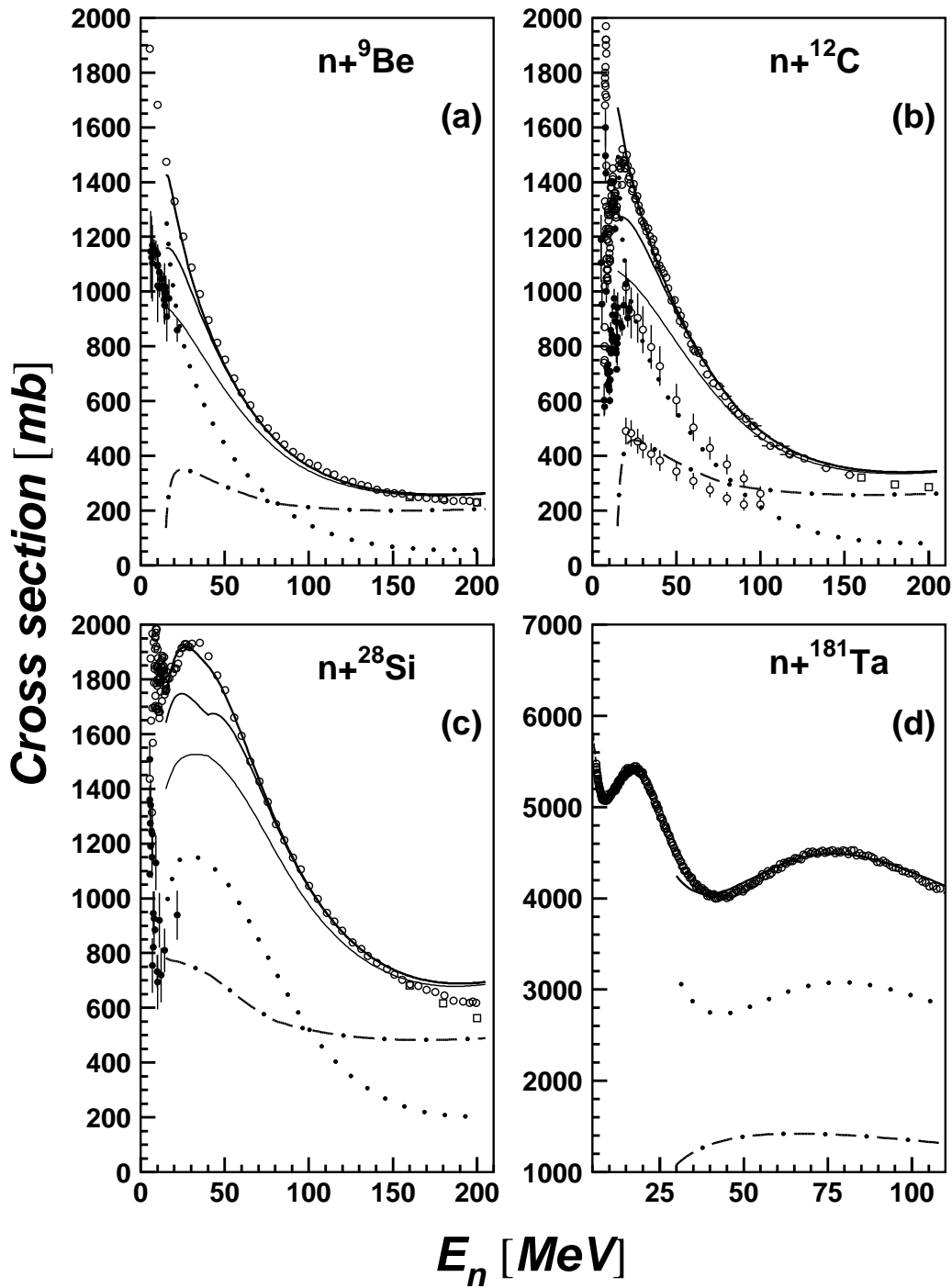,width=15cm}}
\end{center}
\caption{Elastic (dotted), reaction (dash-dotted) and total (solid line) 
neutron cross 
sections for reaction on four targets as a function of incident energy. The 
calculations were
performed using JLM microscopic potentials in the eikonal approximation. Non 
eikonal corrections up to second order were included and the effects are 
shown by thin 
continuous lines (the convention is as per the previous two figures) 
for Be, C and Si targets. 
For the Ta target, the eikonal series 
does not converge in the range of energies shown. The experimental 
data for the total, elastic and reaction 
cross sections (open and closed circles and open squares) were 
taken from refs.~\protect\cite{Web,chad}}
\label{fig:jlm}
\end{figure}

\newpage

% fig  11 

\begin{figure}[!htbp]
\begin{center}
\mbox{\epsfig{file=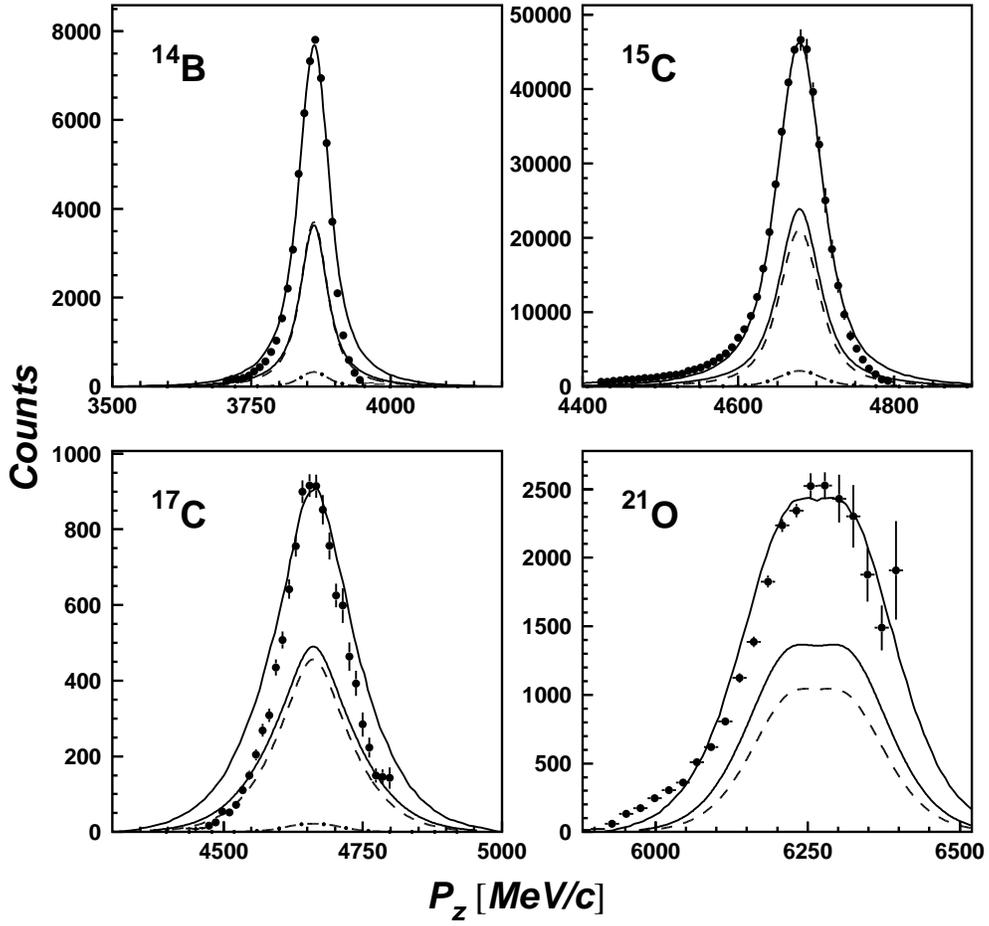,width=15cm}}
\end{center}
\caption{Selected examples (see text) of the core fragment longitudinal 
momentum 
distributions obtained using the Carbon target showing the contributions 
from 
the different reaction mechanisms.
 The calculated distributions (thick solid lines) include the 
absorption (thin solid lines), 
diffraction (dashed) and Coulomb (dash-dotted) components.}
\label{fig:exglau}
\end{figure}

\newpage

%fig 12 

\begin{figure}[h]
\begin{center}
\mbox{\epsfig{file=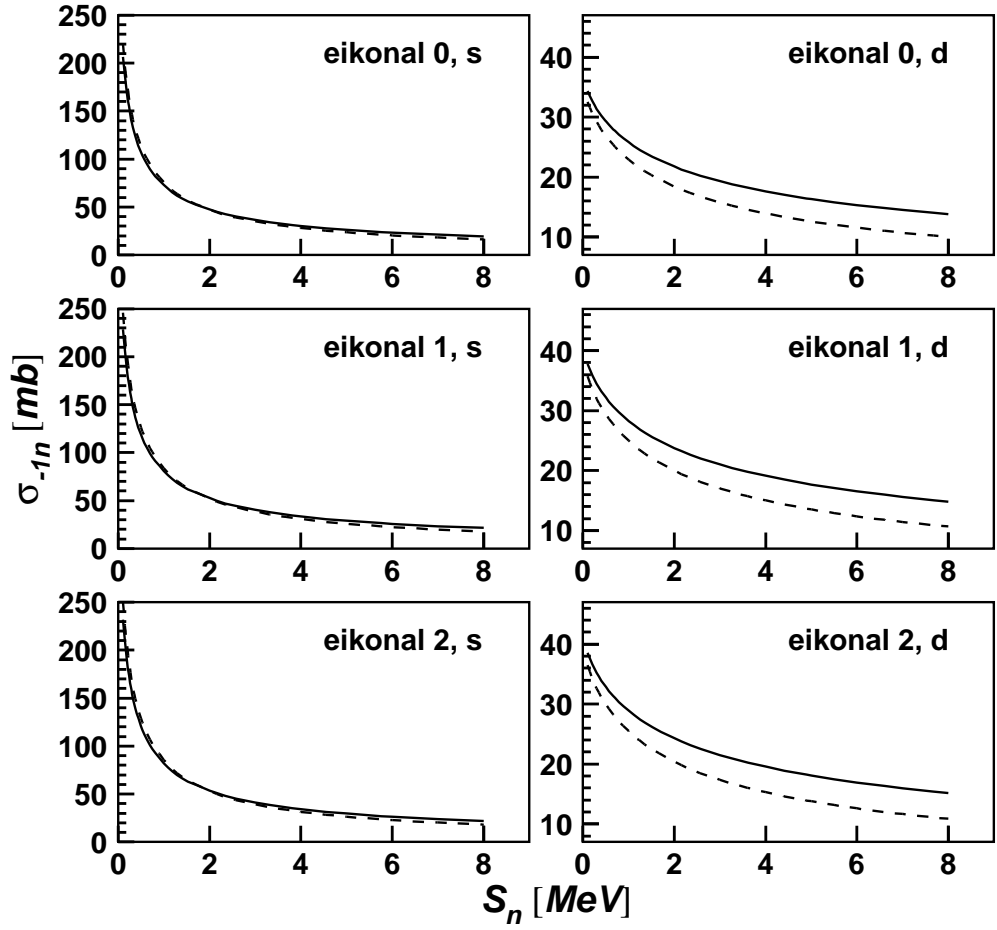,width=15cm}}
\end{center}
\caption{Calculated absorption (solid lines) and diffraction 
(dashed lines) cross sections versus binding energy for $s$ and $d$-wave 
states of an A=17 system at 50~MeV/nucleon on a Carbon target using the 
JLM interaction and various 
orders of eikonal theory.}
\label{fig:testeiko}
\end{figure}

\newpage

%fig 13 

\begin{figure}[h]
\begin{center}
\mbox{\epsfig{file=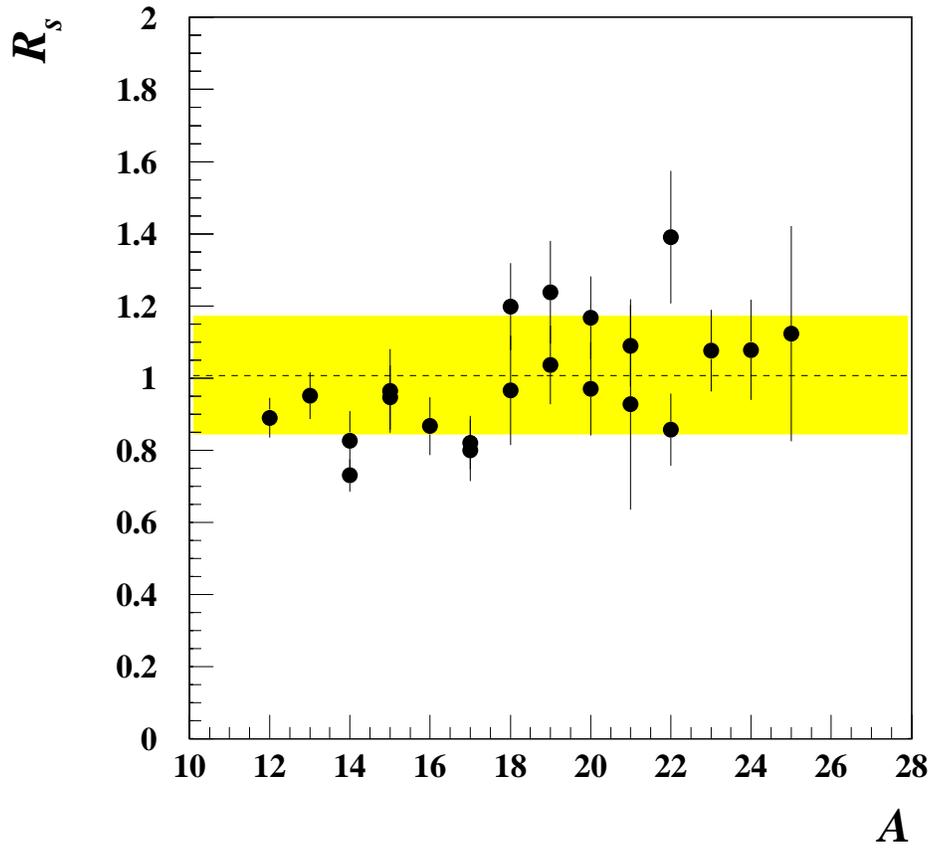,width=16cm}}
\end{center}
\caption{Ratio of experimental to theoretical cross sections ($R_s$) as a 
function of
projectile mass number for the data obtained with the Carbon target. The 
dotted 
line
indicates the mean value and the shadded band the 1$\sigma$ variance.}
\label{fig:rs_flo}
\end{figure}

\newpage

%fig 14 

\begin{figure}[h]
\begin{center}
\mbox{\epsfig{file=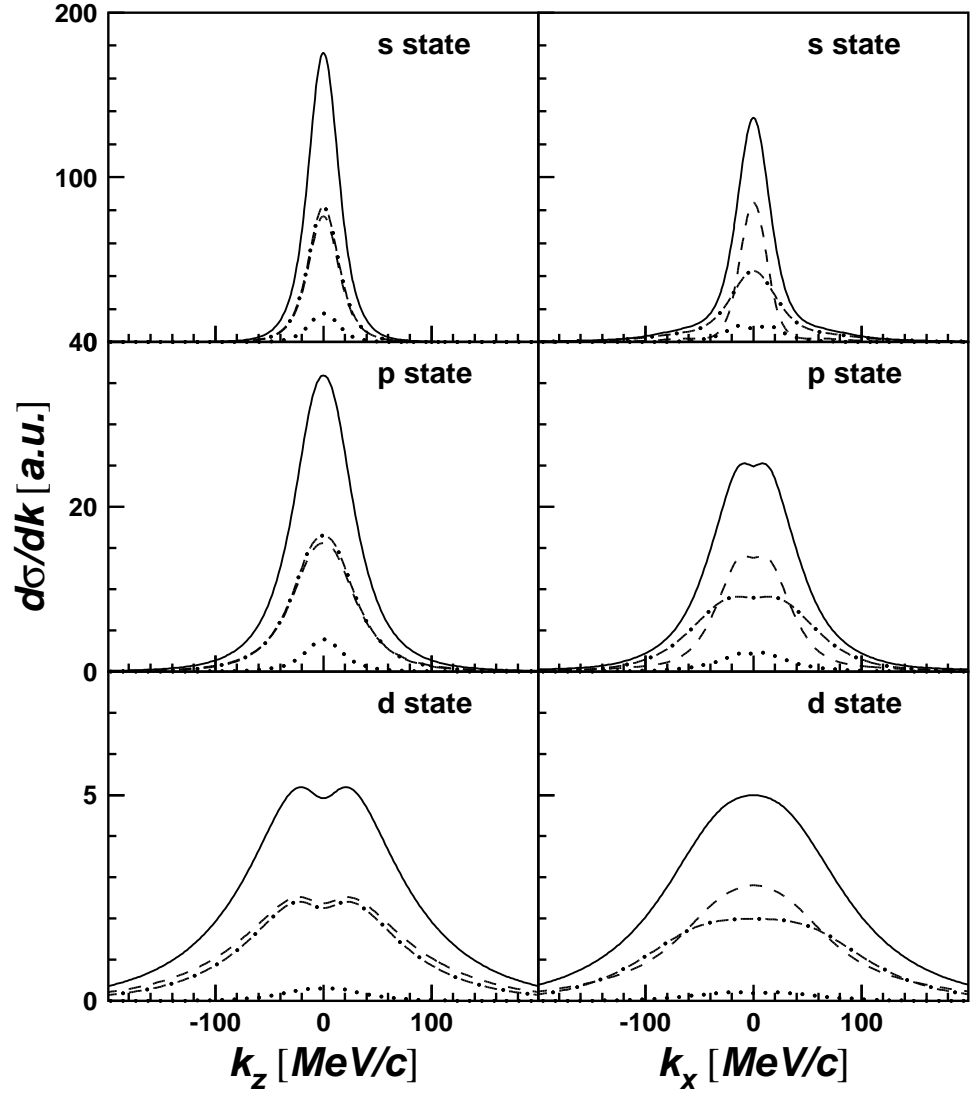,width=15cm}}
\end{center}
\caption{Test case calculations of the longitudinal ($k_z$) and transverse 
($k_x$) 
core fragment momentum distributions for one-neutron removal by a Carbon 
target at 50 MeV/nucleon
assuming an 
$s$, $p$ or $d$-wave state in an A=14 system with $S_n$= 1 MeV. 
The total (solid line), absorption (dashed), diffraction (dash-dotted)
and Coulomb (dotted) components are indicated.} 
\label{fig:testpxpz}
\end{figure}

\newpage

%fig 15 

\begin{figure}[h]
\begin{center}
\mbox{\epsfig{file=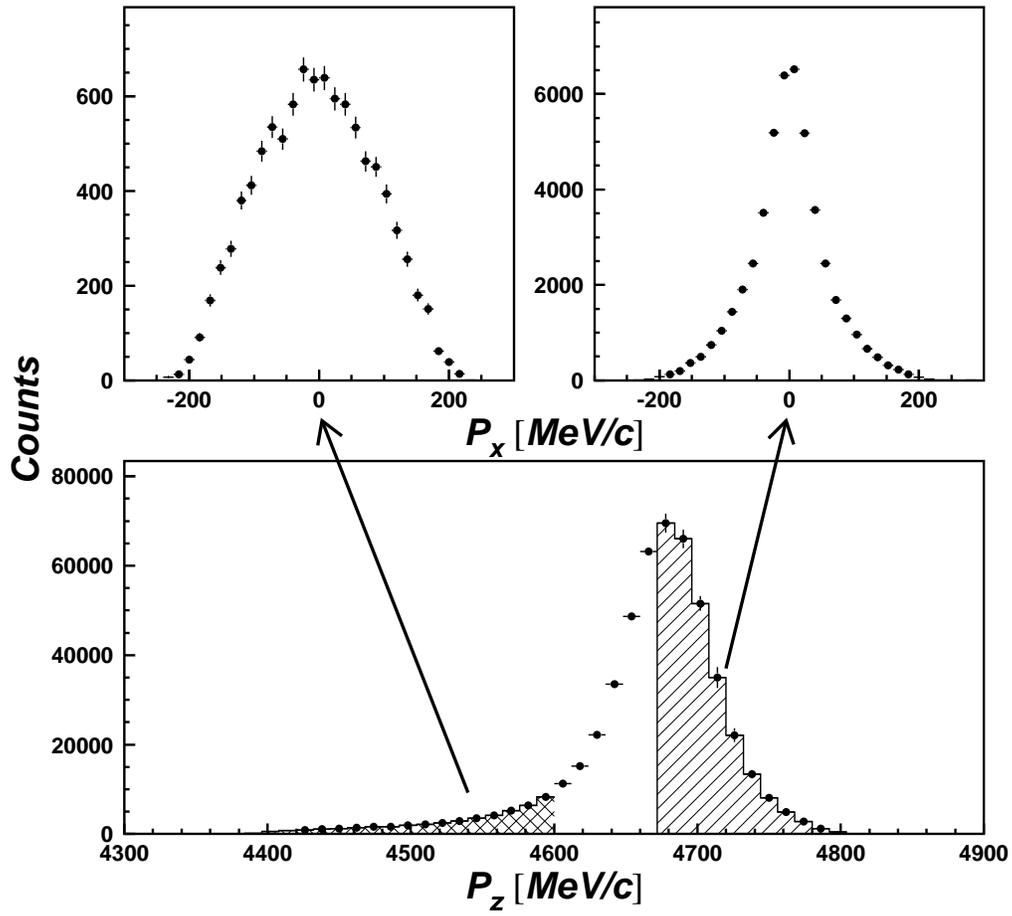,width=15cm}}
\end{center}
\caption{Correlations between the longitudinal and transverse $^{14}$C core 
fragment 
momenta for one-neutron removal from $^{15}$C by the Carbon target.}
\label{fig:trainepx}
\end{figure}

\newpage

%fig 16 

\begin{figure}[h]
\begin{center}
\mbox{\epsfig{file=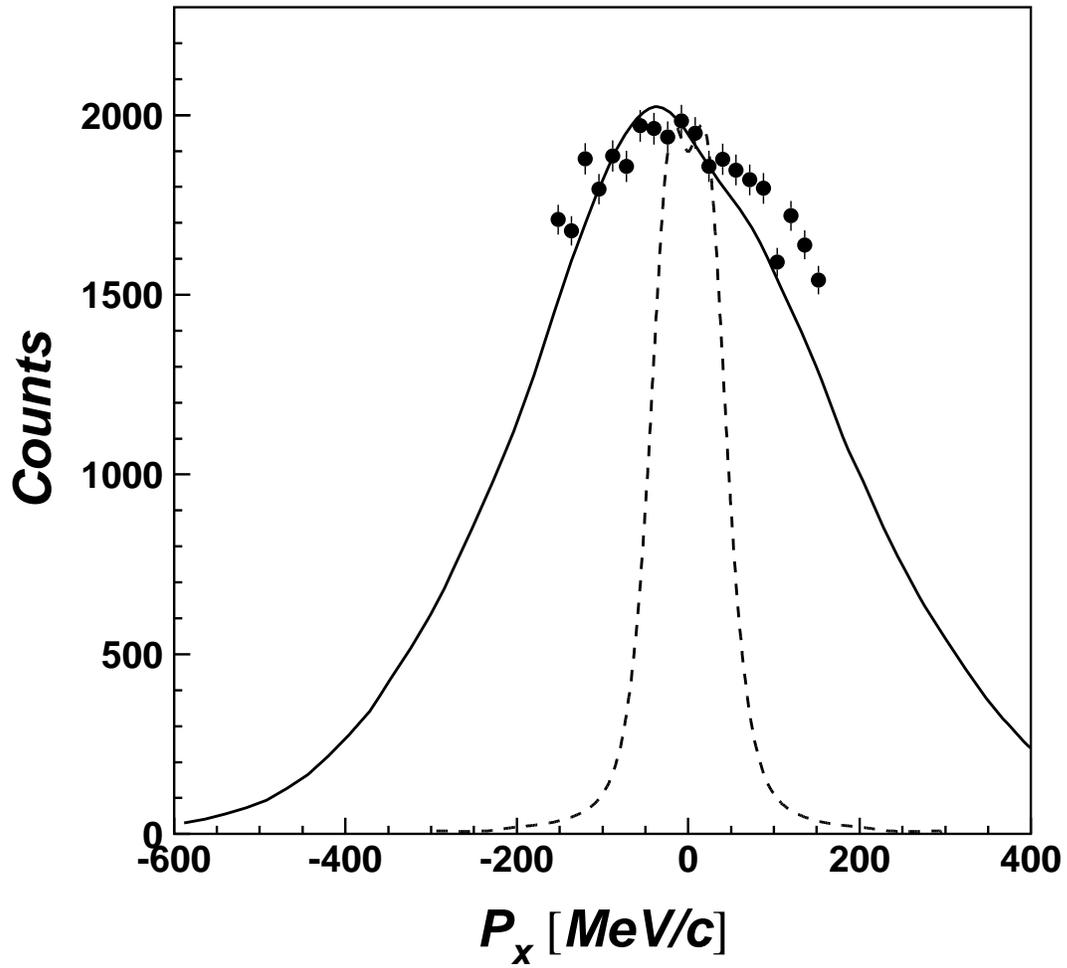,width=15cm}}
\end{center}
\caption{Core fragment transverse momentum distributions for one-neutron 
removal from
$^{15}$C by the Tantalum target (filled circles). Dashed line: calculated 
transverse momentum 
distribution for Coulomb induced breakup. Solid 
line: same 
calculation after convolution with the orbital deflection in the Coulomb 
field of the target (see text for details). }
\label{fig:px_ta}
\end{figure}

\newpage

%fig 17 

\begin{figure}[h]
\begin{center}
\mbox{\epsfig{file=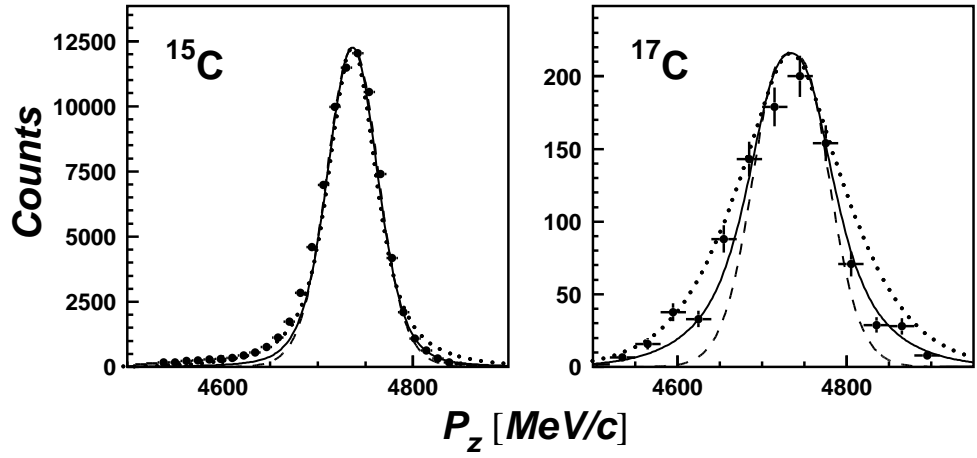,width=15cm}}
\end{center}
\caption{Examples of the core fragment longitudinal momentum distributions 
for 
reactions 
on the Ta target. The filled circles represent the data. 
To aid in the comparison the calculated distributions for the total (solid 
lines), 
nuclear (dotted) and Coulomb dissociation (dashed line) are normalised to 
the 
data.
}
\label{fig:excoul}
\end{figure}

\newpage

%fig 18 

\begin{figure}[!htbp]
\begin{center}
\mbox{\epsfig{file=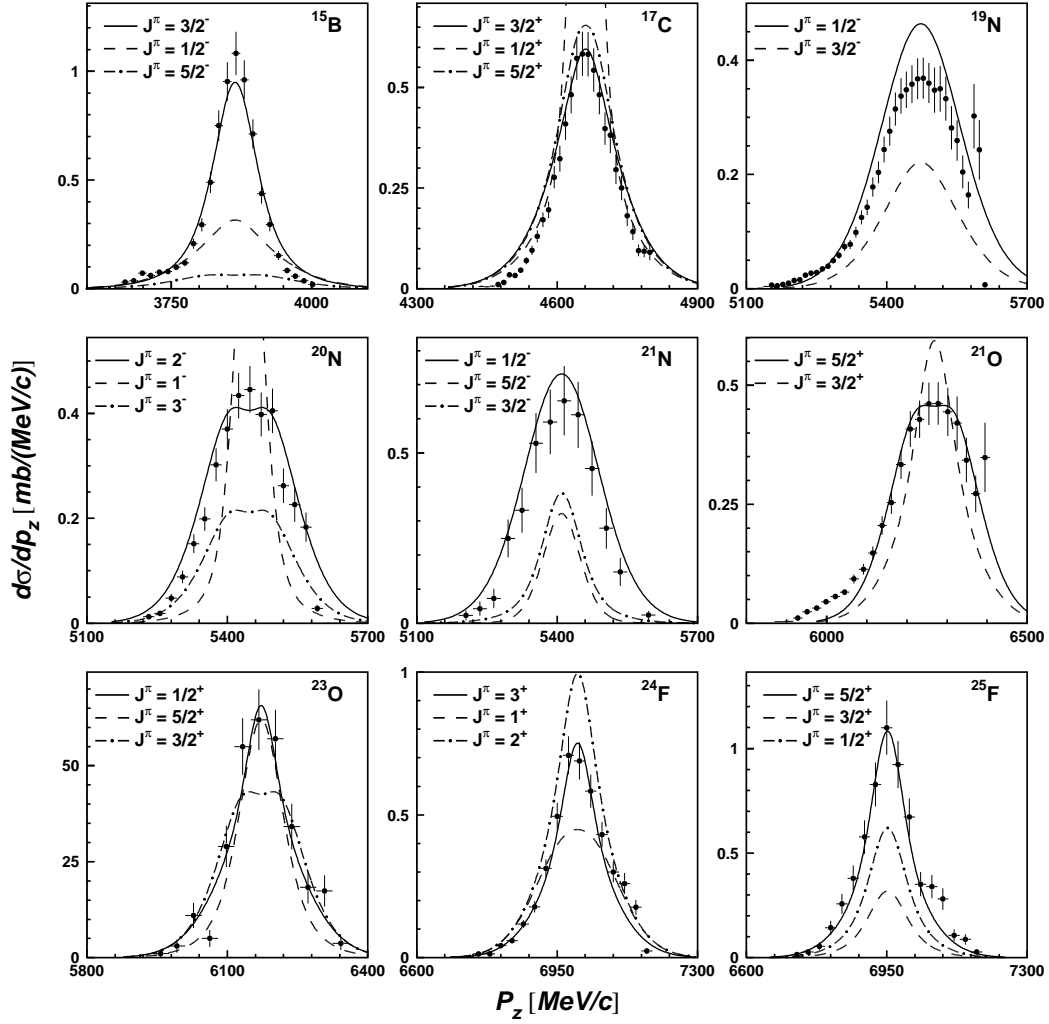,width=15cm}}
\end{center}
\caption{Comparison of the measured core fragment longitudinal momentum
distributions (from reactions on the Carbon target) with those predicted for 
different projectile ground state spin-parity
assignments. The favoured assignements
are displayed as solid lines (see text).}
\label{fig:spectro}
\end{figure}

\newpage

%%%%%%%%%%%%%%%%%%%%%%%%%%%%%%%%%%%%%%
%%% px distrib as a spectroscopic tool
%%%%%%%%%%%%%%%%%%%%%%%%%%%%%%%%%%%%%%

%fig 19 

\begin{figure}[!htbp]
\begin{center}
\mbox{\epsfig{file=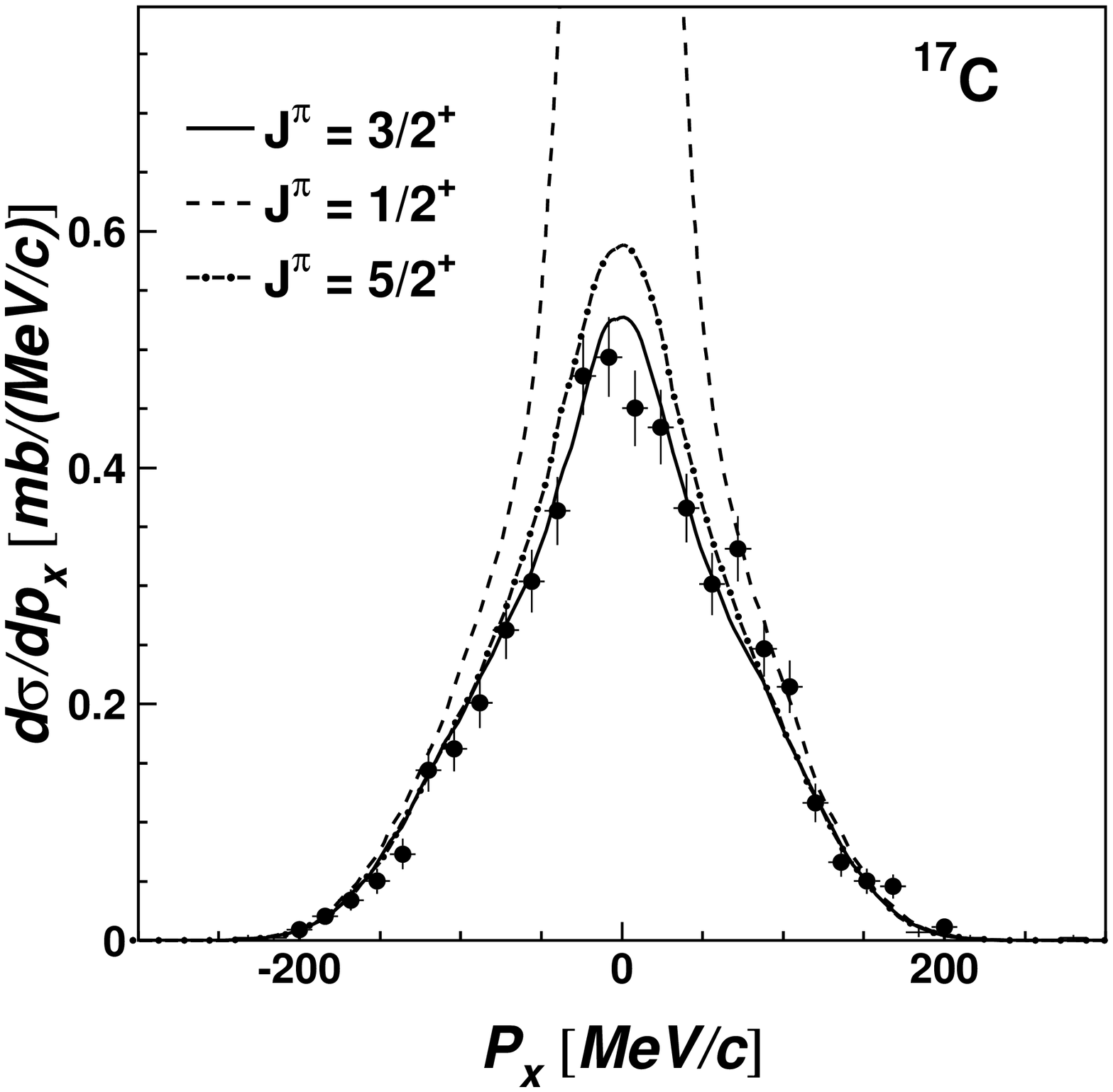,width=15cm}}
\end{center}
\caption{Comparison of the measured $^{16}$C core fragment transverse 
momentum 
distribution 
for reactions of $^{17}$C on the Carbon target with the distributions 
calculated 
for the three
possible ground state spin-parity assignments.}
\label{fig:pxtool}
\end{figure}

%%%%%%%%%%%%%%%%%% end one column
}


\begin{thebibliography}{99}

%1
\bibitem{sco80} D. K. Scott, Prog. in Part. Nucl. Phys. $\bf {4}$, 5 (1980)  
and references therein.
%2
\bibitem{sto84} R. G. Stokstad, Comments Nucl. Part. Phys. $\bf{13}$, 231 
(1984).
%3
\bibitem{gre75} D. E.~Greiner {\etal}, Phys. Rev. Lett. $\bf{35}$, 
152  
(1975).
%4
\bibitem{ols83} D. L. Olson {\etal}, Phys. Rev. $\bf{28}$, 1602 
(1983).

%5
\bibitem{gol74} A. S. Goldhaber, Phys. Lett. $\bf{B53}$, 306 (1974).
%6
\bibitem{Hufner} J. H\"{u}fner and M. C. Nemes, Phys. Rev. $\bf{C23}$, 2538 
(1981).
%7
\bibitem{fri83} W. A. Friedmann, Phys. Rev. $\bf{C27}$, 569 (1983).
%8
\bibitem{orr97} see, for example, N. A. Orr, Nucl. Phys. $\bf{A616}$, 155 
(1997) and
references therein.
%9
\bibitem{Serber} R. Serber, Phys. Rev. $\bf {72}$, 1008 (1947).

%%%%%%%%% Anne's papers
%10
\bibitem{ann93} R. Anne {\em et al.}, Phys. Lett. $\bf{B304}$, 55 (1993).
%11
\bibitem{ann94} R. Anne {\em et al.}, Nucl. Phys. $\bf{A575}$, 125 
(1994).

%%%%%%%%%% Orr's famous papers
%12
\bibitem{orr92} N. A. Orr {\em et al.}, Phys. Rev. Lett. $\bf{69}$, 2050 
(1992).
%13
\bibitem{orr95} N. A. Orr {\em et al.}, Phys. Rev. $\bf{C51}$, 3116 
(1995).
%!4
\bibitem{kel95} J. H. Kelley {\em et al.}, Phys. Rev. Lett. $\bf{74}$, 30 
(1995).

%%%%%%%%%% Reaction theory

%% always ignords Sagawa and Yazaki paper
\bibitem{Sag90} H. Sagawa and K. Yazaki, Phys. Lett. $\bf{B244}$, 149 
(1990).  

\bibitem{Sag94} H. Sagawa and N. Takigawa, Phys. Rev. $\bf{C50}$, 985 
(1994).

%15
\bibitem{han96} P. G. Hansen, Phys. Rev. Lett $\bf{77}$, 1016 (1996).    
%16
\bibitem{Esbensen} H. Esbensen, Phys. Rev. $\bf{C53}$, 2007 (1996); 
H. Esbensen, in Extremes of Nucl. Structure,
Proc. Int. Workshop XXIV, Hirschegg on Extremes of Nuclear Structure, 
eds H. Feldmeyer, J. Knoll, W. Norenberg (GSI, Darmstadt, 1996) p321.
%17
\bibitem{Hussein} M. S. Hussein and K. W. McVoy, Nucl. Phys.
$\bf{A445}$, 123 (1985).
%18
\bibitem{Hencken} K. Hencken, G. Bertsch and H. Esbensen, Phys. Rev. 
$\bf{C54}$, 3043 (1996).                                  
%19
\bibitem{oga97} Y.~Ogawa and I.~Tanihata, Nucl. Phys. $\bf{A616}$, 239c 
(1997).
%20
\bibitem{Nego} F. Negoita {\em et al.}, Phys. Rev. $\bf{C59}$, 2082 
(1999).
%21
\bibitem{Tostev} J. A. Tostevin, J. Phys. {\bf G}: Nucl. Part. Phys. 
$\bf{25}$, 735 (1999).
%22
\bibitem{Yabana} K. Yabana, Y. Ogawa and Y. Suzuki, Nucl. 
Phys. $\bf{A539}$, 295 (1992).

\bibitem{par00} Yu.L. Parfenova, M.V. Zhukov and J.S. Vaagen, 
Phys. Rev. $\bf{C62}$, 044602-1 (2000).

%%%%%%%%%%%% MSU gammas etc
%23
\bibitem{bonacc5} A. Bonaccorso, D. M. Brink, Phys. Rev. 
$\bf{C44}$, 1559 (1991).                                   

%24
\bibitem{Han00} P. G. Hansen and B. M. Sherrill, Nucl. Phys. $\bf{A693}$, 
133 
(2001).

%25
\bibitem{nav98} A. Navin {\em et al.}, Phys. Rev. Lett. $\bf{81}$, 5089 
(1998).

%26
\bibitem{aum00} T. Aumann {\em et al.}, Phys. Rev. Lett. $\bf{84}$, 35 
(2000). 
%27
\bibitem{gui00} V. Guimar\~{a}es {\em et al.}, Phys. Rev. 
$\bf{C61}$, 064609 (2000).
%28
\bibitem{nav00} A. Navin {\em et al.}, Phys. Rev. Lett. $\bf{85}$, 266 
(2000).
%29
\bibitem{mad01} V. Maddalena {\em et al.}, Phys. Rev. {\bf C62}, 024613 
(2001).
%30
\bibitem{Lola02} D. Cortina-Gil {\em et al.}, Phys. Lett. 
{\bf B529}, 36 (2002). 
% 31 bis 
\bibitem{enders02} J. Enders {\em et al.},  Phys. Rev. {\bf C65}, 034318 
(2002).    

\bibitem{Han04} P. G. Hansen and J. A. Tostevin, Ann. Rev. Nucl. Part. Sci. 
{\bf 53}, in press.
                                
\bibitem{enders03} J. Enders {\em et al.},  Phys. Rev. {\bf C67}, 064301 
(2003). 
                                   
%32
\bibitem{Tostev2} J. A. Tostevin, in Proc. of the 2$^{nd}$ International 
Conference on Fission and Neutron Rich Nuclei, ed. W. Philips,  (World 
Scientific) p. 735 (1999).
  
%33
\bibitem{bau98} T. Baumann {\em et al.}, Phys. Lett. $\bf{B439}$, 256 
(1998).


%%%%%%%%%%%% sauvan PLB      
%34
\bibitem{sau00} E. Sauvan {\em et al.}, Phys. Lett. $\bf{B491}$, 1 
(2000).    

%%%%%%%%%%%% Exp. method.
%35
\bibitem{ann97} R. Anne, Nucl. Inst. Meth. $\bf{B126}$, 279 (1997).
%36
\bibitem{bia89} L.  Bianchi {\em et al.}, Nucl. Inst. Meth. 
$\bf{A276}$, 509 (1989).

%%%%%%%%%%%%%%% transverse acceptances
%37
\bibitem{rii93} K. Riisager, in Proc. of the 3rd Int. Conference on 
Radioactive Nuclear Beams, ed. D. J. Morrissey (Editions Fronti\`eres, 
Gif-sur-Yvette, 1993) p281.
%38
\bibitem{riihab} K. Riisager, Habilitation Thesis, University of Aarhus 
(1994).
%39
\bibitem{kel97} J. H. Kelley {\em et al.}, Nucl. Inst. Meth. 
$\bf{A386}$,492 (1997).
%40
\bibitem{sauthese} E. Sauvan, Th\`ese, Universit\'e de Caen (2000), 
LPCC~T00-01.
%41
\bibitem{car00} F. Carstoiu, E. Sauvan and N. A. Orr, to be published.
%42
\bibitem{silk88} J. D. Silk {\em et al.}, Phys. Rev. $\bf C37$, 158 (1988).

%%%%%%%%%% dissipative core-target collisions
%%%%%%% experimental comparisons
%43
\bibitem{baz95} D. Bazin {\em et al.}, Phys. Rev. Lett. $\bf{74}$, 3569 
(1995).
%44
\bibitem{baz98} D. Bazin {\em et al.}, Phys. Rev. $\bf{C57}$, 2156 
(1998).
\bibitem{Yam03} T. Yamaguchi {\em et al.}, Nucl. Phys. {\bf A724}, 3 
(2003).
%45
\bibitem{Kan02} R.~Kanungo {\em et al.}, Phys. Rev. Lett. $\bf{88}$, 
142502 (2002).
%46
%\bibitem{bau99} T. Baumann {\em et al.}, in Experimental Nuclear Physics 
%in 
%Europe, Proc. ENPE99, Sevilla, June 1999 p.29., ed. by B. Rubio, M. Lozano, 
%W. 
%Gelletly.
%47
\bibitem{cor01} D. Cortina-Gil {\em et al.}, Eur. Phys. J. {\bf A10}, 49 
(2001).
%48
\bibitem{priv_lola} D. Cortina-Gil, priv. comm.
%49
\bibitem{Bertsch} G. Bertsch, H. Esbensen and A. Sustich, Phys. Rev. 
$\bf{C42}$, 758 (1990). 

%%%%%%%% px-ppar



%%%%%%%% theoretical part
%50
\bibitem{bonacc3} A. Bonaccorso and G. F. Bertsch, Phys. Rev. 
$\bf{C63}$, 044604 (2001). 
%51
\bibitem{Bertul} C. A. Bertulani and G. Baur, Nucl. Phys. $\bf{A480}$, 615 
(1988).
%52
\bibitem{Winth} A. Winther and K. Adler, Nucl. Phys. $\bf{A319}$, 518 
(1979).
%53
\bibitem{Bonacc} A. Bonaccorso and F. Carstoiu, Phys. Rev. 
$\bf{C61}$, 034605 (2000).
%54
\bibitem{Jeuken} J. P. Jeukenne, A. Lejeune and C. Mahaux, Phys. Rev.
$\bf{C16}$, 80 (1977). 

%55
\bibitem{Trache} L. Trache {\em et al.}, Phys. Rev. $\bf{C61}$, 024612 
(2000).
%56
\bibitem{Beiner} M. Beiner and R. J. Lombard, Ann. Phys. $\bf{86}$, 262 
(1974).

%%%%%%%%% matter radii %%%%%%
%57
\bibitem{liatard} E. Liatard {\em et al.}, Europhys. Lett. 
$\bf{13}$, 401 (1990).

%%%%%%%% matter radii calc.
%58
\bibitem{ren95} Zhongzhou Ren {\em et al.}, Phys. Rev. $\bf{C52}$, R20 
(1995).  
%59
\bibitem{ren96a} Zhongzhou Ren {\em et al.}, Nucl. Phys. $\bf{A605}$, 75 
(1996).  
%60
\bibitem{ren96b} Zhongzhou Ren {\em et al.}, J. Phys. {\bf G}: Nucl. 
Part. 
Phys. {\bf 22}, L1 (1996).  
%61
\bibitem{oza01} A. Ozawa {\em et al.}, Nucl. Phys. $\bf{A691}$, 599 
(2001).


%62
\bibitem{tan88} I. Tanihata {\em et al.}, Phys. Lett. $\bf{B206}$, 592 
(1988).
%63 
\bibitem{Wallace} J. S. Wallace, Ann. Phys. (N.Y.) $\bf{78}$, 190 (1973).
%64
%\bibitem{Carst} F. Carstoiu and R. J. Lombard, Phys. Rev. $\bf{C48}$, 830 
%(1993). 
%65
%\bibitem{Varner} R. L. Varner {\em et al.}, Phys. Rep. $\bf{201}$, 57 
%(1991).
%66
\bibitem{Web}http://www.nndc.bnl.gov 
%67
\bibitem{chad} M. B. Chadwick, L. J. Cox, P. G. Young and A. S. Mengooni,
Nucl. Sci. Eng. {\bf 123}, 17 (1996).

%%%%%% oxbash %%%%%%%%%%%
%68
\bibitem{oxbash} B. A. Brown, A. Etchegoyen, W. D. M. Rae, Report MSUCL-524 
(November 1988);
http://www.nscl.msu.edu/~brown/
%69
\bibitem{war92a} E. K.  Warburton, B. A. Brown, Phys. Rev. 
$\bf{C46}$, 923 (1992).
%70
\bibitem{war92b} E. K. Warburton, B. A. Brown, D. J. Millener, Phys. Lett. 
$\bf{B293}$, 7 (1992).
%71
\bibitem{mil75} D. J. Millener and D. Kurath, Nucl. Phys. $\bf{A255}$, 315 
(1975).
%72
\bibitem{as82} F. Ajzenberg-Selove, Nucl. Phys. $\bf{A375}$, 1 (1982).
%73
\bibitem{as86} F. Ajzenberg-Selove, Nucl. Phys. $\bf{A449}$, 1 (1986).
%74
\bibitem{as87} F. Ajzenberg-Selove, Nucl. Phys. $\bf{A475}$, 1 (1987).
%75
\bibitem{end90} P. M. Endt, Nucl. Phys. $\bf{A521}$, 1 (1990).
%76
\bibitem{mad01c} V. Maddalena and R. Shyam, Phys. Rev. $\bf{C63}$, 051601(R) 
(2001).
\bibitem{orrrnbiii}  N. A. Orr {\em et al.}, in Proc. of the 3rd Int. 
Conference on 
Radioactive Nuclear Beams, ed. D.J. Morrissey (Editions Fronti\`eres, 
Gif-sur-Yvette, 1993) p389.

%%%%%%%%% Jpi assignments exp.
%77
\bibitem{cat89} W. N. Catford {\em et al.}, Nucl. Phys. $\bf{A503}$, 263 
(1989). 
%78
\bibitem{orr89} N. A.~Orr {\em et al.}, Nucl. Phys. {\bf A491}, 457 
(1989).
%79
\bibitem{goo74} D. R.~Goosman Phys. Rev. {\bf 
C10}, 756 (1974).
%80
\bibitem{ree99} A. T. Reed {\em et al.},  Phys. Rev. $\bf{C60}$, 024311 
(1999). 

\bibitem{Oga02} H. Ogawa {\em et al.}, Eur. Phys. J. {\bf A13}, 81 (2002).

%81
\bibitem{mou81} J. Mougey {\em et al.}, Phys. Lett. $\bf{B105}$, 25 
(1981).
%82
\bibitem{JeffCDCC} J. Tostevin {\em et al.}, Phys. Rev. $\bf{C66}$, 024607
(2002). 
%83
\bibitem{oza00} A. Ozawa {\em et al.}, Phys. Rev. Lett. $\bf{84}$, 5493 
(2000).

%85 - superceded by JeffCDCC
%\bibitem{mad01b} V. Maddalena {\em et al.}, Nucl. Phys. $\bf{A682}$, 332c 
%(2001).                                                                     

%86

\bibitem{brown} B. A. Brown, P. G. Hansen, B. M. Sherrill and J. A. Tostevin
Phys. Rev. {\bf C65}, 061601(R) (2002).                                    


\bibitem{panda} V. R. Pandharipande, I. Sick and P. K. A. deWitt Huberts,
Rev. Mod. Phys. $\bf{69}$, 981 (1997).

\bibitem{esb01} H. Esbensen and G. Bertsch, Phys. Rev. $\bf{C64}$, 014608
(2001). 

\bibitem{Bro03} B. A. Brown {\em et al.}, Phys. Rev. Lett. $\bf{90}$, 159201 
(2003).

\bibitem{Kan03} R. Kanungo {\em et al.}, Phys. Rev. Lett. $\bf{90}$, 159202 
(2003).

\bibitem{Dat03} U. Datta Pramanik {\em et al.}, Phys. Lett. {\bf B551}, 63 
(2003).

\bibitem{Sat80} see, for example, G.R. Satchler, Direct Nuclear Reactions (Oxford University 
Press, Oxford, 1983). 

\bibitem{Kra01} G. J. Kramers, H. P. Blok and L. Lapik\'as, 
Nucl. Phys. $\bf{A679}$, 267 (2001) and refs therein.                                                                  

\bibitem{Win01} J. S. Winfield  {\em et al.}, 
Nucl. Phys. $\bf{A683}$, 48 (2001).                 


%%%%%%%%%%%%%%%%%%%%%%%%%%%%%%%%%%%%%%%%%%%%%%%%%%%%%%%%%%%%%%%%%%%%%%


\end{thebibliography}
\end{document}